\documentclass[journal]{IEEEtran}

\usepackage{xcolor} 

\usepackage{cite} 

\usepackage[pdftex]{graphicx}
\graphicspath{{../figures/}}
\DeclareGraphicsExtensions{.pdf} 

\usepackage{tikz}       
\usepackage{pgfplots}   
\pgfplotsset{compat=1.15} 

\usepackage{amsmath}
\usepackage{amsthm}
\usepackage{amsfonts}
\usepackage{amssymb}

\usepackage{algorithm}
\usepackage{algpseudocode}

\usepackage{array} 

\usepackage[caption=false,font=footnotesize]{subfig}

\usepackage{stfloats} 

\usepackage{url} 

\makeatletter
    \@ifpackageloaded{xcolor}{}{\usepackage{xcolor}}
\makeatother
\makeatletter
    \@ifpackageloaded{needspace}{}{\usepackage{needspace}}
\makeatother

\newcommand{\titledParagraph}[1]{\medskip\noindent {\bf #1}}

\newcommand{\uppercaseGreek}[1]{%
    \begingroup
    \ucmathlist\MakeUppercase{#1}%
    \endgroup
}

\newcommand{\ucmathlist}{%
    \def\alpha{\mathrm{A}}%
    \def\beta{\mathrm{B}}%
    \let\gamma=\Gamma
    \let\delta=\Delta
    \def\epsilon{\mathrm{E}}%
    \def\varepsilon{\mathrm{E}}%
    \def\zeta{\mathrm{Z}}%
    \def\eta{\mathrm{H}}%
    \let\theta=\Theta
    \let\vartheta=\Theta
    \def\iota{\mathrm{I}}%
    \def\kappa{\mathrm{K}}%
    \let\lambda=\Lambda
    \def\mu{\mathrm{M}}%
    \def\nu{\mathrm{N}}%
    \let\xi=\Xi
    \let\pi=\Pi
    \let\varpi=\Pi
    \def\rho{\mathrm{P}}%
    \def\varrho{\mathrm{P}}%
    \let\sigma=\Sigma
    \def\tau{\mathrm{T}}%
    \let\upsilon=\Upsilon
    \let\phi=\Phi
    \let\varphi=\Phi
    \def\chi{\mathrm{X}}%
    \let\psi=\Psi
    \let\omega=\Omega
}

\makeatletter
    \@ifpackageloaded{amsthm}{}{

        \usepackage{amsthm}
    }
\makeatother
\theoremstyle{plain}
    \newtheorem{theorem}{Theorem}
    \newtheorem{proposition}[theorem]{Proposition}

\theoremstyle{definition}

\makeatletter
\def\renewtheorem#1{%
    \expandafter\let\csname#1\endcsname\relax
    \expandafter\let\csname c@#1\endcsname\relax
    \gdef\renewtheorem@envname{#1}
    \renewtheorem@secpar
}
\def\renewtheorem@secpar{\@ifnextchar[{\renewtheorem@numberedlike}{\renewtheorem@nonumberedlike}}
\def\renewtheorem@numberedlike[#1]#2{\newtheorem{\renewtheorem@envname}[#1]{#2}}
\def\renewtheorem@nonumberedlike#1{
    \def\renewtheorem@caption{#1}
    \edef\renewtheorem@nowithin{\noexpand\newtheorem{\renewtheorem@envname}{\renewtheorem@caption}}
    \renewtheorem@thirdpar
}
\def\renewtheorem@thirdpar{\@ifnextchar[{\renewtheorem@within}{\renewtheorem@nowithin}}
\def\renewtheorem@within[#1]{\renewtheorem@nowithin[#1]}
\makeatother
\input{auxFiles/symbolDef.sty}
\input{auxFiles/riceColors.sty}
\def\myfactor{0.2}
\input{auxFiles/auxTikzDefs.sty}

\usepackage[framemethod=tikz]{mdframed}
\newenvironment{SPMbox}[1]
{\begin{figure*}
    \vspace{-.3in}
    \fboxsep=3mm
    \fboxrule=2pt
    \begin{tcolorbox}[width=\textwidth, colback = lightgray, colframe = burgundy, arc = 0pt, outer arc = 0pt, boxrule = 1pt]
        \begin{minipage}{6.75in}
            \textbf{#1.}
}
{       \end{minipage}
    \end{tcolorbox}
    \vspace{-.2in}
    \end{figure*}
}
\newenvironment{defBox}[1]
{
    \begin{mdframed}[hidealllines=false,backgroundcolor=lightgray,linecolor=riceblue]\textbf{#1.}
}
{
    \end{mdframed}
}
\usepackage[font=footnotesize]{caption}
\usepackage{tcolorbox}
\usepackage{tabularx}
\usepackage{framed,enumitem}
\usepackage{balance}

\theoremstyle{definition}
\newtheorem{property}{Property}
\newtcolorbox{mybox}[1]{colback=gray!20!white,colframe=red!50!black,fonttitle=\bfseries,title=#1}

\begin{document}



\title{Graph Filters for Signal Processing\\and Machine Learning on Graphs}


\author{Elvin~Isufi, %
    Fer\hspace{0.015cm}nando~Gama,~%
    David~I~Shuman,~%
    and~Santiago~Segarra
    \\ \vskip.25cm\emph{\small{Overview Article}} \vspace{-.75cm}
\thanks{E. Isufi is with the Faculty of Electrical Engineering, Mathematics and Computer Science, Delft University of Technology, Delft, The Netherlands. Email: e-isufi.1@tudelft.nl.}
\thanks{F. Gama was with the Department of Electrical and Computer Engineering, Rice University, Houston, TX 77005 USA. Email: fgama@ieee.org}
\thanks{D. I Shuman is with Franklin W. Olin College of Engineering, Needham, MA 02492 USA. Email: dshuman@olin.edu}
\thanks{S. Segarra is with the Department of Electrical and Computer Engineering, Rice University, Houston, TX 77005 USA. Email: segarra@rice.edu}}



\maketitle


\begin{abstract}
Filters are fundamental in extracting information from data. For time series and image data that reside on Euclidean domains, filters are the crux of many signal processing and machine learning techniques, including convolutional neural networks. Increasingly, modern data also reside on networks and other irregular domains whose structure is better captured by a graph. To process and learn from such 
data, graph filters account for the structure of the underlying data domain.
In this article, we provide a comprehensive overview of graph filters, including the different filtering categories, design strategies for each type, and trade-offs between different types of graph filters.
%
We discuss how to extend graph filters into filter banks and graph neural networks to enhance the representational power; that is, to model a broader variety of signal classes, data patterns, and relationships. We also showcase the fundamental role of graph filters in signal processing and machine learning applications. {Our aim is that this article 
provides a unifying framework for both beginner and experienced researchers, as well as a common understanding that promotes collaborations at the intersections of signal processing, machine learning, and application domains.}

    %
\end{abstract}



%
\IEEEpeerreviewmaketitle


\section{Introduction} \label{sec:intro}



%
%
%
%


Filters are information processing architectures that preserve only the relevant content of the input for the task at hand.
In signal processing (SP), {filtering preserves specific spectral content of input signals and is a common building block in domains including }audio, speech, radar, communication, and multimedia \cite{oppenheim2001discrete}.
In machine learning (ML), filtering is used to extract relevant patterns from the data \, or as an inductive bias for building neural networks\cite{goodfellow2016deep}. For instance, principal component analysis (PCA) can be seen as a low-pass filter in the correlation matrix, where only the parts of the data contributing to the directions of the largest variance are preserved~\cite{marques2017stationary}.
Likewise, the success of convolutional neural networks (CNNs) can 
be attributed to the convolutional filters used in each layer, allowing for easier training and scalability, as well as exploiting structural invariances in the data \cite{goodfellow2016deep,bronstein2021geometric}.

{Conventional filtering applies to signals defined on Euclidean domains, but cannot be directly applied to irregular data structures arising in 
biological, financial, social, economic, power, water, sensor, and multi-agent networks, 
among others \cite{ortega2018graph,dong2020graph}.}
Graph filters are information processing architectures tailored to graph-structured data, generalizing the conventional Euclidean counterparts. 

Graph filters have many similarities with conventional ones; they are linear, shift invariant, parametric functions of the input, they enjoy a spectral interpretation via the spectral graph theory \cite{chung1997spectral}, and their spectral design boils down to function fitting \cite{Segarra2017-GraphFilterDesign,shuman2018distributed}. However, striking differences also arise from the new graph medium; e.g., graph filters are equivariant to permutations in the support, can be implemented distributively, and can have more generalized forms such as node varying \cite{Segarra2017-GraphFilterDesign} or edge varying \cite{Coutino2019-EdgeVariant}.
Due to the wide variety of network-based data and the flexibility of graphs to represent irregular structures, graph filters are used in myriad
SP tasks (signal reconstruction, anomaly detection, image processing, distributed processing) and ML tasks (semi-supervised and unsupervised learning, matrix completion, Gaussian process regression), as well as robotics, point clouds, Internet of Things, biology, and vision applications.

Early formalisms of graph filters find their roots in the 1990's in mesh processing \cite{taubin1995signal,taubin1996optimal}.
In the 2000's, graph filters were used in ML applications, mostly as graph kernels \cite{smola2003kernels, zhou2004regularization}, and later on in graph-based image processing \cite{bougleux2007discrete,zhang2008graph}.
From a SP viewpoint, graph wavelets \cite{crovella2003graph, coifman2006diffusion} can be considered as the first instances. 
The tutorial article \cite{shuman2013emerging} helped provide a unifying mathematical framework for many of the problem-specific efforts that were being carried out in different research communities, encouraging more signal processing researchers to readily dive into problems involving network-based data and develop new filter methods derived from first principles, inspired from the more familiar Euclidean setting.
Simultaneously, a specialization of the algebraic signal processing framework \cite{puschel2008algebraic} to the graph domain paved the way for a structured mathematical framework of graph filtering \cite{sandryhaila2013discrete,sandryhaila2014discrete}.
More recently, with the advent of graph neural networks (GNNs), graph filters play a fundamental role as the key component to learn representations from graph-based data \cite{gama2020graphs, ma2021unified, zhu2021interpreting, ruiz2021graph, isufi2021edgenets}. 

Despite the early roots of graph filtering, and its span across different applications -- often developed in an interdisciplinary fashion -- there is no comprehensive, point-of-entry reference for new researchers interested in either conducting fundamental research or exploring applications related to SP and ML, or both.
{This article has been designed to target this need, thus providing an extensive, principled overview of the fundamental aspects of graph filtering research, as well as highlighting the main application areas in both SP and ML.} 
A number of valuable tutorials and overviews on graph signal processing (GSP) and GNNs that are worth discussing have been written since \cite{shuman2013emerging}.
The survey in \cite{ortega2018graph} and book \cite{ortega2022introduction} provide excellent starting points on graph convolutional filtering and its links to the broader field of GSP.
Since the focus of these works is on the fundamentals of GSP, they only cover a single type of graph filter and only scratch the surface on design methods and applications.
The book chapter \cite{tremblay2018design} provides more detail on filter design strategies and their role in filter banks but leaves out a large portion of other filtering methods and their applications.
Ref. \cite{shuman2020Localized}
focuses specifically on how a single-level graph
convolutional filter bank can be used to build dictionaries of atoms for linear wavelet and vertex-frequency transforms.
The recent tutorial \cite{gama2020graphs} discusses the role of graph filters in building GNNs, whereas \cite{isufi2021edgenets} shows that different filtering solutions lead to fundamentally different GNN architectures, evidencing trade-offs in terms of inductive and transductive learning, locality, and representation capacity. Nevertheless, these two works focus explicitly on the role of graph filters in understanding GNNs.
A recent survey on the application of GSP tools and GNNs in machine learning can be found in \cite{dong2020graph}, and another one discussing the applicability of a particular class of graph filters in \cite{ramakrishna2020user}.
Both works review the applications where such tools can be used, but do not detail 
the particular contribution of the filters, nor the implications of the choice of graph filter type. 
Here, we aim to introduce and compare in detail the different filter types, and subsequently show their role in staple SP and ML applications.

In short, the above works discuss particular forms or properties of graph filters linked to specific case studies or applications.
None of them, however, offers a comprehensive and unifying treatment of the role of graph filtering that showcases the design choices, trade-offs, and applications of graph-based data in both SP and ML .
Our goal in this paper is to give the reader a detailed tour on the different graph filtering forms and their properties, and 
to show how they can be used to develop more expressive solutions via filter banks and neural networks.
We also aim to provide details on filter design and learning strategies.
Another key contribution of this unifying framework is to show that graph filters are crucial in myriad applications in both SP and ML.
While these two fields remain somewhat different in their approach to solving problems, a comprehensive understanding of filtering as the fundamental tool in both fields 
{promotes collaboration and facilitates new research developments.} 


\subsection{Organization}\label{subsec:organiz}

\begin{figure*}[!t]
    \centering\resizebox{\textwidth}{!}{
    \input{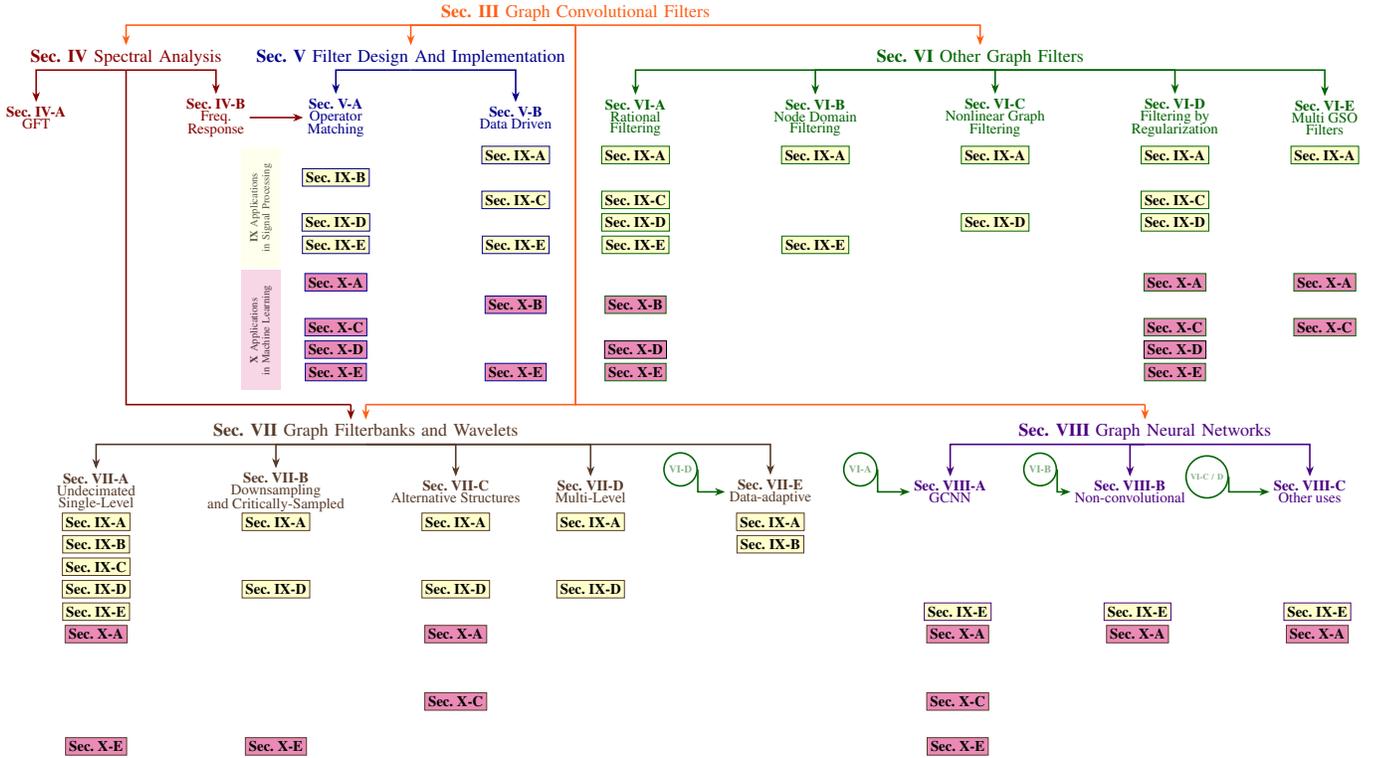}}
    \caption{{A roadmap of this article. Solid arrows prerequisite relationships between sections. For example, Sec. \ref{sec:wavelets} can be mostly understood without reading Sec. \ref{sec:other}, but not without reading Sec. \ref{sec:filt_spect}. The boxes for applications in signal processing and machine learning correspond to specific application examples we discuss in this article, and are not meant to be a comprehensive representation of all work that has been done in the field.} }
    \label{Fig:gcf_taxomony}
    \vspace{-.2in}
\end{figure*}

Sec.~\ref{sec:GSP} sets up the basic concepts about graphs, signals and embeddings, including also a list of landmark applications encountered in SP and ML. 
Sec.~\ref{sec:graphConv} introduces the central piece of this paper, the graph convolutional filter (GCF), which is viewed as a shift-and-sum operation of graph signals with respect to any graph representation matrix (see Figs.~\ref{fig:graphConv} and \ref{fig:directedCycle}), thus generalizing the principle of convolution for discrete-time signals.
Here, we also discuss several properties of this filter from a vertex perspective, such as its invariances and distributed implementation.
In Sec.~\ref{sec:filt_spect}, we characterize the spectral response of the GCF using a notion of graph Fourier transform.
We then discuss the spectral properties of GCFs, including the generalization of the convolution theorem to the graph setting.
Sec.~\ref{sec:filterDesign} is dedicated to filter design and identification strategies, either when a user-specific operator (e.g., a low-pass filter) is given or when the design is based on input-output data pairs.
In Sec.~\ref{sec:other}, we discuss other forms of graph filters and their link with the GCF. Table~\ref{tab:filters} provides an overview of the different architectures, their properties, and a discussion about the advantages and limitations of each form.
Sec.~\ref{sec:wavelets} is dedicated to building graph filter banks and graph wavelet transforms. 
We review different structures for graph filter banks (see, e.g., Figs. \ref{Fig:filter_bank} and \ref{Fig:filter_bank2}) and examine important design considerations.
Then, in Sec.~\ref{sec:GNN}, we show how the popular graph convolutional neural networks (GCNNs) can be seen as no more than a nonlinear graph filter built by nesting a GCF into an activation function (see Fig.~\ref{fig:GNN}).
We also discuss here how to build a non-convolutional GNN by changing the GCF with another filter type from Table~\ref{tab:filters}.
The next two sections are dedicated to staple applications of graph filters in SP (Sec.~\ref{sec:SP}) -- signal interpolation, anomaly detection, image processing, and distributed signal processing -- and in ML (Sec.~\ref{sec:ML}) -- semi-supervised and unsupervised learning on graphs, matrix completion, Gaussian processes, point clouds, and computer vision. 
{Fig.~\ref{Fig:gcf_taxomony} illustrates how the filtering concepts in Sections~\ref{sec:graphConv}-\ref{sec:other} relate to the filterbanks and GNNs, and how each of them has been used in select applications.}
{In Sec. \ref{sec:Start}, we provide some suggestions for researchers and developers new to the area, including a suggested order in which to explore the graph filtering tools.} 
Finally, in Sec. \ref{sec:Future}, we highlight some future directions. 


%

\section{Graphs and Signals} \label{sec:GSP}

This section first reviews ways to construct graphs and the basic terminology for representing them (Sec. \ref{subsec:graphs}). This will bridge the high-level discussion of the previous section with the more detailed mathematical formulations of graph filters in the succeeding sections. Next, it highlights some landmark tasks where graph filters are used (Sec.~\ref{subsec:signals}).

\subsection{Graph Terminology}\label{subsec:graphs}

We denote a weighted graph by $\graph = (\stV,\stE,\fnW)$, where $\stV = \{1, \ldots, N\}$ is the set of nodes, $\stE \subseteq \stV\times\stV$ is the set of edges such that  $(i,j) \in \stE$ if and only if there is an edge from node $i$ to node $j$, and $\fnW: \stE \to \fdR_+$ is a weight function. If all edge weights equal one, the graph is said to be \emph{unweighted}.
A graph is \emph{undirected} if there is no orientation in the edges in $\stE$. For an undirected graph, we denote the neighboring set for a node $i$ by $\stN_i = \{j\in\stV:(i,j)\in\stE	\}$.  
Instead, in a directed graph (or digraph), an edge $(i,j) \in \stE$ has an orientation starting from node $i$ and ending at node $j$.
We say that node $j$ is an out-neighbor of $i$ (and $i$ an in-neighbor of $j$).
The out-neighboring set of node $i$ is denoted by $\stN_i^\text{out} = \{j\in \stV: (i,j) \in \stE\}$ and, likewise, the in-neighboring set by $\stN_i^{\text{in}} = \{j \in \stV: (j,i) \in \stE\}$.

We represent graph $\graph$ via the weighted adjacency matrix $\mtA$, which is an $N \times N$ sparse matrix with nonzero elements $[\mtA]_{ji} = a_{ji} > 0$ representing the strength of edge $(i,j) \in \stE$. Matrix $\mtA$ is symmetric ($a_{ij} = a_{ji}$ for all $i,j$) for an undirected graph, but may be asymmetric 
for a directed one. For undirected graphs, another widely used matrix is the graph Laplacian $\mtL = \mtD - \mtA$, where the diagonal matrix $\mtD = \diag(\mtA\vcOnes)$ has as $i$th diagonal element the sum of all edge weights incident to node $i$.


\begin{defBox}{Graph shift operator (GSO)}
We represent the structure of a graph $\stG$ with a generic matrix $\mtS\in \fdR^{N\times N}$ called the graph shift operator matrix.
The only requirement for $\mtS$ to be a valid GSO is that
$$[\mtS]_{ji} = s_{ji} = 0~~\text{whenever}~~(i,j)\notin\stE~~\text{for}~~i \neq j.$$\vskip-1cm
\end{defBox}

\noindent Both matrices $\mtA$ and $\mtL$ are special cases for $\mtS$ \cite{sandryhaila2013discrete, ortega2018graph}. Other examples include the normalized adjacency matrix $\mtA_{\text{n}} = \mtD^{-1/2}\mtA\mtD^{-1/2}$, the normalized Laplacian matrix $\mtL_{\text{n}} = \mtD^{-1/2}\mtL\mtD^{-1/2}$, and the random walk Laplacian $\mtL_{\text{rw}} = \mtD^{-1}\mtL$. 

Graphs and 
associated GSOs can represent:


\begin{enumerate}
\item \emph{Physical networks:} Here, nodes and edges physically exist. For example, in a sensor network, nodes are sensors and edges are communication links \cite{jablonski2017graph}.
A directed edge indicates the communication direction and the edge weight captures the communication channel properties.
Other examples include: (i)~multi-agent robot systems where nodes are robots and edges are communication channels \cite{gama2022controlGNN}; (ii)~power networks where nodes are buses and edges are power lines \cite{ramakrishna2021grid};
(iii) telecommunication networks where nodes are transceivers and edges are channels \cite{eisen2020wirelessEGNN, chowdhury2021uwmmse}; (iv)~water networks where nodes are junctions and edges are pipes \cite{candelieri2014graph}, and; (v)~road networks where nodes are intersections and edges are roads \cite{jain2014big}.

%
\item \emph{Abstract networks:} These graphs typically represent dependencies between the data points. Consider $N$ data points, each described by a feature vector $\vcf_i \in \fdR^F$, and let $\mathrm{dist}(\vcf_i, \vcf_j)$ be a distance measure (e.g., Euclidean) between data points $i$ and $j$. Each data point is considered as a node and two data points could be connected based on \cite{von2007tutorial}: (i) \emph{$\varepsilon-$neihborhood}, where the edge weight is
\begin{align}\label{eq.graph}
a_{ij} \!=\! \left\{\!\!\!
                \begin{array}{ll}
                  \fnf(\mathrm{dist}(\vcf_i, \vcf_j); \theta)~&\!\!\text{if}~\mathrm{dist}(\vcf_i, \vcf_j) \le \varepsilon,\\
                  0 ~&\text{otherwise},
                \end{array}
              \right.
\end{align}
where $\fnf(\cdot; \theta)$ is a parametric function \big(e.g., a Gaussian kernel $\fnf(\mathrm{dist}(\vcf_i, \vcf_j); \theta) = \exp\left(- {\mathrm{dist}(\vcf_i, \vcf_j)}/{2\theta^2}\right)$\big) and $\varepsilon>0$ is a constant controlling the edge sparsity; (ii) \emph{$k-$nearest neighbor}, where each node is connected only to the $k$ closest data points with respect to $\fnf(\mathrm{dist}(\vcf_i, \vcf_j); \theta)$, which can again be a Gaussian kernel or a Pearson correlation. {The covariance matrix or modifications thereof have also been used to build abstract networks as we elaborate on in Sec.~\ref{subsec_gti}; see also \cite{giannakis2018topology,mateos2019connecting}.}


\smallskip
\smallskip
The above approaches build undirected abstract networks but alternatives for directed or causal dependencies are also possible; see \cite{giannakis2018topology,mateos2019connecting,dong2019learning}. These abstract networks are useful, for example, in: (i) recommender systems, where two items are connected, e.g., if their Pearson correlation is greater than some value \cite{huang2018rating};  (ii) brain networks, where the nodes are brain regions and the edges are given, e.g., by the cross-correlation or mutual information of electroencephalography (EEG) time series in the different regions~\cite{sporns2016networks}; (iii) social networks, where nodes are users and edge weights may represent the number of interactions between them; (iv) economic networks, where nodes are different economic sectors and the edges may represent the input and output production going from one sector to another  \cite{jackson2010social}. 

\end{enumerate}

Since abstract networks represent dependencies between datapoints, they can be manipulated by recomputing edge weights, clustering, or pruning to facilitate representation.
However, this is not typically the case for physical networks, as they often represent the medium with respect to which processing is performed. We shall see in Sec.~\ref{subsec_distSP} that graph filters can leverage such structure for distributed processing.


\subsection{Signals Defined on Graphs { and Common Processing Tasks}}\label{subsec:signals}

{We often encounter data that can be represented as a signal or set of features, with one value associated to each node.}


\begin{defBox}{Graph signal}
A graph signal $\fnx$ is a function from the node set to the field of real numbers; i.e., $\fnx: \stV \to \fdR$. We can represent a graph signal as a vector $\vcx \in \fdR^N$, where the $i$th entry $ [\vcx]_{i}  = x_i$ is the signal value at node $i$~\cite{shuman2013emerging}.
\end{defBox}
{We denote the space of all graph signals defined on graph $\graph$ with node set $\stV$ as $\fdX^{\stV} = \{\fnx : \stV \to \fdR\}$.}
An example of a graph signal is a recording in a brain network, i.e., each brain area corresponds to a node, two nodes share a link based on structural connectivity, and the brain EEG measurement is the signal of a particular node.
We may want to process such a signal to understand, e.g., how different individuals have mastered a specific task \cite{huang2016graph}.


Processing and learning tasks with 
graph signals include: 

\medskip
\begin{enumerate}
\item \emph{Signal reconstruction, including interpolation and denoising:} We often observe a corrupted version of the graph signal, possibly at only a subset of nodes. 
{Examples include} noisy  or subsampled measurements in sensor \cite{jablonski2017graph}, power \cite{ramakrishna2021grid}, and water networks \cite{xing2022graph}. The goal is to denoise the signal or interpolate the missing values by leveraging the neighboring signal values and the graph structure. 
\item \emph{Signal compression:} When graph signals have similar values at neighboring vertices, it is possible to \emph{compress} the signal by developing representations that require fewer coefficients, and storing those coefficients rather than the original signal \cite{shuman2020Localized}. 
\item \emph{Signal classification:}
This task consists of classifying different graph signals observed over a common underlying graph. One such example is classifying patients based on their brain recordings, as discussed above \cite{huang2016graph}.
\item \emph{Node classification:}
This task consists of classifying a subset of nodes in the graph given the class labels on another subset. When node features are available, we can treat them as a collection of graph signals and leverage their coupling with the underlying connectivity to infer the missing labels. The state-of-the-art for this task is achieved by GNNs, which, as we shall see in Section~\ref{sec:GNN},
rely heavily on graph filters \cite{gama2020graphs}. When node features are unavailable, we treat the available labels as graph signals and transform node classification into a label interpolation task that can be solved with graph filters \cite{sandryhaila2014discrete}.
\item \emph{Graph classification / regression:} These tasks start with a collection of different graphs and (optionally) graph signals. The classification task assigns a label to the whole graph (e.g.,
classifying molecules into different categories such as soluble vs. non-soluble), whereas the regression task assigns a continuous number to each graph (e.g., the degree of solubility) \cite{wu2020comprehensive}. 
\item \emph{Link prediction:} Here, the goal is to infer if two nodes are connected given the current structure and the respective graph signals \cite{lu2011link}. This is the case of friend recommendation in social networks, whereby leveraging the friendship connectivity between users based on their feature signals (e.g., geo-position, workplace) we can infer missing links.
\item \emph{Graph identification:} This task extends the link prediction to that of inferring the whole graph structure given only the graph signals \cite{mateos2019connecting}. Graph filters play a role in modeling the relationships between candidate graph structures and the observed signals. We detail this problem in Sec.~\ref{subsec_gti}.
\item \emph{Distributed processing:} Here the graph topology represents the structure of a sensor network and we want to distributively solve a task related to graph signals \cite{shuman2018distributed}. Graph filters lend themselves naturally to this setup because they rely only on local information. 
In~Sec.~\ref{subsec_distSP}, we discuss their use for different distributed tasks. 
\end{enumerate}



\section{Graph Convolutional Filters} \label{sec:graphConv}

{The convolution is a key operation in SP as it helps to define filtering operations and to understand linear, time-invariant systems. In ML, convolutional filters are the building block of CNNs, and their computational efficiency and parameter-sharing property tackle the curse of dimensionality. Convolutions also leverage the symmetries in the domain (such as translations in space) and allow for a degree of mathematical tractability with respect to domain perturbation \cite{mallat2012scattering}. We present here a now standard generalization of the convolutional filter to the graph domain, with the goal of inheriting the above properties. Then, in Sec.~\ref{sec:filt_spect} we analyze the filter behavior in the graph spectral domain, akin to the Fourier analysis for temporal filters, and in Sec.~\ref{sec:filterDesign} we discuss strategies to design the filter parameters. 
}




\subsection{Definition}\label{subsec_gConv}

A convolutional filter is a \emph{shift-and-sum} operation of the input signal \cite{willsky1997signals}. While a shift in time implies a delay, a graph signal shift requires taking into account the topological structure.


\begin{defBox}{Graph signal shift}
A graph signal shift is a linear transformation $\fnS:\fdX^{\stV} \to \fdX^{\stV}$ obtained from applying a GSO $\mtS$ to a signal $\vcx$, i.e. $\fnS(\vcx) = \mtS\vcx$. The shifted signal at node $i$ is computed as
\begin{equation} \label{eq:graphShift}
[\mtS \vcx]_{i} = \sum_{j=1}^{N} [\mtS]_{ij} \vcx = \sum_{j \in \stN_{i}^{\text{in}} \cup \{i\}} s_{ij}x_{j},
\end{equation}
which is a local linear combination of the signal values at neighboring nodes.
\end{defBox}

\noindent If the GSO is the adjacency matrix $\mtA$, the shifted signal represents a one-step propagation. {Instead, if the GSO is the graph Laplcaian $\mtL$, the shifted signal is a weighted difference of the signals at neighboring nodes $[\mtL\vcx]_i = \sum_{j \in \stN_{i}}a_{ij}(x_i-x_j)$.}



\begin{figure*}[!t]
    {\def \thisplotscale {0.2}
        \begin{flushleft}

\def \unit {\thisplotscale cm}

\def\ypos{1.478}
\def\xpos{1.25}
\def\lbl{1.25}
\def\tanAngle{0.577} 
\def\signalLength{3}

\tikzstyle{small blue node} = [ blue node,
                                inner sep = 0,
                                minimum size = 1.2*\unit]

\tikzstyle{small red node} = [ red node,
                               inner sep = 0,
                               minimum size = 1.2*\unit]

\tikzstyle{overlay edge} = [tight edge,
                            shorten >= 3]

\def \radius   {3}

{\tiny
\begin{tikzpicture}[scale = \thisplotscale]


    \node at (0,0) (start) {};

    \path (start) ++ (0*\xpos + 0*\ypos*\tanAngle, 0*\ypos) node [small highlight node] (1) {$1$};

    \path (start) ++ (2.5*\xpos - 2.1*\ypos*\tanAngle, -2.1*\ypos) node [small base node] (4) {$4$};
    \path (start) ++ (2*\xpos - 0.7*\ypos*\tanAngle, -0.7*\ypos) node [small base node] (2) {$2$};
    \path (start) ++ (2*\xpos+ 0.7*\ypos*\tanAngle, 0.7*\ypos) node [small base node] (6) {$6$};
    \path (start) ++ (2.5*\xpos+ 2.1*\ypos*\tanAngle, 2.1*\ypos) node [small base node] (3) {$3$};

    \path (start) ++ (4*\xpos + 0*\ypos*\tanAngle, 0*\ypos) node [small base node] (5) {$5$};
    \path (start) ++ (5.25*\xpos - 0.8*\ypos*\tanAngle, -0.8*\ypos) node [small base node] (8) {$8$};
    \path (start) ++ (5.25*\xpos+ 0.8*\ypos*\tanAngle, 0.8*\ypos) node [small base node] (10) {$10$};

    \path (start) ++ (7*\xpos - 2.1*\ypos*\tanAngle, -2.1*\ypos) node [small base node] (7) {$7$};
    \path (start) ++ (8*\xpos + 0*\ypos*\tanAngle, 0*\ypos) node [small base node] (9) {$9$};
    \path (start) ++ (7*\xpos + 2.1*\ypos*\tanAngle, 2.1*\ypos) node [small base node] (11) {$11$};

    \path(1) ++ (0, .4*\signalLength) node [highlight dot] (x1) {};
    \path (x1) edge[signal bar,  draw = figHighlightColor] (x1|-1);

    \path(2) ++ (0, 0.4*\signalLength) node [base dot] (x2) {};
    \path (x2) edge[signal bar,  draw = figBaseColor] (x2|-2);

    \path(3) ++ (0, 1.2*\signalLength) node [base dot] (x3) {};
    \path (x3) edge[signal bar,  draw = figBaseColor] (x3|-3);

    \path(4) ++ (0, 0.4*\signalLength) node [base dot] (x4) {};
    \path (x4) edge[signal bar,  draw = figBaseColor] (x4|-4);

    \path(5) ++ (0, 0.4*\signalLength) node [base dot] (x5) {};
    \path (x5) edge[signal bar,  draw = figBaseColor] (x5|-5);

    \path(6) ++ (0, 0.4*\signalLength) node [base dot] (x6) {};
    \path (x6) edge[signal bar,  draw = figBaseColor] (x6|-6);

    \path(7) ++ (0, 0.4*\signalLength) node [base dot] (x7) {};
    \path (x7) edge[signal bar,  draw = figBaseColor] (x7|-7);

    \path(8) ++ (0, 0.4*\signalLength) node [base dot] (x8) {};
    \path (x8) edge[signal bar,  draw = figBaseColor] (x8|-8);

    \path(9) ++ (0, 0.4*\signalLength) node [base dot] (x9) {};
    \path (x9) edge[signal bar,  draw = figBaseColor] (x9|-9);

    \path(10) ++ (0, .4*\signalLength) node [base dot] (x10) {};
    \path (x10) edge[signal bar,  draw = figBaseColor] (x10|-10);

    \path(11) ++ (0, 0.4*\signalLength) node [base dot] (x11) {};
    \path (x11) edge[signal bar,  draw = figBaseColor] (x11|-11);

    \path (1)  edge [tight edge] node {} (2);
    \path (1)  edge [tight edge] node {} (6);

    \path (2) edge [tight edge] node {} (6);
    \path (2) edge [tight edge] node {} (4);
    \path (2) edge [tight edge] node {} (5);
    \path (6) edge [tight edge] node {} (5);
    \path (6) edge [tight edge] node {} (3);

    \path (5) edge [tight edge] node {} (8);
    \path (5) edge [tight edge] node {} (10);

    \path (3) edge [tight edge] node {} (10);
    \path (3) edge [tight edge] node {} (11);

    \path (4) edge [tight edge] node {} (8);
    \path (4) edge [tight edge] node {} (7);

    \path (10) edge [tight edge] node {} (11);
    \path (10) edge [tight edge] node {} (8);
    \path (10) edge [tight edge] node {} (9);

    \path (8) edge [tight edge] node {} (7);
    \path (8) edge [tight edge] node {} (9);

    \path (9) edge [tight edge] node {} (11);
    \path (9) edge [tight edge] node {} (7);

\end{tikzpicture}}\hspace{30mm}
    {\tiny
\begin{tikzpicture}[scale = \thisplotscale]


    \node at (0,0) (start) {};

    \path (start) ++ (0*\xpos + 0*\ypos*\tanAngle, 0*\ypos) node [small highlight node] (1) {$1$};

    \path (start) ++ (2.5*\xpos - 2.1*\ypos*\tanAngle, -2.1*\ypos) node [small base node] (4) {$4$};
    \path (start) ++ (2*\xpos - 0.7*\ypos*\tanAngle, -0.7*\ypos) node [small highlight node] (2) {$2$};
    \path (start) ++ (2*\xpos+ 0.7*\ypos*\tanAngle, 0.7*\ypos) node [small highlight node] (6) {$6$};
    \path (start) ++ (2.5*\xpos+ 2.1*\ypos*\tanAngle, 2.1*\ypos) node [small base node] (3) {$3$};

    \path (start) ++ (4*\xpos + 0*\ypos*\tanAngle, 0*\ypos) node [small base node] (5) {$5$};
    \path (start) ++ (5.25*\xpos - 0.8*\ypos*\tanAngle, -0.8*\ypos) node [small base node] (8) {$8$};
    \path (start) ++ (5.25*\xpos+ 0.8*\ypos*\tanAngle, 0.8*\ypos) node [small base node] (10) {$10$};

    \path (start) ++ (7*\xpos - 2.1*\ypos*\tanAngle, -2.1*\ypos) node [small base node] (7) {$7$};
    \path (start) ++ (8*\xpos + 0*\ypos*\tanAngle, 0*\ypos) node [small base node] (9) {$9$};
    \path (start) ++ (7*\xpos + 2.1*\ypos*\tanAngle, 2.1*\ypos) node [small base node] (11) {$11$};

    \path(1) ++ (0, .4*\signalLength) node [highlight dot] (x1) {};
    \path (x1) edge[signal bar,  draw = figHighlightColor] (x1|-1);

    \path(2) ++ (0, 0.4*\signalLength) node [highlight dot] (x2) {};
    \path (x2) edge[signal bar,  draw = figHighlightColor] (x2|-2);

    \path(3) ++ (0, 1.2*\signalLength) node [base dot] (x3) {};
    \path (x3) edge[signal bar,  draw = figBaseColor] (x3|-3);

    \path(4) ++ (0, .4*\signalLength) node [base dot] (x4) {};
    \path (x4) edge[signal bar,  draw = figBaseColor] (x4|-4);

    \path(5) ++ (0, .4*\signalLength) node [base dot] (x5) {};
    \path (x5) edge[signal bar,  draw = figBaseColor] (x5|-5);

    \path(6) ++ (0, .17*\signalLength) node [highlight dot] (x6) {};
    \path (x6) edge[signal bar,  draw = figHighlightColor] (x6|-6);

    \path(7) ++ (0, 0.4*\signalLength) node [base dot] (x7) {};
    \path (x7) edge[signal bar,  draw = figBaseColor] (x7|-7);

    \path(8) ++ (0, .4*\signalLength) node [base dot] (x8) {};
    \path (x8) edge[signal bar,  draw = figBaseColor] (x8|-8);

    \path(9) ++ (0, .4*\signalLength) node [base dot] (x9) {};
    \path (x9) edge[signal bar,  draw = figBaseColor] (x9|-9);

    \path(10) ++ (0, .19*\signalLength) node [base dot] (x10) {};
    \path (x10) edge[signal bar,  draw = figBaseColor] (x10|-10);

    \path(11) ++ (0, .13*\signalLength) node [base dot] (x11) {};
    \path (x11) edge[signal bar,  draw = figBaseColor] (x11|-11);

    \path (1)  edge [tight edge] node {} (2);
    \path (1)  edge [tight edge] node {} (6);

    \path (2) edge [tight edge] node {} (6);
    \path (2) edge [tight edge] node {} (4);
    \path (2) edge [tight edge] node {} (5);
    \path (6) edge [tight edge] node {} (5);
    \path (6) edge [tight edge] node {} (3);

    \path (5) edge [tight edge] node {} (8);
    \path (5) edge [tight edge] node {} (10);

    \path (3) edge [tight edge] node {} (10);
    \path (3) edge [tight edge] node {} (11);

    \path (4) edge [tight edge] node {} (8);
    \path (4) edge [tight edge] node {} (7);

    \path (10) edge [tight edge] node {} (11);
    \path (10) edge [tight edge] node {} (8);
    \path (10) edge [tight edge] node {} (9);

    \path (8) edge [tight edge] node {} (7);
    \path (8) edge [tight edge] node {} (9);

    \path (9) edge [tight edge] node {} (11);
    \path (9) edge [tight edge] node {} (7);

\end{tikzpicture}}\hspace{12mm}
    {\tiny
\begin{tikzpicture}[scale = \thisplotscale]


    \node at (0,0) (start) {};

    \path (start) ++ (0*\xpos + 0*\ypos*\tanAngle, 0*\ypos) node [small highlight node] (1) {$1$};

    \path (start) ++ (2.5*\xpos - 2.1*\ypos*\tanAngle, -2.1*\ypos) node [small highlight node] (4) {$4$};
    \path (start) ++ (2*\xpos - 0.7*\ypos*\tanAngle, -0.7*\ypos) node [small highlight node] (2) {$2$};
    \path (start) ++ (2*\xpos+ 0.7*\ypos*\tanAngle, 0.7*\ypos) node [small highlight node] (6) {$6$};
    \path (start) ++ (2.5*\xpos+ 2.1*\ypos*\tanAngle, 2.1*\ypos) node [small highlight node] (3) {$3$};

    \path (start) ++ (4*\xpos + 0*\ypos*\tanAngle, 0*\ypos) node [small highlight node] (5) {$5$};
    \path (start) ++ (5.25*\xpos - 0.8*\ypos*\tanAngle, -0.8*\ypos) node [small base node] (8) {$8$};
    \path (start) ++ (5.25*\xpos+ 0.8*\ypos*\tanAngle, 0.8*\ypos) node [small base node] (10) {$10$};

    \path (start) ++ (7*\xpos - 2.1*\ypos*\tanAngle, -2.1*\ypos) node [small base node] (7) {$7$};
    \path (start) ++ (8*\xpos + 0*\ypos*\tanAngle, 0*\ypos) node [small base node] (9) {$9$};
    \path (start) ++ (7*\xpos + 2.1*\ypos*\tanAngle, 2.1*\ypos) node [small base node] (11) {$11$};

    \path(1) ++ (0, .48*\signalLength) node [highlight dot] (x1) {};
    \path (x1) edge[signal bar,  draw = figHighlightColor] (x1|-1);

    \path(2) ++ (0, 0.46*\signalLength) node [highlight dot] (x2) {};
    \path (x2) edge[signal bar,  draw = figHighlightColor] (x2|-2);

    \path(3) ++ (0,1.41*\signalLength) node [highlight dot] (x3) {};
    \path (x3) edge[signal bar,  draw = figHighlightColor] (x3|-3);

    \path(4) ++ (0, .4*\signalLength) node [highlight dot] (x4) {};
    \path (x4) edge[signal bar,  draw = figHighlightColor] (x4|-4);

    \path(5) ++ (0, 0.5*\signalLength) node [highlight dot] (x5) {};
    \path (x5) edge[signal bar,  draw = figHighlightColor] (x5|-5);

    \path(6) ++ (0, -0.06*\signalLength) node [highlight dot] (x6) {};
    \path (x6) edge[signal bar,  draw = figHighlightColor] (x6|-6);

    \path(7) ++ (0, 0.4*\signalLength) node [base dot] (x7) {};
    \path (x7) edge[signal bar,  draw = figBaseColor] (x7|-7);

    \path(8) ++ (0, 0.44*\signalLength) node [base dot] (x8) {};
    \path (x8) edge[signal bar,  draw = figBaseColor] (x8|-8);

    \path(9) ++ (0, 0.52*\signalLength) node [base dot] (x9) {};
    \path (x9) edge[signal bar,  draw = figBaseColor] (x9|-9);

    \path(10) ++ (0, 0.06*\signalLength) node [base dot] (x10) {};
    \path (x10) edge[signal bar,  draw = figBaseColor] (x10|-10);

    \path(11) ++ (0, -.08*\signalLength) node [base dot] (x11) {};
    \path (x11) edge[signal bar,  draw = figBaseColor] (x11|-11);

    \path (1)  edge [tight edge] node {} (2);
    \path (1)  edge [tight edge] node {} (6);

    \path (2) edge [tight edge] node {} (6);
    \path (2) edge [tight edge] node {} (4);
    \path (2) edge [tight edge] node {} (5);
    \path (6) edge [tight edge] node {} (5);
    \path (6) edge [tight edge] node {} (3);

    \path (5) edge [tight edge] node {} (8);
    \path (5) edge [tight edge] node {} (10);

    \path (3) edge [tight edge] node {} (10);
    \path (3) edge [tight edge] node {} (11);

    \path (4) edge [tight edge] node {} (8);
    \path (4) edge [tight edge] node {} (7);

    \path (10) edge [tight edge] node {} (11);
    \path (10) edge [tight edge] node {} (8);
    \path (10) edge [tight edge] node {} (9);

    \path (8) edge [tight edge] node {} (7);
    \path (8) edge [tight edge] node {} (9);

    \path (9) edge [tight edge] node {} (11);
    \path (9) edge [tight edge] node {} (7);

\end{tikzpicture}}\hspace{12mm}
    {\tiny
\begin{tikzpicture}[scale = \thisplotscale]


    \node at (0,0) (start) {};

    \path (start) ++ (0*\xpos + 0*\ypos*\tanAngle, 0*\ypos) node [small highlight node] (1) {$1$};

    \path (start) ++ (2.5*\xpos - 2.1*\ypos*\tanAngle, -2.1*\ypos) node [small highlight node] (4) {$4$};
    \path (start) ++ (2*\xpos - 0.7*\ypos*\tanAngle, -0.7*\ypos) node [small highlight node] (2) {$2$};
    \path (start) ++ (2*\xpos+ 0.7*\ypos*\tanAngle, 0.7*\ypos) node [small highlight node] (6) {$6$};
    \path (start) ++ (2.5*\xpos+ 2.1*\ypos*\tanAngle, 2.1*\ypos) node [small highlight node] (3) {$3$};

    \path (start) ++ (4*\xpos + 0*\ypos*\tanAngle, 0*\ypos) node [small highlight node] (5) {$5$};
    \path (start) ++ (5.25*\xpos - 0.8*\ypos*\tanAngle, -0.8*\ypos) node [small highlight node] (8) {$8$};
    \path (start) ++ (5.25*\xpos+ 0.8*\ypos*\tanAngle, 0.8*\ypos) node [small highlight node] (10) {$10$};

    \path (start) ++ (7*\xpos - 2.1*\ypos*\tanAngle, -2.1*\ypos) node [small highlight node] (7) {$7$};
    \path (start) ++ (8*\xpos + 0*\ypos*\tanAngle, 0*\ypos) node [small base node] (9) {$9$};
    \path (start) ++ (7*\xpos + 2.1*\ypos*\tanAngle, 2.1*\ypos) node [small highlight node] (11) {$11$};

    \path(1) ++ (0, .62*\signalLength) node [highlight dot] (x1) {};
    \path (x1) edge[signal bar,  draw = figHighlightColor] (x1|-1);

    \path(2) ++ (0, 0.52*\signalLength) node [highlight dot] (x2) {};
    \path (x2) edge[signal bar,  draw = figHighlightColor] (x2|-2);

    \path(3) ++ (0, 1.79*\signalLength) node [highlight dot] (x3) {};
    \path (x3) edge[signal bar,  draw = figHighlightColor] (x3|-3);

    \path(4) ++ (0, 0.37*\signalLength) node [highlight dot] (x4) {};
    \path (x4) edge[signal bar,  draw = figHighlightColor] (x4|-4);

    \path(5) ++ (0, 0.67*\signalLength) node [highlight dot] (x5) {};
    \path (x5) edge[signal bar,  draw = figHighlightColor] (x5|-5);

    \path(6) ++ (0, -.42*\signalLength) node [highlight dot] (x6) {};
    \path (x6) edge[signal bar,  draw = figHighlightColor] (x6|-6);

    \path(7) ++ (0, 0.35*\signalLength) node [highlight dot] (x7) {};
    \path (x7) edge[signal bar,  draw = figHighlightColor] (x7|-7);

    \path(8) ++ (0, 0.46*\signalLength) node [highlight dot] (x8) {};
    \path (x8) edge[signal bar,  draw = figHighlightColor] (x8|-8);

    \path(9) ++ (0, 0.73*\signalLength) node [base dot] (x9) {};
    \path (x9) edge[signal bar,  draw = figBaseColor] (x9|-9);

    \path(10) ++ (0, -.14*\signalLength) node [highlight dot] (x10) {};
    \path (x10) edge[signal bar,  draw = figHighlightColor] (x10|-10);

    \path(11) ++ (0, -.36*\signalLength) node [highlight dot] (x11) {};
    \path (x11) edge[signal bar,  draw = figHighlightColor] (x11|-11);

    \path (1)  edge [tight edge] node {} (2);
    \path (1)  edge [tight edge] node {} (6);

    \path (2) edge [tight edge] node {} (6);
    \path (2) edge [tight edge] node {} (4);
    \path (2) edge [tight edge] node {} (5);
    \path (6) edge [tight edge] node {} (5);
    \path (6) edge [tight edge] node {} (3);

    \path (5) edge [tight edge] node {} (8);
    \path (5) edge [tight edge] node {} (10);

    \path (3) edge [tight edge] node {} (10);
    \path (3) edge [tight edge] node {} (11);

    \path (4) edge [tight edge] node {} (8);
    \path (4) edge [tight edge] node {} (7);

    \path (10) edge [tight edge] node {} (11);
    \path (10) edge [tight edge] node {} (8);
    \path (10) edge [tight edge] node {} (9);

    \path (8) edge [tight edge] node {} (7);
    \path (8) edge [tight edge] node {} (9);

    \path (9) edge [tight edge] node {} (11);
    \path (9) edge [tight edge] node {} (7);

\end{tikzpicture}}\hspace{9mm}    \\
        \vspace{-4mm}
    \end{flushleft}}
    \begin{flushright}
        {\def \thisplotscale {0.2}

\def \unit {\thisplotscale cm}

\def\ypos{1.478}
\def\xpos{1.25}
\def\lbl{1.25}
\def\tanAngle{0.577} 
\def\signalLength{3}

\tikzstyle{small blue node} = [ blue node,
                                inner sep = 0,
                                minimum size = 1.2*\unit]

\tikzstyle{small red node} = [ red node,
                               inner sep = 0,
                               minimum size = 1.2*\unit]

\tikzstyle{overlay edge} = [tight edge,
                            shorten >= 3]

\def \radius   {3}

{\tiny
\begin{tikzpicture}[scale = \thisplotscale]


    \node at (0,0) (start) {};

    \path (start) ++ (0*\xpos + 0*\ypos*\tanAngle, 0*\ypos) node [small base node] (1) {$1$};

    \path (start) ++ (2.5*\xpos - 2.1*\ypos*\tanAngle, -2.1*\ypos) node [small base node] (4) {$4$};
    \path (start) ++ (2*\xpos - 0.7*\ypos*\tanAngle, -0.7*\ypos) node [small base node] (2) {$2$};
    \path (start) ++ (2*\xpos+ 0.7*\ypos*\tanAngle, 0.7*\ypos) node [small base node] (6) {$6$};
    \path (start) ++ (2.5*\xpos+ 2.1*\ypos*\tanAngle, 2.1*\ypos) node [small base node] (3) {$3$};

    \path (start) ++ (4*\xpos + 0*\ypos*\tanAngle, 0*\ypos) node [small base node] (5) {$5$};
    \path (start) ++ (5.25*\xpos - 0.8*\ypos*\tanAngle, -0.8*\ypos) node [small base node] (8) {$8$};
    \path (start) ++ (5.25*\xpos+ 0.8*\ypos*\tanAngle, 0.8*\ypos) node [small base node] (10) {$10$};

    \path (start) ++ (7*\xpos - 2.1*\ypos*\tanAngle, -2.1*\ypos) node [small base node] (7) {$7$};
    \path (start) ++ (8*\xpos + 0*\ypos*\tanAngle, 0*\ypos) node [small base node] (9) {$9$};
    \path (start) ++ (7*\xpos + 2.1*\ypos*\tanAngle, 2.1*\ypos) node [small base node] (11) {$11$};

    \path(1) ++ (0, .43*\signalLength) node [base dot] (x1) {};
    \path (x1) edge[signal bar,  draw =  figBaseColor] (x1|-1);

    \path(2) ++ (0, 0.43*\signalLength) node [base dot] (x2) {};
    \path (x2) edge[signal bar,  draw = figBaseColor] (x2|-2);

    \path(3) ++ (0, .66*\signalLength) node [base dot] (x3) {};
    \path (x3) edge[signal bar,  draw = figBaseColor] (x3|-3);

    \path(4) ++ (0, 0.41*\signalLength) node [base dot] (x4) {};
    \path (x4) edge[signal bar,  draw = figBaseColor] (x4|-4);

    \path(5) ++ (0, 0.44*\signalLength) node [base dot] (x5) {};
    \path (x5) edge[signal bar,  draw = figBaseColor] (x5|-5);

    \path(6) ++ (0, 0.49*\signalLength) node [base dot] (x6) {};
    \path (x6) edge[signal bar,  draw = figBaseColor] (x6|-6);

    \path(7) ++ (0, 0.41*\signalLength) node [base dot] (x7) {};
    \path (x7) edge[signal bar,  draw = figBaseColor] (x7|-7);

    \path(8) ++ (0, 0.43*\signalLength) node [base dot] (x8) {};
    \path (x8) edge[signal bar,  draw = figBaseColor] (x8|-8);

    \path(9) ++ (0, 0.44*\signalLength) node [base dot] (x9) {};
    \path (x9) edge[signal bar,  draw = figBaseColor] (x9|-9);

    \path(10) ++ (0, .5*\signalLength) node [base dot] (x10) {};
    \path (x10) edge[signal bar,  draw = figBaseColor] (x10|-10);

    \path(11) ++ (0, 0.51*\signalLength) node [base dot] (x11) {};
    \path (x11) edge[signal bar,  draw = figBaseColor] (x11|-11);

    \path (1)  edge [tight edge] node {} (2);
    \path (1)  edge [tight edge] node {} (6);

    \path (2) edge [tight edge] node {} (6);
    \path (2) edge [tight edge] node {} (4);
    \path (2) edge [tight edge] node {} (5);
    \path (6) edge [tight edge] node {} (5);
    \path (6) edge [tight edge] node {} (3);

    \path (5) edge [tight edge] node {} (8);
    \path (5) edge [tight edge] node {} (10);

    \path (3) edge [tight edge] node {} (10);
    \path (3) edge [tight edge] node {} (11);

    \path (4) edge [tight edge] node {} (8);
    \path (4) edge [tight edge] node {} (7);

    \path (10) edge [tight edge] node {} (11);
    \path (10) edge [tight edge] node {} (8);
    \path (10) edge [tight edge] node {} (9);

    \path (8) edge [tight edge] node {} (7);
    \path (8) edge [tight edge] node {} (9);

    \path (9) edge [tight edge] node {} (11);
    \path (9) edge [tight edge] node {} (7);

\end{tikzpicture}}   }
    \vspace{-17mm}
    \end{flushright}
    {\def \thisplotscale {1.7}

\def \unit {\thisplotscale cm}

\def \deltax {1.6}
\def \deltay {0.75}
\def \sumshift {0.5}

{\footnotesize
\begin{tikzpicture}[x = 1*\unit, y = 1.3*\unit]

\node (origin) [] {};
\path (origin) ++ (0.1*\deltax, 0) node (first) [] {};

\path (first) ++ (1.3*\deltax, 0) node (0) [Phi] {$\mtS$};
\path (0)     ++ (1.3*\deltax, 0) node (1) [Phi] {$\mtS$};
\path (1)     ++ (1.3*\deltax, 0) node (2) [Phi] {$\mtS$};

\path (2.east) ++ (0.7*\sumshift*\deltax, 0) node [anchor=west] (last) [] {};

\path (first.east) ++ (1.5*\sumshift*\deltax, -\deltay) node (sum0) [sum] {$+$};
\path (0.east) ++ (\sumshift*\deltax, -\deltay) node (sum1) [sum] {$+$};
\path (1.east) ++ (\sumshift*\deltax, -\deltay) node (sum2) [sum] {$+$};
\path (2.east) ++ (\sumshift*\deltax, -\deltay) node (sum3) [sum] {$+$};

\path[-stealth] (first) edge [very near start, above] node {$\vcx$}               (0);
\path[-stealth] (0)     edge [above] node {$\mtS\vcx$}     (1);
\path[-stealth] (1)     edge [above] node {$\mtS^{2}\vcx$} (2);
\path[-]        (2)     edge [above] node {$\mtS^{3}\vcx$} (sum3 |- last);

\path[-stealth, draw] (sum0 |- first) -- (sum0) node [midway, right] {$h_{0}$};
\path[-stealth, draw] (sum1 |- 0)     -- (sum1) node [midway, right] {$h_{1}$};
\path[-stealth, draw] (sum2 |- 1)     -- (sum2) node [midway, right] {$h_{2}$};
\path[-stealth, draw] (sum3 |- 2)     -- (sum3) node [midway, right] {$h_{3}$};

\path[-stealth, draw] (sum0) -- (sum1);
\path[-stealth, draw] (sum1) -- (sum2);
\path[-stealth, draw] (sum2) -- (sum3);

\path[-stealth] (sum3) edge [above] node
{$\mtH(\mtS)\vcx$} ++ (1.3*\deltax, 0);

\end{tikzpicture}}}
    \caption{The graph convolutional filter as a shift register. Highlighted are the nodes that reach node $1$ on each consecutive shift; that is, the nodes $j$ whose signal value $x_j$ contributes to $[\mtS^k\vcx]_i$. The resulting summary of each communication $\mtS^{k}\vcx$ is correspondingly weighted by a filter parameter $h_{k}$. For each $k$, the parameter $h_k$ is the same for all nodes. {In this example, $\mtS=\mtL_n$ and $\mtH(\mtS)=1\mtL_n^0-1.5\mtL_n^1+1\mtL_n^2-0.25\mtL_n^3$ is a lowpass filter that smooths the input signal.}}\vskip.25cm
    \label{fig:graphConv}
    \vspace{-.15in}
\end{figure*}
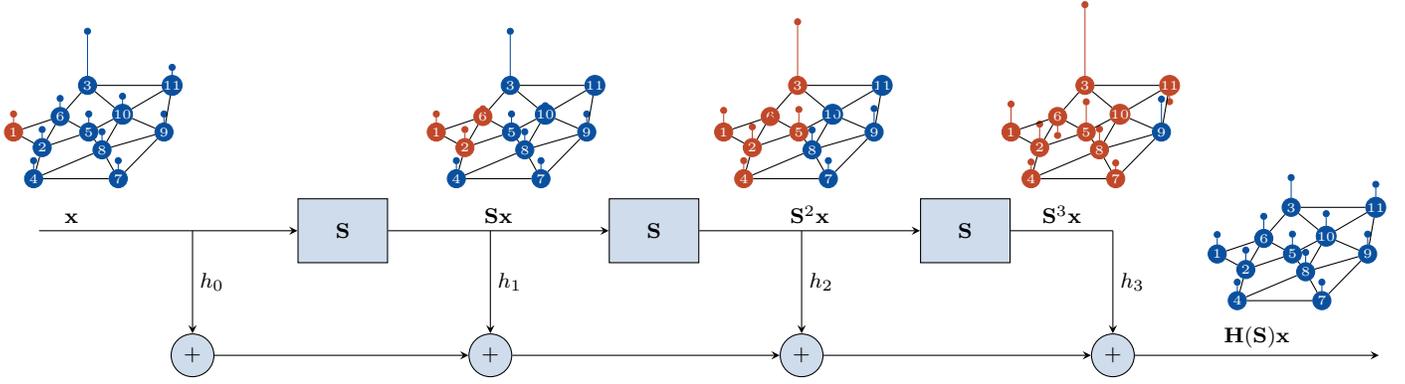

\begin{defBox}{Graph convolutional filter}
Given a set of parameters $\vch = [h_0, \ldots, h_K]^\top$, a graph convolutional filter of order $K$ is a linear mapping $\fnH: \fdX^{\stV} \to \fdX^{\stV}$ comprising a linear combination of $K$ shifted signals
\begin{equation} \label{eq:graphConv}
   \fnH(\vcx) = \sum_{k=0}^{K} h_{k} \mtS^{k} \vcx = \mtH(\mtS) \vcx
\end{equation}
where $\mtH(\mtS) = \sum_{k=0}^{K} h_{k} \mtS^{k}$ is the $N \times N$ polynomial filtering matrix.
\end{defBox}
%
\noindent The output at node $i$ is $y_i = h_0x_i + h_1[\mtS\vcx]_i + \ldots + h_K[\mtS^K\vcx]_i$, which is a linear combination of signal values located at most up to $K-$hops away. {This is because $[\mtS^k]_{ji} \neq 0$ implies that there exists at least one path of length $k$ between nodes $i$ and $j$ through which the signals can diffuse.}
%
These signals are shifted repeatedly over the graph as per \eqref{eq:graphShift}; see also Fig.~\ref{fig:graphConv}. {The term \emph{convolution} for \eqref{eq:graphConv} is rooted in the algebraic extension of the convolution operation \cite{puschel2008algebraic} and the discrete-time counterpart can be seen as a particular case over a cyclic graph; see Box 1.} 


\subsection{Properties}\label{subsec_prop}

Graph convolutional filters satisfy the following properties.



\begin{property}[Linearity]\label{prop_linear} For two inputs $\vcx_1$, $\vcx_2$, scalars $\alpha, \beta$, and filter $\mtH(\mtS)$, it holds that
$$
\alpha\mtH(\mtS)\vcx_1 + \beta\mtH(\mtS)\vcx_2 = \mtH(\mtS)(\alpha\vcx_1 + \beta\vcx_2).
$$
%
\end{property}



\begin{property}[Shift invariance]\label{prop:shift_invatiance}
The graph convolution is invariant to shifts, i.e., $\mtS \mtH(\mtS) = \mtH(\mtS) \mtS$. This implies that given two filters $\mtH_1(\mtS)$ and $\mtH_2(\mtS)$ and an input signal $\vcx$, it holds that we can switch the order of the filters:
$$
\mtH_1(\mtS)\mtH_2(\mtS)\vcx = \mtH_2(\mtS)\mtH_1(\mtS)\vcx.
$$
%
\end{property}




\begin{property}[Permutation equivariance]\label{prop:perm_equiv}
Denote the set of permutation matrices by
\begin{equation*} \label{eq:permutationSet}
    \stP = \big\{ \mtP \in \{0,1\}^{N \times N} : \mtP \vcOnes = \vcOnes \ , \ \mtP^{\Tr} \vcOnes = \vcOnes\}.
\end{equation*}
Then for a graph with GSO $\mtS$ and $\mtP \in \stP$, the permuted graph {(the graph obtained by permuting the node indices by $\mtP^{\Tr}$)} has the GSO $\mthS = \mtP^{\Tr} \mtS \mtP$, which describes the same topology but with a reordering of the nodes.
Likewise, the permuted signal corresponding to the ordering in $\mthS$ is $\vchx = \mtP^{\Tr} \vcx$. Permutation equivariance for filter \eqref{eq:graphConv} implies
\begin{equation*} \label{eq:permutationEquivariance}
\mtH(\mthS) \vchx = \mtP^{\Tr} \mtH(\mtS) \vcx;
\end{equation*}
i.e., the filter output operating on the permuted graph $\mthS$ with the permuted signal $\vchx$ is the permuted output of the same filter operating on the original graph $\mtS$ with the original signal $\vcx$.
\end{property}


Thus, graph convolutions are independent of the arbitrary ordering of the nodes. {Moreover, the permutation equivariance shows that the graph convolutional filter exploits the signal patterns with respect to the underlying graph structure. This is a direct analogue of translation equivariance in temporal or spatial signals, where the respective convolutional filters are translation equivariant functions.} This is key to their success in learning from a few training samples \cite{bronstein2021geometric}.


\begin{property}[Parameter sharing]\label{prop:param_share}
All the nodes share the parameters among them. For two nodes $i, j$, the respective outputs are $y_i = h_0x_i + h_1[\mtS\vcx]_i + \ldots + h_K[\mtS^K\vcx]_i$ and $y_j = h_0x_j + h_1[\mtS\vcx]_j + \ldots + h_K[\mtS^K\vcx]_j$, which shows that the $k$-shifted signal $\mtS^k\vcx$ is weighted by the same parameter $h_k$.
\end{property}

Props.~\ref{prop:perm_equiv}-\ref{prop:param_share} imply that graph convolutions are inductive processing architectures. They can be designed or trained over a graph $\stG$ and transferred to another graph $\widehat{\stG}$ (with possibly a different number of nodes) without redesigning or retraining. This is particularly useful, e.g., when using graph filters for distributed SP tasks, as the physical channel graph may change.
In Sec.~\ref{sec:filt_spect} (Prop.~\ref{prop:LipCont}), we discuss the degree of transference. 


\begin{property}[Locality]\label{prop:loc} Graph convolutions are local architectures. 
To see this, set $\vcz^{(0)} = \mtS^0\vcx$. The one shifted signal $\vcz^{(1)} = \mtS\vcx = \mtS\vcz^{(0)}$ is local by definition. The $k>1$ shift $\vcz^{(k)} = \mtS^k\vcx$ can be computed recursively as $\vcz^{(k)} = \mtS(\mtS^{(k-1)}\vcx) = \mtS\vcz^{(k-1)}$, which implies that the $(k-1)$st shift $\vcz^{(k-1)}$ needs to be shifted locally to the neighbors. Hence, to compute the output, each node exchanges locally with neighbors all $K$ shifts $\vcz^{(0)}, \ldots, \vcz^{(K-1)}$.
\end{property}

Locality of computation makes the graph convolutional filters readily distributable, as we discuss in Sec.~\ref{subsec_distSP}.



\begin{property}[Linear computation cost]\label{prop:compCost}
Graph convolutions have a computational complexity of order $\bigOh(K|\stE| + KN)$; i.e., linear in the number of edges and filter order.
\end{property}

Props.~\ref{prop:param_share}-\ref{prop:compCost} imply that graph convolutions tackle the curse of dimensionality in large graphs.
The parameter sharing makes them suitable architectures to learn input-output mappings from a few training samples, irrespective of the graph dimensions (i.e., $\bigOh(K)$ number of parameters);
their locality allows them to extract patterns in the data in the surrounding of a node, and; their linear computational complexity facilitates scalability. 
In Sec.~\ref{sec:GNN}, we discuss how to learn more expressive representations via neural networks while preserving these benefits in a form akin to CNNs for time series and images.

\section{Spectral Analysis} \label{sec:filt_spect}



\begin{SPMbox}{(Box 1) Discrete-time circular convolution}
The graph signal shift \eqref{eq:graphShift}, the graph convolutional filter \eqref{eq:graphConv}, and their spectral equivalents in Sec.~\ref{sec:filt_spect} generalize the respective concepts developed for discrete-time periodic signals.
\begin{center}
    \begin{minipage}{0.45\textwidth}
        \begin{center}
        \def \thisplotscale {0.55}

\def \unit {\thisplotscale cm}

\def \xdisplaced{8}

\def\ypos{1.5}
\def\xpos{1.5}
\def\lbl{0.75}
\def\mylinewidth{0.75}

{\footnotesize
\begin{tikzpicture}[scale = \thisplotscale]

	\node at (0,0) (DC1) {};
	\node at (\xdisplaced,0) (DC2) {};


	\path (0,\ypos) node[dot,fill=riceblue] (x11) {} ++ (0,\lbl) node {${
	 x_{1}}$};
	\path (0.866*\xpos,0.5*\ypos) node[dot,fill=grassgreen] (x12) {} ++ (0.866*\lbl,0*\lbl) node {${
	 x_{2}}$};
	\path (0.866*\xpos,-0.5*\ypos) node[dot,fill=warmyellow] (x13) {} ++ (0.866*\lbl,-0*\lbl) node {${
	 x_{3}}$};
	\path (0,-1*\ypos) node[dot,fill=brickred] (x14) {} ++ (0,-1*\lbl) node {${
	 x_{4}}$};
	\path (-0.866*\xpos,-0.5*\ypos) node[dot,fill=burgundy] (x15) {} ++ (-0.866*\lbl,-0*\lbl) node {${
	 x_{5}}$};
	\path (-0.866*\xpos,0.5*\ypos) node[dot,fill=shadowpurple] (x16) {} ++ (-0.866*\lbl,0*\lbl) node {${
	 x_{6}}$};

	\path (x11) edge[bend left=20,-stealth, line width=\mylinewidth] (x12);
	\path (x12) edge[bend left=20,-stealth, line width=\mylinewidth] (x13);
	\path (x13) edge[bend left=20,-stealth, line width=\mylinewidth] (x14);
	\path (x14) edge[bend left=20,-stealth, line width=\mylinewidth] (x15);
	\path (x15) edge[bend left=20,-stealth, line width=\mylinewidth] (x16);
	\path (x16) edge[bend left=20,-stealth, line width=\mylinewidth] (x11);


	\path (\xdisplaced,0) ++ (0,\ypos) node[dot,fill=shadowpurple] (x21) {} ++ (0,\lbl) node {${x_{6}}$};
	\path (\xdisplaced,0) ++(0.866*\xpos,0.5*\ypos) node[dot,fill=riceblue] (x22) {} ++ (0.866*\lbl,1*\lbl) node {${x_{1}}$};
	\path (\xdisplaced,0) ++(0.866*\xpos,-0.5*\ypos) node[dot,fill=grassgreen] (x23) {} ++ (0.866*\lbl,-1*\lbl) node {${x_{2}}$};
	\path (\xdisplaced,0) ++(0,-1*\ypos) node[dot,fill=warmyellow] (x24) {} ++ (0,-1*\lbl) node {${x_{3}}$};
	\path (\xdisplaced,0) ++(-0.866*\xpos,-0.5*\ypos) node[dot,fill=brickred] (x25) {} ++ (-0.866*\lbl,-1*\lbl) node {${x_{4}}$};
	\path (\xdisplaced,0) ++(-0.866*\xpos,0.5*\ypos) node[dot,fill=burgundy] (x26) {} ++ (-0.866*\lbl,1*\lbl) node {${x_{5}}$};

	\path (x21) edge[bend left=20,-stealth, line width=\mylinewidth] (x22);
	\path (x22) edge[bend left=20,-stealth, line width=\mylinewidth] (x23);
	\path (x23) edge[bend left=20,-stealth, line width=\mylinewidth] (x24);
	\path (x24) edge[bend left=20,-stealth, line width=\mylinewidth] (x25);
	\path (x25) edge[bend left=20,-stealth, line width=\mylinewidth] (x26);
	\path (x26) edge[bend left=20,-stealth, line width=\mylinewidth] (x21);


	\draw (DC1.center) edge[-stealth, line width=3*\mylinewidth, shorten <= 35, shorten >= 35] node[midway,above] {$\mtS = \mtA_{c}$}(DC2.center);






\end{tikzpicture}
}

        \end{center}
    \end{minipage}
    \hspace{0.2cm}
    \begin{minipage}[c]{0.45\textwidth}
        \begin{center}
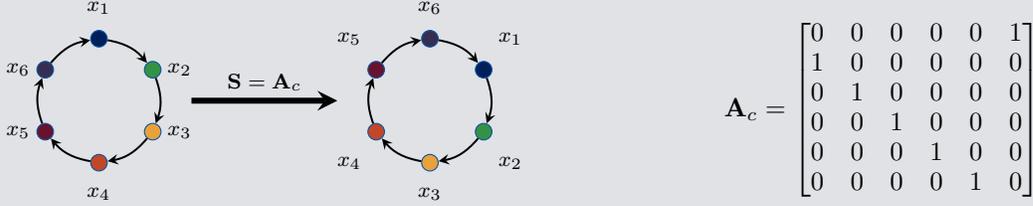

        $$ \mtA_{c} = \begin{bmatrix}
            0 & 0 & 0 & 0 & 0 & 1 \\
            1 & 0 & 0 & 0 & 0 & 0 \\
            0 & 1 & 0 & 0 & 0 & 0 \\
            0 & 0 & 1 & 0 & 0 & 0 \\
            0 & 0 & 0 & 1 & 0 & 0 \\
            0 & 0 & 0 & 0 & 1 & 0
        \end{bmatrix} $$
        \end{center}
    \end{minipage}
    \captionof{figure}{Discrete-time periodic signals as graph signals over a directed cycle graph. Each node $\stV_c = \{1, \ldots, 6\}$ is a time instant with adjacencies captured in the matrix $\mtA_c$. The temporal signal forms the graph signal $\vcx = [x_1, \ldots, x_6]^{\Tr}$ and the shift $\mtA_c \vcx$ acts as a delay operation that moves the signal to the next time instant node. 
    }
    \label{fig:directedCycle}
\end{center}

\noindent
We can represent an arbitrary discrete-time periodic signal as a graph signal $\vcx = [x_1, \ldots, x_{N}]^{\Tr}$ residing over the vertices of a directed cyclic graph $\graph_c = (\stV_c, \stE_c)$ in which each node is a time instant and directed edges connect adjacent time instances of the form $(n, 1+n \mod N)$, $n = 1, \ldots, N$, as shown in Fig.~\ref{fig:directedCycle}. The adjacency matrix of this graph is a cyclic matrix $\mtA_c$ such that $[\mtA_{c}]_{1+n \mod N, n} = 1$ and zeros everywhere else.

\smallskip
\noindent \emph{Signal shift:} Setting the GSO $\mtS = \mtA_c$, operation \eqref{eq:graphShift} shifts the signal cyclically and acts as a delay operation, i.e., $[\mtA_c \vcx]_{1+n\mod N} = x_{n}$.

\smallskip
\noindent \emph{Convolutional filter:} The graph convolutional filter \eqref{eq:graphConv} for graph $\graph_c$ reduces to the circular convolution, i.e., the output at the temporal node $i$ is $y_{n} = [\mtH(\mtA_{c})\vcx]_{n} = \sum_{k=0}^{K} h_{k} [\vcx]_{1+(n-k+1) \mod N}$. 

\smallskip
\noindent \emph{Signal variation:} Using the total variation in \eqref{eq.totV_dir}, we measure how much the signal changes from its delayed version.
This is a key building block for developing filters in standard signal processing \cite{schafer2010discreteSP}.

\smallskip
\noindent \emph{Fourier transform:} The cyclic adjacency matrix can be eigendecomposed as $\mtA_c = \textbf{DFT}_N \diag(\vclambda) \textbf{DFT}_N^{-1}$ with eigenvectors $[\textbf{DFT}_N]_{kn} = (1/\sqrt{N})\exp^{\imu 2\pi (k-1)(n-1)/N}$ forming the discrete Fourier transform (DFT) matrix and eigenvalues $\vclambda = [\exp(-\imu 2\pi 0/N, \ldots, -\imu 2\pi (N-1)/N)	]$ containing the frequencies.
The DFT for signal $\vcx$ is $\vctx = \textbf{DFT}_N\vcx$, which coincides with the graph Fourier transform (GFT) for this particular graph.
\end{SPMbox}

{While in the previous section we discussed the graph convolutional filter in the vertex domain, we now shift attention to the graph spectral domain to characterize the frequency response of these filters. This frequency response is key to interpreting the filter behavior, and it facilitates design when we want to achieve a desired spectral response, as we detail in Sec.~\ref{sec:filterDesign}.}


\subsection{Graph Fourier Transform} \label{subsec:gft}

The DFT can be seen as the projection of a temporal signal onto the eigenvectors of the cyclic graph adjacency matrix (see Fig.~\ref{fig:directedCycle}). We define similarly the graph Fourier transform (GFT). 

\begin{defBox}{Graph Fourier transform (GFT)} \label{def:GFT}
{Consider the eigendecomposition of the diagonlizable GSO $\mtS = \mtV\mtLambda\mtV^{-1}$ with eigenvectors $\mtV = [\vcv_1, \ldots, \vcv_N]$, and where $\mtLambda = \diag(\vclambda)$ is a diagonal matrix with the corresponding eigenvalues $\vclambda = [\lambda_1, \ldots, \lambda_N]$.} The GFT of a signal $\vcx$ is defined as the signal projection onto the GSO eigenspace
\begin{equation} \label{eq:GFT}
    \vctx = \mtV^{-1}\vcx.
\end{equation}
The inverse GFT is defined as $\vcx = \mtV\vctx$.
\end{defBox}
\noindent In the definition of the GFT, we are assuming the GSO $\mtS$ is diagonalizable. While definitions of GFT for nondiagonalizable GSOs exist \cite{sandryhaila2014discrete,sardellitti2017graph,shafipour2018directed}, we hold to the diagonalizability assumption for a consistent and simple exposition. 
{Furthermore, in some specified examples -- particularly those with a spectral interpretation -- we additionally assume that $\mtS$ is Hermitian; i.e.,  $\mtS$ is equal to $\mtS^{\fnH}$, its conjugate transpose. Common choices of shift operators such as the combinatorial and normalized graph Laplacians on undirected graphs satisfy this condition. Such operators have the nice property that $\mtS=\mtV \mtLambda \mtV^{\fnH}$, where the entries of the diagonal matrix $\mtLambda$ are real, and  $\mtV$ is a unitary matrix satisfying $\mtV^{\fnH} \mtV = \mtI$. Refer, for example, to \cite{girault2018irregularity} for more details on the choice of the GSO and its spectral consequences.}

The eigenvectors $\mtV$ serve as the basis expansion for the GFT.
In the discrete-time case, the complex exponentials fulfill this role.
The GFT coefficients $\vctx$ are the weights each of these eigenvectors contribute to represent the signal. Following again this analogy, the vector $\vclambda$ contains the so-called graph frequencies. Interpreting these graph frequencies $\vclambda$ and the respective GFT coefficients $\vctx$ requires understanding how the signal varies over the graph. In turn, measuring the signal variability requires accounting for the graph structure. We review two basic criteria used for undirected \cite{shuman2013emerging} and directed graphs \cite{sandryhaila2013discrete,sandryhaila2014discrete}. 

\titledParagraph{Undirected graphs.} The variability of a signal over an undirected graph is measured via the quadratic variation (QV)
\begin{equation}\label{eq:quad_und}
\QV(\vcx) := \vcx^{\fnH}\mtL\vcx = \frac{1}{2}\sum_{i \in \stV,j \in \stN_i} a_{ij}(x_i - x_j)^2,
\end{equation}
which quantifies how much a signal at a node is different from that of the strong connected ones \cite{shuman2013emerging}. The lower $\QV(\vcx)$, the smoother signal $\vcx$ is with respect to the underlying graph. In fact, the constant graph signal $\vcx = c\vcOnes$ has a zero variability.

Using \eqref{eq:quad_und}, we can interpret the variability of the Laplacian eigenvectors $\mtL = \mtV\diag(\vclambda)\mtV^{\fnH}$ so as to provide a Fourier basis. Treating each eigenvector $\vcv_i$ as a graph signal, we have
\begin{equation}\label{eq.Eigvar}
\QV(\vcv_i) = \vcv_{i}^{\fnH}\mtL\vcv_i = \lambda_i.
\end{equation}
Thus, we can sort eigenvectors based on their variability $0 = \QV(\vcv_1) \le \QV(\vcv_2) \le \ldots \le \QV(\vcv_N)$, which implies that the respective eigenvalues $0 \le \lambda_1 \le \lambda_2 \le \ldots \le \lambda_N$ carry the notion of frequency in the graph setting. We refer to the eigenvalues $\lambda_i$ close to $0$ as low frequencies and to those $\lambda_i \gg 0$ as high frequencies. The lowest graph frequency is $\lambda_1 = 0$ which corresponds to a constant eigenvector {for a connected graph}. Accordingly, {when $\mtS=\mtL$,} the GFT coefficient $\tilde{x}_i$ indicates how much eigenvector $\vcv_i$ contributes to the variability of signal $\vcx$ over $\graph$.

\titledParagraph{Directed graphs.} To measure the signal variability for directed graphs, we use again the analogy with the cyclic graph representing time signals, shown in Fig.~\ref{fig:directedCycle}. We measure how much the diffused signal $\mtS\vcx$ changes from the signal $\vcx$ via the total variation (TV)
\begin{equation}\label{eq.totV_dir}
\TV({\vcx}) = \|\vcx -  \mtS\vcx\|_1, 
\end{equation}
%
Expression \eqref{eq.totV_dir} attains a high value if the shifted signal differs more from the original one. However, unlike the quadratic measure for undirected graphs \eqref{eq:quad_und}, the total variation in \eqref{eq.totV_dir} may be non-zero for constant signals.

{If the shift operator is $\mtS = |\lambda_{\max}|^{-1}\mtA$ and $\lambda_{\max}$ is the eigenvalue of maximum amplitude, then $\TV({\vcx})=  \|\vcx -   |\lambda_{\max}|^{-1}\mtA\vcx\|_1$. In this case,}
we can  measure the variability of the adjacency matrix eigenvectors $\mtA = \mtV\diag(\vclambda)\mtV^{-1}$; we have that $\TV({\vcv_i}) \le \TV({\vcv_j})$ iff $|\lambda_\text{max} - \lambda_i| < |\lambda_\text{max} - \lambda_j|$. That is, the eigenvector associated with the largest eigenvalue has the lowest variability, while the eigenvector associated with the eigenvalue farthest from $\lambda_{\max}$ has the highest variability. Since the eigenvalues may be complex, the distances have to be computed in the complex plane. The order of the eigenvalues according to increasing variability is $\Re\{\lambda_{1}\} \geq \Re\{\lambda_{2}\} \geq \cdots \geq \Re\{\lambda_{N}\}$, see \cite[Figs. 2, 3]{sandryhaila2014discrete}.
In this case, the eigenvalues located (in a complex-plane sense) closest to the largest real eigenvalue are the ones corresponding to lower frequencies, while the eigenvalues located farthest from it correspond to the highest frequency.


Either on a directed or an undirected graph, the variability of a graph signal $\vcx$ can often be expressed by only a few $N^\prime \ll N$ GFT basis vectors.
In this case, we say the graph signal is $N^\prime-$bandlimited and expand it as
\begin{equation} \label{eq:bandlimited}
\vcx = \mtV_{N^\prime} \vctx_{N^\prime},
\end{equation}
with $\mtV_{N^\prime} = [\vcv_{1},\ldots,\vcv_{N^\prime}]$ and $ \vctx_{N^\prime} \in \fdR^{N^\prime}$. 

\subsection{Frequency Response} \label{subsec:freqResponse}

By subsituting the eigendecomposition $\mtS = \mtV\mtLambda\mtV^{-1}$ into \eqref{eq:graphConv}, we can write the filter output as
\begin{align}
\begin{split}
\vcy = \sum_{k=0}^{K} h_{k} \mtS^{k} \vcx = \sum_{k=0}^{K} h_{k}\mtV\mtLambda^{k}\mtV^{-1}\vcx.
\end{split}
\end{align}
Using \eqref{eq:GFT} and defining the GFT of the output $\vcty := \mtV^{-1}\vcy$, we can write the filter input-output spectral relation as
\begin{equation} \label{eq:GFToutput}
\vcty = \sum_{k=0}^{K} h_{k} \mtLambda^{k} \vctx.
\end{equation}
%

\begin{defBox}{Convolution theorem for graph filters}
It follows from \eqref{eq:GFToutput} that a \emph{shift-and-sum graph convolutional filter} of the form \eqref{eq:graphConv} operates in the spectral domain as a \emph{pointwise multiplication}
$\scty_{i} = \fnth(\lambda_{i})\sctx_{i}$ between the input signal GFT $\vctx = \mtV^{-1}\vcx$ and the filter frequency response
\begin{equation} \label{eq:freqResponse}
\fnth(\lambda) = \sum_{k=0}^{K} h_{k} \lambda^{k}.
\end{equation}
\end{defBox}

\begin{figure}
    \centering

\def \thisplotscale {3}
\def \unit {\thisplotscale cm}

\def \frequencyresponse 
     { 0.9 - 0.7*exp(-(0.7*(x-1.6))^2) }

\hspace{-2.9mm}
\begin{tikzpicture}[x = 1*\unit, y=1*\unit]

\def \factorx {2.4/8}
\def \deltax  {0.5*\factorx}
\def \shadeshift  {0.05}

\begin{axis}[scale only axis,
             width  = 2.4*\unit,
             height = 1*\unit,
             xmin = -0.5, xmax=7.5,
             xtick = {-0.01, 3.9393, 6.9048},
             xticklabels = {{\color{figBaseColor}$\lambda_1\ \ $}, 
                            {\color{figBaseColor}$\lambda_i$},
                            {\color{figBaseColor}$\lambda_N$}},
             ymin = -0, ymax = 1.15,
             ytick = {0.8193},
             yticklabels={$\tilde{\fnh}(\lambda)$},
             enlarge x limits=false]

\addplot+[samples at = {-0.01, 0.0819, 1.1413, 
                        2.6367, 3.9393, 4.5059, 
                        5.04, 6.9048}, 
          color = figBaseColor, 
          ycomb, 
          mark=oplus*, 
          mark options={figBaseColor}]
         {\frequencyresponse};

\addplot[ domain=-0.5:7.5, 
          samples = 80, 
          color = black,
          line width = 0.8]
         {\frequencyresponse};

\end{axis}
\end{tikzpicture}

    \caption{The frequency response of the filter \eqref{eq:freqResponse}, given in the black solid line, is completely characterized by the values of the filter parameters $\vch$. Given a graph, this frequency response gets instantiated on the specific eigenvalues of that graph, determining the 
    effect the filter will have on an input \eqref{eq:GFToutput}.} 
    \label{fig:freqResponse}
    \vspace{-.1in}
\end{figure}
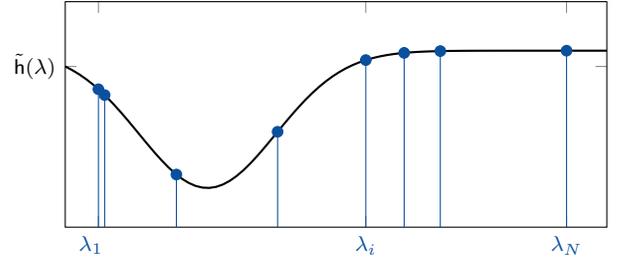

Such a result is reminiscent of the convolution theorem \cite{schafer2010discreteSP}, whereby the convolution in the graph domain corresponds to multiplication in the frequency domain. The filter frequency response is an analytic polynomial in $\lambda$ and it is independent of the graph. 
The specific filter effect on a given graph is on the positions where the frequency response is instantiated; see Fig.~\ref{fig:freqResponse}. 

In this context, graph convolutional filters satisfy the following spectral properties.


{
\begin{property}[GFT of the filter]
Eq. \eqref{eq:GFToutput} can be rewritten as
\begin{equation} \label{eq:GFTfilter}
\vcty = \diag(\vcth) \vctx \ \text{with}\ \vcth = \mtPsi \vch,
\end{equation}
where $\mtPsi \in \fdC^{N \times (K+1)}$  is a Vandermonde matrix such that $[\mtPsi]_{ik} = \lambda_{i}^{k-1}$.
{The vector $\vcth \in \fdC^{N}$ is known as the GFT of the filter parameters, which depends on the eigenvalues of the GSO (cf. \eqref{eq:GFTfilter}), unlike that of the signal, which depends on the eigenvectors (cf. \eqref{eq:GFT}). Refer to \cite{Segarra2017-GraphFilterDesign} for more detail.}

\end{property}
}

From~\eqref{eq:GFTfilter}, it follows that a graph convolutional filter defined as in~\eqref{eq:graphConv} has the same frequency response for two frequencies with the same eigenvalues, as this results in $\mtPsi$ having repeated rows.
Alternatively, one can define graph convolutional filters through their action in the frequency domain, potentially accommodating different responses for repeated eigenvalues.

\begin{property}[Lipschitz continuity to changes in $\mtS$]\label{prop:LipCont}
Let $\mtS, \mthS \in \fdR^{N \times N}$ be two GSOs, potentially corresponding to different graphs with the same number of nodes $N$.
Define the relative difference of $\mthS$ with respect to $\mtS$ as
\begin{equation} \label{eq:relDiff}
    \fnd(\mthS; \mtS) = \min_{\mtE \in \stR(\mthS;\mtS)} \| \mtE \|
\end{equation}
for $\stR(\mthS;\mtS) = \{\mtE: \mtP^{\Tr} \mthS \mtP = \mtS+ (\mtE \mtS+ \mtS \mtE) \ , \ \mtP \in \stP\}$, the set containing all the relative difference matrices $\mtE$.
Let the frequency response of the filter (cf. \eqref{eq:freqResponse}) satisfy $|\lambda \fnth'(\lambda)| \leq C$ for some $C < \infty$ and where $\fnth'(\lambda)$ is the derivative of \eqref{eq:freqResponse}.
Then, it holds that
\begin{equation} \label{eq:stabilityGraphFilter}
    \big\| \mtH(\mthS) \vcx - \mtH(\mtS)\vcx \big\|_{2} \leq {\fnd(\mthS; \mtS)} (1+8\sqrt{N}) C \| \vcx\|_{2} + \bigOh({\fnd^{2}(\mthS; \mtS)}).
\end{equation}
%
Thus, if the relative difference between two GSOs is small, the outputs of the filters with the same input signal will also be small. 
 {
 This is a scenario that arises often when learning on graph-structured data. The graph observed at training time is often different than the graph observed at testing time. Thus, we would like to obtain certain guarantees that the learned filters will still be useful at inference time. Property \ref{prop:LipCont} provides one type of guarantee.}
{Refer to \cite{gama2020stability} for an extensive discussion on the choice of \eqref{eq:relDiff} as the function to capture differences between $\mtS$ and $\mthS$.}
\end{property}

Filters whose frequency response satisfies $|\lambda \fnth'(\lambda)| \leq C$ are known as integral Lipschitz filters.
These filters may exhibit high variability for low values of $\lambda$ (because its derivative can be high), but they have to be approximately constant for high values of $\lambda$ (because its derivative has to be small). An example is shown in Fig.~\ref{fig:freqResponse}.
For finite graphs with finite edge weights, all convolutional filters (cf. \eqref{eq:graphConv}) are integral Lipschitz within the spectrum interval of interest, but the constant $C$ may be large.
This constant depends only on the filter parameters and, thus, filters can be designed or learned to have a small value of $C$, guaranteeing a tighter bound; see \cite{gama2020stability} for details.

\section{Filter Design and Identification} \label{sec:filterDesign}




{In this section, we discuss strategies to find the graph filter parameters to solve a specific task. 
We split our discussion into two common scenarios, each of which arises in the applications in Sec.~\ref{sec:SP} and Sec.~\ref{sec:ML}. 
First, we consider designing the filter \eqref{eq:graphConv} to match (or approximately match) a given operator $\mtB \in \fdR^{N\times N}$; i.e., find $\mtH$ such that $\mtH(\mtS) = \mtB$ (or $\mtH(\mtS) \approx \mtB$)  (Sec.~\ref{Ss:design_opMatch}). Second, we seek a filter \eqref{eq:graphConv} that represents a data-driven mapping between input-output signal pairs (Sec.~\ref{Ss:design_dataDriven}). 

}

%
%

\subsection{Operator Matching}\label{Ss:design_opMatch}

{Many SP applications on graphs can be formulated as a linear operator $\mtB$ on the signal $\vcx$. Such an operator may arise from the solution to a denoising problem~\cite{Isufi2017-ARMA}, be the consensus operator~\cite{Kruzick2018-filter}, or implement a specific spectral response that can be useful for graph wavelets (Sec.~\ref{sec:wavelets}) or spectral clustering (Sec.~\ref{subsec:unsuperv}). We want to represent the operator as a graph filter to reduce the computational cost (Property~\ref{prop:compCost}) if the matrix $\mtB$ is dense, or to implement $\mtB$ distributively over a sensor network (Property~\ref{prop:loc}).} In the following, we distinguish between exactly and approximately matching the operator $\mtB$  with a graph filter.

%

\smallskip
\noindent\textbf{Exact match.} 
Denote by $\mtH(\vch, \mtS)$ the convolutional filtering matrix in \eqref{eq:graphConv}, where we make explicit the dependency on parameters $\vch$. The following holds.

\begin{proposition}[\cite{Segarra2017-GraphFilterDesign}]\label{prop_filter_perfect_design}
	Given the three following conditions:
	\begin{enumerate}
	\item Matrices $\mtB$ and $\mtS$ are simultaneously diagonalizable; i.e., $\mtS = \mtV\mtLambda\mtV^{-1}$ and $\mtB = \mtV\diag(\vcbeta)\mtV^{-1}$ with eigenvalues $\vcbeta = [\beta_1, \beta_2, \ldots, \beta_N]^\top$.
	\item For all $(k_1,k_2)$ such that $\lambda_{k_1} = \lambda_{k_2}$ , it holds $\beta_{k_1} = \beta_{k_2}$.
	{\item The order of $\mtH(\vch, \mtS)$ is such that $K \geq D$, where $D$ denotes the number of distinct eigenvalues of $\mtS$.}
	\end{enumerate}
	Then, $\mtB = \mtH(\vch^*, \mtS)$ where $\vch^* = \mtPsi^\dag \vcbeta$, $\mtPsi$ is the Vandermonde matrix defined after \eqref{eq:GFTfilter}, and $(\cdot)^\dag$ denotes the pseudo-inverse. 
\end{proposition}

Condition 1 implies that transformation $\mtB$ is diagonalized by the GFT matrix. This implies that we can specify the spectral response of the operation and implement it via the filter in \eqref{eq:graphConv}. Since obtaining the eigendecomposition of $\mtS$ has a cost $\fnO(N^3)$, such an operation is not important for a centralized solution, as we could perfectly filter the signal in the spectral domain. However, it is important for distributed processing, because of the filter locality (Property~\ref{prop:loc}). We shall detail this in Sec.~\ref{subsec_distSP}.

\smallskip
\noindent\textbf{Approximate match.} 
%
{When the conditions of Proposition~\ref{prop_filter_perfect_design} are too stringent (especially the first one), we resort to optimally approximating the desired operator. If Condition 1 holds, it might be that we want to approximate $\mtB$ with a low-degree polynomial (violating Condition 3) or that we do not have access to the specific eigenvalues of $\mtB$ due to the computational cost of obtaining them. In either case, we can perform the spectral approximation described below. If Condition 1 does not hold (i.e., $\mtB$ does not have the same eigenvectors as the shift operator), we can approximate $\mtB$ directly in the vertex domain via the non-spectral approximation described below.}


\smallskip
\emph{Spectral approximation:} Consider {the common situation where $\mtS$ is a real, symmetric GSO, the desired operator $\mtB$ is jointly diagonalizable with $\mtS$, and} the spectral response $\tilde{\fnbeta}(\vclambda)$ of  $\mtB$ is a real-valued function. The goal is to find a low-order filter $\fnth(\lambda)$ that approximates this spectral response. If we have access to the specific eigenvalues of $\mtS$, we can easily obtain the spectral response $[\tilde{\fnbeta}(\lambda_1), \tilde{\fnbeta}(\lambda_2), \ldots, \tilde{\fnbeta}(\lambda_N)]$ on those eigenvalues. Then, Proposition~\ref{prop_filter_perfect_design} offers the least squares approximate solution \cite{sandryhaila2014discrete}. If, on the other hand, we only know the analytic expression of $\tilde\beta(\lambda)$ on a graph frequency interval $[\lambda_{\min}, \lambda_{\max}]$, the problem reduces to a one-dimensional polynomial approximation problem, such as the least squares problem 
\begin{equation}\label{eq:univDesign}
\begin{aligned}
\fnth ~=~& \underset{\vch}{\text{argmin}}
&  \int_{\lambda_{\min}}^{\lambda_{\max}} \bigg|\tilde{\fnbeta}(\lambda) -  \sum_{k = 0}^Kh_k\lambda^k 	\bigg|^2 {d\lambda}, \\
\end{aligned}
\end{equation}
or the minimax problem
\begin{equation}\label{eq:univDesign2}
\begin{aligned}
\fnth ~=~& \underset{\vch}{\text{argmin}}
& \underset{\lambda \in [\lambda_{\min},\lambda_{\max}]}{\text{sup}}\left\{ \bigg|\tilde{\fnbeta}(\lambda) -  \sum_{k = 0}^Kh_k\lambda^k 	\bigg| \right\}.
\end{aligned}
\end{equation}
The approaches in \eqref{eq:univDesign} and \eqref{eq:univDesign2} are referred to as  \emph{universal design}, because the approximating filters are designed over the interval  $[\lambda_{\min}, \lambda_{\max}]$, as opposed to specific eigenvalues. Thus, if the same filter is used on a different graph with the same spectral bounds, the approximating filter will also be the same. 
%
%
%
The solution to \eqref{eq:univDesign} can be found by orthogonally projecting $\tilde{\fnbeta}(\lambda)$ onto the span of the first $K+1$ Legendre polynomials. A near-optimal solution to \eqref{eq:univDesign2} can be found by truncating the expansion of $\tilde{\fnbeta}(\lambda)$ into shifted Chebyshev polynomials \cite{druskin1989two} (see Box 2). Alternative minimax approximations are investigated in \cite{tseng2021minimax,pakiyarajah2022minimax}.
For more details on the trade-offs involved in polynomial approximations, see, e.g, \cite[Sec. V.C]{shuman2018distributed} and \cite{trefethen2019approximation}.


\begin{SPMbox}{(Box 2) Chebyshev polynomial approximation} {Stretched and shifted Chebyshev polynomials offer an orthogonal basis for approximating a desired frequency response $\beta(\vclambda)$ via a low-order polynomial graph convolutional filter 
\cite{shuman2018distributed,trefethen2019approximation}. Moreover, they offer a closed-form solution for the approximating polynomial filter coefficients.} 
%
Formally, our goal is to approximate  {$\mtB\vcx=\mtV\diag(\tilde{\fnbeta}(\lambda))\mtV^{\fnH}\vcx$ by $\mtV\diag(\fnth(\vclambda))\mtV^{\fnH}\vcx$}, where $\fnth$ is a degree $K$ polynomial. Let $\{\fnT_k(x)\}_{k=0,1,\ldots}$ be Chebyshev polynomials of the first kind, which form an orthogonal basis for the function space $L^2\left([-1,1],\frac{dx}{\sqrt{1-x^2}}\right)$. 
Since, our frequency response $\tilde{\fnbeta}(\lambda)$ is defined on the interval $[0, \lambda_{\max}]$ (for positive semidefinite GSOs), we consider the change of variable $\lambda = \frac{1}{2} \lambda_{\max} (x+1)$. This leads to the stretched and shifted Chebyshev polynomials
\begin{equation}\label{eq.chebPoly}
\overline{\fnT}_k(\lambda):= \fnT_k\left(\frac{\lambda-\gamma}{\gamma}\right)~~\text{and}~~\gamma:= \frac{\lambda_{\max}}{2},
\end{equation}
which can be used to expand the desired frequency response as
\begin{equation}\label{eq:ChebInf}
\tilde{\fnbeta}(\lambda) = \frac{1}{2}c_0 + \sum_{k = 1}^\infty c_k\overline{\fnT}_k(\lambda),~~\forall \lambda\in [0, \lambda_{\max}],
\end{equation}
where each parameter $c_k$ can be found in closed-form by solving the integral
\begin{equation}\label{eq.ChebCoeff}
c_k := \frac{2}{\pi}\int_0^\pi\cos(k\theta) \tilde{\fnbeta}\big(\gamma(\cos(\theta) + 1)	\big)d\theta.
\end{equation}
For computational efficiency, we truncate the summation in \eqref{eq:ChebInf} to a finite $K$, resulting in an approximation \cite{druskin1989two}
$$
\tilde{\fnbeta}(\lambda) \approx \fnth(\lambda) := \frac{1}{2}c_0 + \sum_{k = 1}^K c_k\overline{\fnT}_k(\lambda),~~\forall \lambda\in [0, \lambda_{\max}].
$$
This Chebyshev polynomial approximation yields a $K$th order graph convolutional filter \eqref{eq:graphConv}; i.e., for any graph signal $\vcx$, we have
\begin{equation}
   \mtB\vcx  \approx \mtH(\mtS)\vcx = \sum_{k=0}^{K} c_{k} \overline{\fnT}_{k}(\mtS)\vcx,
\end{equation}
where we can compute the $k$th term recursively as $ \overline{\fnT}_{k}(\mtS)\vcx = \frac{2}{\gamma}\left(\mtS -\gamma \mtI\right)\overline{\fnT}_{k-1}(\mtS)\vcx -  \overline{\fnT}_{k-2}(\mtS)\vcx$, with initial values $ \overline{\fnT}_{0}(\mtS)\vcx = \vcx$ and $ \overline{\fnT}_{1} (\mtS)\vcx = \frac{1}{\gamma}\mtS\vcx-\vcx$ \cite{druskin1989two}.
The closed-form parameters can therefore be computed offline ahead of time and because of the recursive implementation, these filters can also be implemented distributively (cf. Property~\ref{prop:loc}). 
%
\end{SPMbox}
%

While these approximations are constructed for the entire interval $[\lambda_{\min}, \lambda_{\max}]$, it is only the approximation error at the (unknown) eigenvalues of $\mtS$ that  affects the spectral approximation error. Thus, additional partial knowledge of the spectrum can be leveraged to improve the polynomial approximation. In this regard, \cite{coutino2020fast} proposes a fast spectrum approximation method for specific families of graphs, whereas~\cite{Fan2020-Spectrum} leverages a fast estimation of the eigenvalue density for any graph shift operator to achieve a lower error in the high density regions of the spectrum. Alternatively, \cite{Kruzick2017-filter, Kruzick2018-filter} estimate the spectral distribution of frequencies via random matrix theory for random graphs (e.g., Erd\H{o}s-R\'enyi). All these approaches are developed for symmetric GSOs with real eigenvalues, while extensions for directed graphs with complex eigenvalues are discussed in \cite{liu2018filter,Sakiyama2017-polynomial}.


%

\smallskip
\emph{Non-spectral approximation:} When the desired operation $\mtB$ is not jointly diagonalized with $\mtS$, we can instead approximate it directly in the vertex domain, as stated by the following result.
%
%
\begin{proposition}[\cite{Segarra2017-GraphFilterDesign}]\label{prop_filter_approx_design}
Define the $N^2 \times (K+1)$ matrix $\mtTheta = [\mathrm{vec}(\mtI), \mathrm{vec}(\mtS), \ldots,\mathrm{vec}(\mtS^K)]$. The optimal filter parameters $\vch^* = \argmin_{\vch} \| \mtB - \mtH(\vch, \mtS) \|_\mathrm{F}$ are $\vch^* = \mtTheta^\dag \mathrm{vec}(\mtB)$.
\end{proposition}
%

%
%

\subsection{Data-driven}
\label{Ss:design_dataDriven}

{In many cases, we do not know the exact operator but rather have input-output realizations of a graph-based system. This is for instance the case of opinion formation and source identification in social networks, biological signals supported on graphs, and modeling and estimation of diffusion processes in multi-agent networks. The assumption here is that the data input-output relation can be modeled as a graph filter map and our goal is to identify the filter parameters.
Formally, consider the input $\vcx$ and output $\vcy$ satisfy
%
%
%
%
\begin{equation}\label{eqn_lms_model}
\vcy = \mtH(\vch, \mtS) \vcx + \vcn,
\end{equation}
where $\vcn$ is a zero-mean measurement noise.
Different variants of the problem have been studied depending on whether $\vcx$ and $\vcy$ are fully observed or not, and whether we have additional side information (statistical or structural) about the input~\cite{Segarra2017-Blind, Ramirez2021-Blind}. Particularly, we distinguish between (i) \emph{system identification}, where we estimate the parameters $\vch$ from  $\mtS$, $\vcx$, and a partial or complete observation of $\vcy$;  and (ii)  \emph{blind deconvolution}, where we jointly estimate $\vch$ and $\vcx$ from $\vcy$ and $\mtS$.} 

\smallskip
\noindent\textbf{System identification.} The goal is to use the (partial) observation of $\vcy$ to recover the unobserved elements of $\vcy$ and the filter parameters $\vch$. Consider only $M \le N$ nodes are observed and define the sampling matrix $\mtM \in \{0,1\}^{M \times N}$, which has one $1$ in each row  $m$ corresponding to the $m$th observed node and zero elsewhere. The filter identification problem comprises solving
\begin{equation}\label{eqn_system_id}
	\vch^* = \argmin_{\vch} \| \mtM\big(\vcy - \mtH(\vch, \mtS) \vcx\big) \|_2^2 + \gamma \| \mathrm{diag}(\vcomega) \vch \|_1,
\end{equation}
where $\vcomega \in \fdR_+^{K+1}$ is a weighting vector~\cite{Ramirez2021-Blind}. The first term quantifies the fitting loss between the observed $\mtM\vcy$ and its prediction generated by $\vch$. The second term is a sparsity-promoting regularizer on $\vch$. Since we often do now know the filter degree, we can overestimate it and then penalize higher degrees to promote simpler filters. Consequently, we can select the weights in $\vcomega$ to increase with the entry index. So, the parameters associated with higher powers of $\mtS$ are more heavily penalized, thus promoting a low-complexity and numerically stable model. The scalar $\gamma >0$ is the regularizer weight relative to the fitting loss. 
Extended versions of problem~\eqref{eqn_system_id} consider also estimating the topology in addition to the filter parameters; see e.g., \cite{Natali2020-Topology,rey2022robust,Rey2021-Filter}. 


\smallskip
\noindent\textbf{Blind deconvolution.} This problem arises when both the input $\vcx$ and the filter parameters $\vch$ are unknown. To formalize this problem, 
for a given $\mtS$, $\vcy$ is a bilinear function of $\vch$ and $\vcx$, which we can denote as $\vcy = \fnA(\vcx \vch^\top)$. The linear operator $\fnA(\cdot)$ depends on the eigenvalues and eigenvectors of $\mtS$ and acts on the outer product of the sought vectors. More precisely, $\fnA(\vcx \vch^\top) = (\mtLambda^\top \otimes (\mtV^{\top})^{-1})^\top \mathrm{vec}(\vcx \vch^\top)$, where $\otimes$ denotes the Khatri-Rao (i.e., columnwise Kronecker) product. 
While in principle we can jointly estimate $\vcx$ and $\vch$, this leads to a non-convex problem with few theoretical guarantees. Instead, using the classical idea of lifting~\cite{Ahmed2014-Deconvolution}, we can derive a convex relaxation of the blind deconvolution problem by noting that $\vcy$ is a linear function of the entries of the rank one matrix $\mtZ = \vcx \vch^\top$. Assuming further that $\vcx$ is a sparse vector (only a few nodes inject a signal into the filter),\cite{Segarra2017-Blind} proposes solving the convex problem
\begin{equation}\label{eqn_blind_deconvolution}
	\mtZ^* = \argmin_{\mtZ} \| \vcy - \fnA(\mtZ) \|_2^2 + \gamma_1 \| \mtZ \|_* + \gamma_2 \| \mtZ \|_{2,1}.
\end{equation}
The nuclear norm regularizer $\| \cdot \|_*$ promotes a low-rank solution since the true $\mtZ$ has rank one. The mixed norm $\| \mtZ \|_{2,1} = \sum_{i=1}^N \| \vcz_i\|_2$ is the sum of the $\ell_2$-norms of the rows of $\mtZ$, thus promoting a row-sparse structure in $\mtZ$.
This is aligned with the sparse assumption on $\vcx$, since each zero entry in $\vcx$ generates a whole row of zeros in the outer product $\mtZ = \vcx \vch^{\top}.$
Upon solving for $\mtZ^*$, we can recover $\vcx$ and $\vch$ from, e.g., a rank one decomposition of $\mtZ^*$.
Alternative relaxations to the row-sparsity and rank minimization have been proposed.
For instance, \cite{Ramirez2021-Blind} proposes a majorization-minimization procedure that yields lower-rank solutions compared to the convex relaxation in~\eqref{eqn_blind_deconvolution}, whereas~\cite{Iwata2020-Dconvolution} provides a handle to control the row-sparsity of the recovered matrix. 
%
%

Extensions to multiple input-output pairs (with a common filter) along with theoretical guarantees when the GSO $\mtS$ is normal (i.e., $\mtS \mtS^{\fnH} = \mtS^{\fnH} \mtS $) are investigated in~\cite{Segarra2017-Blind}. 
The related case of a single graph signal as input to multiple filters (generating multiple outputs) is studied in~\cite{Yu2020-Blind}, thus generalizing the classical blind multi-channel identification problem in DSP.
This setting is relevant when studying a common stimulus to several systems (e.g., the same image shown to several patients while their brain activity is recorded) or the effect of a single stimulus measured at different points in time (e.g., several snapshots of the spread of a rumor).
The work in~\cite{Iglesias2020-Demixing} addresses blind demixing when a single observation formed by the sum of multiple outputs is available, and it is assumed that these outputs are generated by different sparse inputs diffused through different graph filters.
This setting is relevant when the observations are given by the superposition of several concurrent processes. 
For example, we can model a brain state as the result of the simultaneous reaction to several stimuli that we want to separate.

\section{Other Graph Filters} \label{sec:other}




Graph convolutional filters implement a polynomial frequency response. {Their descriptive power increases as we grow
the filter order $K$. However, using higher orders implies handling higher matrix powers $\mtS^k$, which introduces numerical instabilities and in turn leads to poor interpolatory and extrapolatory performance \cite{trefethen2019approximation}.} While orthogonal polynomials (e.g., Chebyshev polynomials) can alleviate this issue, they still require a high number of parameters to implement the desired filtering function. Another limiting aspect of convolutional filtering is that its functions lie in the graph spectrum, {meaning that there may not exist a GCF that is a sufficiently good approximation to a general operator.}
%
%

In this section, we look at alternative graph filters to overcome these issues. We start with the family of filters that implement a rational response in Sec.~\ref{subsec_ratGF}. Then, in Sec.~\ref{subsec_ndFilt} we discuss linear filters that go beyond the spectral analogy, a.k.a. node domain filtering, and in Sec.~\ref{subsec_nlinFilt} we discuss nonlinear graph filtering forms. Sec.~\ref{subsec_regFilt} shows how graph-based regularization techniques behave as graph filters, and Sec.~\ref{subsec_multGSO} discusses filtering with multiple graph shift operators. {Table~\ref{tab:filters} provides a more extensive discussion of the properties in Sec.~\ref{subsec_prop} and Sec.~\ref{subsec:freqResponse}, as well as recommendations as to where to use them. }

\subsection{Rational Graph Filters}\label{subsec_ratGF}

A rational graph filter implements the frequency response
\begin{equation}\label{eq.ratFresp}
\fnth(\lambda) =  \bigg(\sum_{q = 0}^Q b_q \lambda^q	\bigg) \bigg/ \bigg(1 + \sum_{p = 1}^Pa_p\lambda^p	\bigg),
\end{equation}
which is the ratio of two polynomials of orders $Q$ and $P$ that control the number of zeros and poles, respectively. This form achieves similar frequency responses as the convolutional filters but with fewer parameters; because rational functions have better interpolatory and extrapolatory properties than polynomials and require a lower order to achieve a similar approximation~\cite{trefethen2019approximation}. However, rational filters have stability issues.
A rational graph filter is stable if the roots of its denominator are different from the GSO eigenvalues, i.e.,
%
%
\begin{equation}\label{eq.stabRat}
\fntp(\lambda) := 1 + \sum_{p = 1}^Pa_p\lambda^p \neq 0,~\forall\lambda\in \{\lambda_1, \ldots, \lambda_N\}.
\end{equation}
%
If we do not have access to the specific eigenvalues, we can also impose \emph{universal stability} by requiring condition~\eqref{eq.stabRat} to hold for all potential eigenvalues in the interval $[\lambda_{\min}, \lambda_{\max}]$ \cite{shi2015infinite,Isufi2017-ARMA}.

Given a stable filter and defining $\fntq(\lambda) = \sum_{q = 0}^Q b_q \lambda^q$, we can write \eqref{eq.ratFresp} as $\fnth(\lambda) = \fntq(\lambda) / \fntp(\lambda)$. Then the filter input-output relation in the spectral domain has the form $\scty_{i} = \fntq(\lambda_i) / \fntp(\lambda_i)\sctx_{i}$ at each graph frequency $\lambda_i$. In the node domain, the rational filtering matrix has the form
\begin{equation}\label{eq.ratFilter}
\mtH(\mtS) = \bigg(\mtI + \sum_{p = 1}^Pa_p\mtS^p	\bigg)^{-1}\bigg(\sum_{q = 0}^Qb_p\mtS^q	\bigg):=\mtP^{-1}(\mtS)\mtQ(\mtS),
\end{equation}
where we define $\mtP(\mtS):= \mtI + \sum_{p = 1}^Pa_p\mtS^p	$ and $\mtQ(\mtS):= \sum_{q = 0}^Qb_p\mtS^q$ with respective frequency responses $\fntp(\lambda)$ and $\fntq(\lambda)$. When applied to a graph signal $\vcx$, we get the input-output relationship in the vertex domain
\begin{equation}\label{eq.ratOut}
\vcy = \mtP^{-1}(\mtS)\mtQ(\mtS)\vcx \Longleftrightarrow \mtP(\mtS)\vcy = \mtQ(\mtS)\vcx.
\end{equation}
%

%

{Expressions \eqref{eq.ratFresp} and \eqref{eq.ratOut} show the two main challenges of rational graph filters. First, obtaining the output $\vcy$ from \eqref{eq.ratOut} requires solving a system of equations, which has a cubic order computational complexity $\fnO(N^3)$, making the filter impractical.\footnote{The inversion order cost matches that of the eigendecomposition of the shift operator. Consequently, there is no need to design a rational filter and apply it in the vertex domain as once we get the GFT of a signal we can implement exactly any desired spectral response.} Second, designing a rational filter is more challenging than fitting a polynomial filter because of the nonlinear nature of the problem and the stability issues. In the remainder of this section, we discuss strategies to approach the latter.}

\smallskip
\noindent\textbf{Implementation.} {To reduce the computational cost of solving \eqref{eq.ratOut}, we resort to iterative solvers that are fast and computationally efficient. If a centralized implementation is targeted, conjugate gradient approaches that exploit the graph structure are of interest \cite{liu2018filter}. They have a computational cost of $\bigOh((PT+Q)|\stE|)$, where $T$ is the number of iterations. Because of the fast convergence of the conjugate gradient, we can stop it in a few tens of iterations. And since a rational function achieves good approximation with low orders $P$ and $Q$, we expect the cost of obtaining the rational filter output to be low and comparable with that of graph convolutional filters (cf. \eqref{eq:graphConv}). 
Other works have considered the Jacobi method \cite{levie2018cayleynets,isufi2021edgenets}, quasi-Newton methods \cite{jiang2019decentralised}, or pre-conditioned gradient descent \cite{cheng2020preconditioned} to speed-up the computation in particular cases or have lighter per-iteration computation cost.

Instead, if a distributed implementation is needed, the algorithm for solving \eqref{eq.ratOut} need also enjoy a local computation. Most strategies rely on first-order methods based either on ARMA-like recursions \cite{loukas2015distributed,Isufi2017-ARMA,isufi2017autoregressive} or gradient-descent \cite{shi2015infinite}. When the graph has a small diameter,
the quasi-Newton method in \cite{jiang2019decentralised} or the pre-conditioned gradient descent  \cite{cheng2020preconditioned} could be a choice since they can be implemented locally with little effect on the performance.}

\smallskip
\noindent\textbf{Design.} 
There are two streams of rational filter design, both reminiscent of rational fitting and filter design in DSP: \emph{optimization-based} approaches and \emph{change of variable} approaches.

\smallskip
\emph{Optimization-based:} These methods can be cast as solving the constrained optimization problem
\begin{equation}\label{eq.ratFitt}
\begin{aligned}
& \underset{\{a_p, b_q\}}{\text{minimize}}
& & \int_{\lambda_{\min}}^{\lambda_{\max}} \bigg| \tilde{\fnbeta}({\lambda}) - \frac{\fntq(\lambda)}{\fntp(\lambda)}	\bigg|^2 {d\lambda}\\
& \text{subject to}
& &\fntp(\lambda) \neq 0 {,~\forall  \lambda \in \left[\lambda_{\min},\lambda_{\max}\right],}
\end{aligned}
\end{equation}
where $ \tilde{\fnbeta}({\lambda})$ is the desired response. 
Driven by their success in DSP \cite{hayes2009statistical}, simple design approaches such as Prony's and Shanks' methods have been extended to the graph setting in \cite{loukas2015distributed,Isufi2017-ARMA,isufi2017autoregressive,liu2018filter}. Such methods focus on the modified error $\tilde{\fne}(\lambda) = \tilde{\fnbeta}({\lambda}){\fntp(\lambda)} - \fntq(\lambda)$ and ignore the stability constraint. It has been consistently observed that these simple approaches offer good fitting and stable filters. 
Instead, stability-enforcing solutions are devised in \cite{aittomaki2019graph,jiang2019stable,pakiyarajah2021wls}, which use, respectively, a sum-of-squares, partial factorization, and constrained weighted least squares.

\smallskip
\emph{Change of variable:} The above strategies require solving an optimization problem 
that may be computationally demanding. To overcome this, 
 some works 
 extend the techniques that utilize Chebyshev polynomials to design convolutional filters to the rational filter design setting, ultimately yielding closed-form solutions and stable filters. Essentially, these methods: $i)$ map graph frequencies $\lambda \in [0, \lambda_{\text{max}}]$ into angular frequencies $\omega \in [0, \pi]$ via the transformation variable $\omega = {\pi\lambda}/{\lambda_{\text{max}}}$; $ii)$ design the filter for $\omega$ via standard DSP techniques, and; $iii)$ generate the graph counterpart as $\fnth(\lambda) = |\fnth(\omega)|^2\big|_{\omega = {\pi\lambda}/{\lambda_{\text{max}}}}$. {References} \cite{desbrun1999implicit} and \cite{shi2015infinite} use the Butterworth method for the design, while \cite{rimleanscaia2020rational} considers rational Chebyshev design of the first kind. A link with the rational filtering design in DSP is also discussed in \cite{tseng2020rational}, while \cite{cheng2019iterative} proposes an iterative design via Chebyshev polynomials (cf. \eqref{eq.chebPoly}) to approximate the inverse response. While having closed-form design, these approaches are often limited to ideal step responses in contrast to the optimization-based methods, which can be used for any response.
\subsection{Node Domain Filtering}\label{subsec_ndFilt}
{Convolutional and rational filters implement locally an operator that has a spectral response. However, in many cases, the desired operator or the data input-output mapping is more complex than a spectral response, which makes these solutions suboptimal (see also Sec.~\ref{sec:filterDesign}). Thus, it is of interest to develop filters  from a node domain perspective and potentially go beyond the spectral duality.} Generally speaking, a graph filter of order $K$ computes the output $[\vcy]_i$ at node $i$ as a linear combination of the input signal localized in the $K-$hop neighbors $\stN(i,K)$, i.e.,
\begin{equation}\label{eq.vx_filt}
[\vcy]_i = h_{ii}[\vcx]_i + \sum_{j \in \stN(i,K)} h_{ij}[\vcx]_j,
\end{equation}
where $\{h_{ij}\}$ are the parameters. {These parameters account also for the graph structure (edge weights) underlying the signal $[\vcx]_j$, seen locally from node $i$.} Here, we first relate the convolutional filtering \eqref{eq:graphConv} with operation \eqref{eq.vx_filt} and then discuss extensions to node varying \cite{Segarra2017-GraphFilterDesign} and edge varying versions \cite{Coutino2019-EdgeVariant}. 

\smallskip
\noindent\textbf{Convolutional filtering.} Leveraging locality (Property~\ref{prop:loc}), the convolutional filter obtains the information from $k$-hop neighbors 
as $\vcz^{(k)} = \mtS^k\vcx.$ 
This can be written as
\begin{align}\label{eq.conF_nodei}
[\vcy]_i = [\mtH(\mtS)\vcx]_i &= h_0[\vcx]_i + \sum_{k = 1}^{K}h_k [\vcz^{(k)}]_i  \nonumber \\ 
&= h_0[\vcx]_i +  \sum_{k = 1}^{K}h_k [\mtS^{k}\vcx]_i.
\end{align}
That is, the same parameter $h_0$ is applied to $[\vcx]_i$ by each node $i\in \stV$, and the same parameters $\{h_k\}$ weight the $k-$hop neighboring signal locally percolated via the GSO, $ [\mtS^{k}\vcx]_i$. 

\smallskip
\noindent\textbf{Node varying filtering.} A \emph{node varying graph filter} applies node-specific parameters $h_{ki}$ to $[\vcx^{(k)}]_i$ and each $[\mtS^{k}\vcx]_i$;  
i.e.,
\begin{equation}\label{eq.NV_nodei}
[\vcy]_i = h_{0i}[\vcx]_i +  \sum_{k = 1}^{K}h_{ki} [\mtS^{k}\vcx]_i.
\end{equation}
Collecting the different parameters applied at shift $k$ into the vector  $\vch_k = [h_{k1}, \ldots, h_{kN}]^\top$,
we can write such a filter as
\begin{equation}\label{eq.NVfilt}
   \fnH(\vcx) = \sum_{k=0}^{K} \diag({\vch_k}) \mtS^{k} \vcx. 
\end{equation}
%
This increased flexibility allows implementing more general operators than the convolutional filter, while still maintaining the local implementation.
Results akin to those in Section~\ref{Ss:design_opMatch} for exact and approximate operator matching have also been derived in~\cite{Segarra2017-GraphFilterDesign} for node varying filters.
To illustrate the result for exactly matching an operator $\mtB$, let us define $\vcdelta_i$ as the $N \times 1$ canonical vector with a $1$ in position $i$ and $0$ elsewhere, and $\vcu_i = \mtV^\top \vcdelta_i$, $\vcb_i = \mtB^\top \vcdelta_i$, and $\bar{\vcb}_i = \mtV^\top \vcb_i$.
With this notation in place, the following holds.
\begin{proposition}[\cite{Segarra2017-GraphFilterDesign}]\label{prop_nv_filter_perfect_design}
	If the following conditions hold for all $i$
	\begin{enumerate}
		\item $[\bar{\vcb}_i]_j = 0$ for those $j$ such that $[\vcu_i]_j = 0$
		\item For all $(j_1,j_2)$ such that $\lambda_{j_1} = \lambda_{j_2}$ , it holds that $[\bar{\vcb}_i]_{j_1} / [\vcu_i]_{j_1} = [\bar{\vcb}_i]_{j_2} / [\vcu_i]_{j_2}$
		\item The degree of $\mtH(\mtS)$ is such that $K \geq D$, where $D$ denotes the number of distinct eigenvalues in $\mtS$
	\end{enumerate}
	then $\mtB$ can be perfectly implemented using a node varying graph filter as defined in~\eqref{eq.NVfilt}.
\end{proposition}
A direct comparison of Propositions~\ref{prop_filter_perfect_design} and~\ref{prop_nv_filter_perfect_design} reveals the added expressivity of node varying graph filters since the stringent requirement of simultaneous diagonalization in Proposition~\ref{prop_filter_perfect_design} is replaced by {the milder Condition 1 in Proposition~\ref{prop_nv_filter_perfect_design}}.
Similarly, data-driven design, modifications to the adaptive methodologies presented in Section~\ref{Ss:design_dataDriven} have also been extended to node varying filters in~\cite{hua2020online, alinaghi2021graph}.

Due to the node-specific nature of node varying filters, we can find frequency representations for every row of $\mtH(\mtS)$ as
\begin{equation}\label{eq_frequency_nv}
	\vcdelta_i^\top \mtH(\mtS) = \sum_{k=0}^K [\vch_k]_i \vcu_i^\top \mtLambda^k \mtV^{-1} = \vcu_i^\top \diag(\tilde{\vch}^{(i)}) \mtV^{-1},
\end{equation}
where $\tilde{\vch}^{(i)} = \mtLambda \vch^{(i)}$ and $\vch^{(i)} = [ [\vch_0]_i, [\vch_1]_i, \ldots, [\vch_K]_i ]$ collects the filter parameters associated to node $i$.
The output at node $i$ is the elementwise product of the input Fourier transform $\mtV^{-1}\vcx$ and the filter being implemented at $i$, $\diag(\tilde{\vch}^{(i)})$, and then combined with node-specific weights $\vcu_i^\top$ that encode how strong each frequency is represented by node $i$.

\smallskip
\noindent\textbf{Edge varying filtering.} 
We can further improve the filter flexibility by allowing each node to weight differently the information of its different neighbors. 
Then the diagonal parameter matrix in \eqref{eq.NVfilt} becomes a matrix $\mtH_k$ with the same support as the GSO $\mtS$, leading to the \emph{edge varying graph filter}:
%
\begin{equation}\label{eq.EVfilt}
   \fnH(\vcx) = \sum_{k=0}^{K} \mtH_k \mtS^{k} \vcx = \mtH(\mtS)\vcx,
\end{equation}
where $[\mtH_k]_{ij} = h_{kij}$ is the parameter node $i$ applies to the signal of neighbor $j$ at iteration $k$. For $k = 0$, we have $\mtH_0 := \diag({\vch_0})$ since each node only weights its own signal \cite{Coutino2019-EdgeVariant}. By weighting differently the shifted information of the neighbors, the edge varying graph filter has an even higher flexibility and still preserves the local implementation. The latter is because the original signal values from the neighbors up to $k$ hops away are still propagated through the graph via the GSO $\mtS^{k} \vcx$ and only then weighted locally by $\mtH_k$. It is shown in \cite{Coutino2019-EdgeVariant} that filter \eqref{eq.EVfilt} can better approximate a defined operator compared with the convolutional and rational filters. In addition, it also enjoys a spectral representation, but that requires long derivations and we refer the reader to \cite{Coutino2019-EdgeVariant}. Other forms of edge varying filters are developed in \cite{Coutino2019-EdgeVariant, coutinominguez2021cascaded}, which use a parametric GSO and a cascaded form, respectively.

{One of the main advantages of node domain filters is their increased degrees of freedom (DoFs) while preserving linearity and locality of implementation (Properties~\ref{prop_linear} and \ref{prop:loc}). As a result, both filters have a computational complexity of order $\fnO(K|\stE|)$, even though the node varying filter \eqref{eq.NVfilt} has $N(K+1)$ parameters and the edge varying filter \eqref{eq.EVfilt} has $N + (N+|\stE|)K$ parameters. 
The node varying graph filter is also proven Lipschitz stable (Property~\ref{prop:LipCont}) in \cite{gama2021node}. }

{Since these filters are neither shift invariant nor permutation equivariant (i.e., Properties~\ref{prop:shift_invatiance}-\ref{prop:perm_equiv} do not hold), we need to account for the order in which we cascade them, as well as the node labeling. The lack of permutation equivariance also implies that we cannot transfer the learned filters across different graphs. Another challenge is that there may be too many DoFs (parameters) to estimate from limited data. In these instances, we can regularize the problem to penalize some norm of the parameters or develop a hybrid filter where edge varying parameters are applied only for a few representative nodes \cite{isufi2021edgenets}. But when operating on a fixed graph and with a reasonable amount of data or a fixed operator, node domain filters can substantially improve the performance, especially in a distributed implementation.}

\subsection{Nonlinear Graph Filtering}\label{subsec_nlinFilt}





Nonlinear filters have been proposed to overcome the limitations of node domain filters (large DoFs and lack of transferability across graphs), but still be more flexible than GCFs.
{To introduce nonlinear filters, we first focus on the graph convolutional filter \eqref{eq:graphConv} output $\vcy_i$ at node $i$. Specifically, we collect the $K-$shifted signals at node $i$ in the vector $\vcx_i^{(K)} = \big[[\vcx]_i, [\mtS\vcx]_i, \ldots, [\mtS^K\vcx]_i	\big]^\top$ and the filter parameters in the set $\stH = \{\vch = [h_0,\ldots, h_K]^\top \}$. Then, we can write the filter output at node $i$ as
\begin{equation}\label{eq.linearGF}
y_i = \fnf(\vcx_i^{(K)}; \stH):=\vch^\top\vcx_i^{(K)},~\forall~i=1, \ldots, N;
\end{equation}
i.e., it is a multivariate linear regression in $\vcx_i^{(K)}$ with parameters $\vch$ that are shared among the nodes. We also see from \eqref{eq.NV_nodei} and \eqref{eq.EVfilt} that node varying filtering is a linear variation of \eqref{eq.linearGF} but with different parameters $\vch_i$ for each node.}
In contrast, nonlinear graph filters can be built by considering a nonlinear function $\fnf(\vcx_i^{(K)}; \stH)$ in \eqref{eq.linearGF} with the same parameters $ \stH$ for all nodes. While the function $\fnf(\cdot)$ can be arbitrary, it has been studied for two models inspired by traditional signal processing: the Volterra graph filter \cite{xiao2021distributed} and the median graph filter \cite{segarra2016center, segarra2017design}.

\smallskip
\noindent\textbf{Volterra filter.} The natural generalization of \eqref{eq.linearGF} is to consider a multivariate polynomial regressor in variables $\vcx_i^{(K)}$:
\begin{equation}\label{eq.Volt_nodei}
y_i = \fnf(\vcx_i^{(K)}; \stH):= \textnormal{poly}_{L_0, \ldots, L_K} (\vcx_i^{(K)}; \stH),
\end{equation}
where $\textnormal{poly}_{L_0, \ldots, L_K}(\cdot)$ denotes a multivariate polynomial of orders $L_0, \ldots, L_K$ in $[\vcx]_i, [\mtS\vcx]_i, \ldots, [\mtS^K\vcx]_i$, respectively, and the set $\stH$ collects the respective parameters. For instance, for a shift order $K = 1$ and polynomial orders $L_0 = 2, L_1 = 3$, we have $\vcx_i^{(1)} = \big[[\vcx]_i, [\mtS\vcx]_i	\big]^\top$ and \eqref{eq.Volt_nodei} becomes
\begin{align}
\begin{split}
y_i = \sum_{l_0 = 0}^{L_0 = 2}\sum_{l_1 = 0}^{L_1 = 3}h_{l_0 l_1}[\vcx]_i^{l_0}[\mtS\vcx]_i^{l_1}~\text{and}~\stH = \{h_{l_0 l_1}\}. 
\end{split}
\end{align}
Expressed compactly, this filter has the input-output relation
\begin{equation}
\vcy = \sum_{l_0 = 0}^{L_0 = 2}\sum_{l_1 = 0}^{L_1 = 3}h_{l_0, l_1} \big[\vcx^{\odot l_0} \odot (\mtS\vcx)^{\odot l_1}\big]~\text{and}~\stH = \{h_{l_0 l_1}\},
\end{equation}
where $\vcx^{\odot a} = \vcx \odot \ldots \odot \vcx$ is the element-wise $a-$th power of $\vcx$. Then, a nonlinear graph filter of order $K$ has the form
\begin{align}\label{eq.polyGF}
\begin{split}
\vcy = \sum_{l_0 = 0}^{L_0}\ldots\sum_{l_K = 0}^{L_K}h_{l_0\ldots l_K} \big[\vcx^{\odot l_0}\odot (\mtS\vcx)^{\odot l_1}\odot \ldots \odot (\mtS^K\vcx)^{\odot l_K}			\big],
\end{split}
\end{align}
where set $\stH = \{h_{l_0\ldots l_K} \}$ 
collects all the parameters of order $\fnO(KL_{\text{max}})$ with $L_{\text{max}} = \max\{L_0, \ldots, L_K\}$. Because data are gathered locally, these filters generate the output with the same order of computational complexity. Such an increased flexibility allows us to represent more complex nonlinear relationships in graph input-output data. But, at the same time, multivariate polynomial regression can overfit the data and may suffer from ill-conditioning. 
{Differently from the node or edge varying filter, the nonlinear filter in \eqref{eq.polyGF} shares the parameters across nodes, which allows transferring it across graphs.} The Volterra graph filter is the particular case of \eqref{eq.polyGF} with reduced DoFs ($L_0\le L_1\le\ldots\le L_K$) and it has been shown that even if the input is a bandlimited signal (cf. \eqref{eq:bandlimited}), the output can have frequency content in the entire graph spectrum \cite{xiao2021distributed}.

\smallskip
\noindent\textbf{Median filter.} All the above filters rely on signal propagation over the graph.
When a particular node is anomalous and has, e.g., a large signal value, it will affect all the neighbors and the filter output.
Median graph filters have been proposed as robust alternatives that can tackle such an issue.

Consider an integer $h\ge0$ and a real scalar $x$. We define the replication operation $h\diamond x = [x, \ldots, x]^\top \in \fdR^h$. Then, the median graph filter output at node $i$ can be obtained as
\begin{equation}\label{eq.Med_filt}
y_i = \fnf(\vcx_i^{(K)}; \stH) :=\text{Med} (h_0\diamond[\vcx]_i; \ldots; h_K \diamond [\mtS^K\vcx]_i),
\end{equation}
where the median operation $\text{Med}(\cdot)$ sorts its arguments in ascending order and outputs the middle one \cite{segarra2017design}. In obtaining $[\mtS \vcx]_i$, we compute a weighted linear combination of the entries in $\vcx$, where the weights are given by the values in the $i$th row of $\mtS$. If a node has a particularly high value, it can be amplified via the shift operator $\mtS$ and be present in $\vcx_i^{(K)}$ for almost all nodes. The median operator attenuates such influence.
The data-driven design of the parameters $\vch$ is discussed in \cite{segarra2017design}.
{Cases where the weights in $\mtS$ can be designed are also studied in~\cite{segarra2017design}.}
As with the Volterra filter, the median filter is nonlinear and local; however, it does not enjoy a spectral equivalence.
Alternative expressions to \eqref{eq.Med_filt} are also proposed in~\cite{segarra2016center, segarra2017design}. They differ in how the data are gathered at the nodes (either linearly via shifting or nonlinearly via the median operator) and how these gathered data are processed (again linearly or via a median operator). 
Lastly, we remark that the local median operator in \eqref{eq.Med_filt} is only one choice and other nonlinear functions such as max or min can be used \cite{ruiz2019invariance,iancu2020graph}.

\subsection{Filtering by Regularization}\label{subsec_regFilt}

The graph filters discussed above can be seen as graph-based parametric functions to model input-output mappings. When the spectral specifications of this mapping are unclear or when the amount of data is limited, these parametric filters can be difficult to design or can easily overfit the data. In these cases, we may want to implement graph filtering via regularization, leveraging prior information about particular properties that graph signals exhibit. For simplicity, we consider graph regularization to denoise graph signals, which is crucial in data processing; however, similar observations extend also to interpolating missing values, as we shall see in Sec.~\ref{subsec_denInterp}.

Consider the task of recovering a graph signal $\vcz$ from a single noisy obervation $\vcx = \vcz + \vcn$, with $\vcn$ being additive Gaussian noise. This can be addressed by solving
\begin{equation}\label{eq.Tikproblem}
 \underset{\vcy \in \fdR^N}{\text{argmin}}~\fnf(\vcx, \vcy) + \gamma \fnr(\vcy,\stG),
\end{equation}
where $\fnf(\vcx, \vcy)$ is the fitting-term, typically $\fnf(\vcx, \vcy) = \|\vcx - \vcy\|_2^2$, and $\fnr(\vcy,\stG)$ imposes a graph-based prior about the true signal.
Depending on the signal behavior with respect to the underlying graph, we discuss three regularization techniques: (i) smooth filtering; (ii) sparsity filtering; and (iii) Wiener filtering.

\smallskip
\noindent \textbf{Smooth filtering.} These approaches consider a regularizer that imposes a low signal variation between adjacent nodes. For undirected graphs, two popular approaches are the Tikhonov regularizer and the Sobolev regularizer, while for directed graphs, the total variation regularizer is commonly applied.

\emph{Tikhonov \cite{zhou2004regularization,shuman2013emerging}:} A smooth signal $\vcy$ over an undirected graph has a low quadratic form $\LQ(\vcy) = \vcy^\Tr\mtL\vcy$ (cf. \eqref{eq:quad_und}). 
Considering $\LQ$ as a regularizer, we obtain
\begin{equation}\label{eq.Tikproblem}
 \underset{\vcy \in \fdR^N}{\text{argmin}}~\|\vcx - \vcy	\|_2^2 + \gamma \vcy^\top\mtL\vcy.
\end{equation}
The more we increase $\gamma \gg 0$, the more we prioritise smoothness on the solution. Problem~\eqref{eq.Tikproblem} is a quadratic convex problem and has the closed-form solution
\begin{equation}\label{eq.Tiksol}
\vcy^* = (\mtI + \gamma\mtL)^{-1}\vcx.
\end{equation}
Comparing \eqref{eq.Tiksol} with \eqref{eq.ratOut}, we see that the Tikhonov filter is an order one rational filter with frequency response $\fnth(\lambda) = (1+\gamma\lambda)^{-1}$. This frequency response also helps understanding the role of parameter $\gamma$; the optimal solution in \eqref{eq.Tiksol} is a low-pass graph filter and the higher $\gamma$, the more low-pass the filter.

\emph{Sobolev \cite{giraldo2022reconstruction}:} The Sobolev regularizer increases the flexibility by allowing for a more expressive rational frequency response. Specifically, it focuses on solving
\begin{equation}\label{eq.Sobproblem}
\underset{\vcy \in \fdR^N}{\text{argmin}}~\|\vcx - \vcy	\|_2^2 + \gamma \vcy^\top(\mtL + \epsilon\mtI)^\beta\vcy,
\end{equation}
with $\epsilon \ge 0$ and $\beta \in \fdR_+$. The closed-form solution is
\begin{equation}\label{eq.Sobsol}
\vcy^* = \big(\mtI + \gamma(\mtL + \epsilon\mtI)^\beta\big)^{-1}\vcx,
\end{equation}
which corresponds to a rational filter with the frequency response $\fnth(\lambda) = (1 + \gamma(\lambda+\epsilon)^\beta)^{-1}$. Here, the scalar $\beta$ controls the expressivity order of this function (cf. $P$ in \eqref{eq.ratFresp}) and $\gamma, \epsilon$ are the parameters of such a rational response.

\emph{Quadratic shift variation \cite{sandryhaila2014discrete}:} When the graph is directed, we measure the variability as the change between the signal $\vcy$ and its shifted version $\mtS\vcy$. Thus, we can recover a smooth signal over a directed graph by solving
\begin{equation}\label{eq.Variation2}
 \underset{\vcy \in \fdR^N}{\text{argmin}}~\|\vcx - \vcy	\|_2^2 + \gamma \| \vcy - \mtS\vcy\|_2^2,
\end{equation}
which has a closed-form solution
%
\begin{equation}\label{eq.totVsol}
\vcy^* = \big(\mtI + \gamma(\mtI - \mtS -\mtS^\top + \mtS^\top\mtS)		\big)^{-1}\vcx.
\end{equation}
This is again an inverse graph filtering, but it does not admit a straightforward spectral analogy as \eqref{eq.Tiksol} and \eqref{eq.Sobsol}.
The role of the regularization parameter $\gamma$ is discussed from a bias-variance perspective in \cite{chen2017bias} and from a graph-kernel perspective in \cite{romero2016kernel}. Differently, \cite{yang2021node} generalizes \eqref{eq.Tiksol} to the case where each node has its own regularization parameters (i.e. a vector of parameters $\vcgamma$), yielding a rational node varying filter (cf. \eqref{eq.NVfilt}). 

\smallskip
\noindent \textbf{Sparsity filtering.} 
These approaches leverage the prior that the signal has discontinuities across neighbors or shifts; e.g., a piecewise smooth signal that has homogeneous values within a group of nodes but can have arbitrarily large variations between groups. 
For undirected graphs, sparsity filtering is implemented via graph trend filtering (GTF), while for directed graphs it is implemented via the total variation in \eqref{eq.totV_dir}. 

\emph{Trend filtering \cite{wang2015trend}:} Let $\mtDelta \in \fdR^{N\times |\stE|}$ be the oriented incidence matrix of an undirected graph $\stG$, whose rows are indexed by the nodes and columns by the edges. 
The operation $\mtDelta^\top\vcx$ computes the pairwise difference between signal values on each edge; hence, $\mtDelta^\top$ can be interpreted as a graph difference operator. In fact, since $\mtL = \mtDelta\mtDelta^\top$, the regularizer in \eqref{eq.Tikproblem} can be written as $\LQ(\vcy) = \vcy^\Tr\mtL\vcy = \|\mtDelta^\top\vcy	\|_2^2$, i.e., the squared $\ell_2-$norm of the difference vector. Instead, the GTF works with regularized problems of the form
\begin{equation}\label{eq.GTF1}
 \underset{\vcy \in \fdR^N}{\text{argmin}}~\|\vcx - \vcy	\|_2^2 + \gamma\|\mtDelta^\top\vcy	\|_1,
\end{equation}
which penalizes the absolute difference of the signal variation in connected nodes. Problem~\eqref{eq.GTF1} is an order $K = 1$ GTF and estimates a signal $\vcy$ whose differences are nonzero only at a few edges. Higher-order GTFs substitute the indicence matrix $\mtDelta^{(1)}:=\mtDelta$ in \eqref{eq.GTF1} with the higher-order versions $ K \ge 1$
%
%
  \[
    \mtDelta^{(K+1), \top}=\left\{
                \begin{array}{ll}
                  \mtDelta\mtDelta^{(K),\top} = \mtL^{\frac{K+1}{2}}, &\text{for odd }K\\
                  \mtDelta^\top\mtDelta^{(K),\top} = \mtDelta^\top\mtL^{\frac{K}{2}}, &\text{for even }K\\
                \end{array}
              \right. .
  \]
For an odd $K$, the GTF recovers a signal that has sparse diffused versions $\mtL\vcy, \mtL^2\vcy$, while for an even $K$, it recovers a signal that has sparse differences of the shifted versions $\mtDelta^\top\mtL\vcy, \mtDelta^\top\mtL^2\vcy$ etc. These sparsity constraints capture discontinuities in the graph signal and recover piecewise constant signals better than smooth filtering methods.
One of the challenges of the GTF is that solving problem \eqref{eq.GTF1} requires running iterative algorithms. In addition, the $\ell_1-$norm in \eqref{eq.GTF1} may often penalize towards zero when the signal components are large. To overcome the latter, the work in \cite{varma2019vector} proposes a GTF with a non-convex regularizer.

\emph{Total variation \cite{sandryhaila2014discrete}:} This is the straightforward extension of \eqref{eq.Variation2} that uses the regularizer $\TV_{1}(\vcy)$; i.e., it solves
\begin{equation}\label{eq.TV1}
 \underset{\vcy \in \fdR^N}{\text{argmin}}~\|\vcx - \vcy	\|_2^2 + \gamma\| \vcy - \mtS\vcy\|_1.
\end{equation}
This means that we are penalizing shifted variations that are substantial only at a few nodes. Similar to the GTF, this is also a convex problem that can be solved with iterative algorithms.

\smallskip
\noindent \textbf{Wiener filtering.} The above regularizers do not consider any statistical behavior of the true signal. When this signal exhibits a graph wide sense stationary behavior \cite{girault2015stationary, perraudin2017stationary,marques2017stationary} or when it is generated by a Gaussian-Markov random field \cite{zhang2015graph}, we can incorporate such a prior to recover the optimal signal in a Wiener filtering sense. 

First, consider a signal from a distribution $\vcy \sim\stD(\vcZeros, \mtSigma_y)$ with covariance matrix $\mtSigma_y$ and let the noise be additive with zero mean and covariance matrix $\mtSigma_n$. Given $\vcy$ and $\vcn$ are mutually independent, the Wiener filter comprises solving
%
\begin{align}\label{eq.WeinFilt}
\begin{split}
 \mtH^* \!=\! \underset{\mtH \in \fdR^{N\times N}}{\text{argmin}} 	\fdE\| \mtH(\vcy + \vcn) - \vcy		\|_2^2 \!=\! \mtSigma_y(\mtSigma_y + \mtSigma_n)^{-1},
 \end{split}
\end{align}
and setting the solution to $\vcy = \mtH^*\vcx$. 
When the process $\vcy$ is defined over a graph and the covariance matrices have the same eigenvectors as the GSO -- i.e., $\mtSigma_y = \mtV\diag(\sigma_y^2(\vclambda))\mtV^{\fnH}$, $\mtSigma_n =  \mtV\diag(\sigma_n^2(\vclambda))\mtV^{\fnH}$ (independent noise) -- the Wiener filter in \eqref{eq.WeinFilt} reduces to a graph Wiener filter $\mtH(\mtS)$ \cite{girault2014semi}. This graph Wiener filter has the frequency response
%
%
\begin{equation}\label{eq.Wienresp}
\fnth(\lambda) = \frac{\sigma_d^2(\lambda)}{\sigma_d^2(\lambda) + \sigma_n^2(\lambda)} = \frac{1}{1 + \frac{\sigma_n^2(\lambda)}{\sigma_d^2(\lambda)}},
\end{equation}
which is a rational filter, and the response at frequency $\lambda$ is controlled by the inverse signal-to-noise (SNR) ratio $\text{SNR}^{-1}(\lambda):={\sigma_n^2(\lambda)}/{\sigma_d^2(\lambda)}$. Contrasting \eqref{eq.Wienresp} with the other regularized filters, we see that the Wiener filter does not imply a constant regularization weight $\gamma$ for each frequency $\lambda$, but rather a frequency-adaptive regularizer given by the inverse $\text{SNR}$. Similar to the rational graph filters discussed above, the output of the Wiener filter can be obtained with conjugate gradient methods; however, the matrices $\mtSigma_y, \mtSigma_n$ are typically dense. One way to tackle this is to approximate $\fnth(\lambda)$ with polynomial or rational filters and then implement it via iterative recursions \cite{isufi2018distributed,zheng2022wiener}.

The main challenge of these filters is to identify a good regularizer or a combination thereof that represents the data. Often this may be a challenging task requiring domain expertise, hence advocating for the easier solution to use more general graph filters as input-output mappings. 

\subsection{Multi-GSO Filters}\label{subsec_multGSO}

Unlike classical signal processing where the shift operation is a time delay, in the graph setting, different choices of the graph shift operator for the data are often possible, especially in abstract networks (Sec.~\ref{subsec:graphs}).
 Designing or learning both the filter coefficients and the GSO is challenging because of the powers of $\mtS$ appearing in the filter expression (e.g., \eqref{eq:graphConv}), and because of the high DoFs that can cause overfitting. A way to circumvent these challenges is to build a graph filter operating on multiple pre-specified GSOs~\cite{sevi2018harmonic,Hua2019-combination, fan2019global,emirov2022polynomial}. Given $Q$ GSOs $\{ \mtS_q\}_{q=1}^Q$, we define a multi-GSO graph filter as
\begin{equation}\label{eq:multiGSOfilt}
   \fnH(\vcx) = \sum_{q =1}^Q \sum_{k = 0}^K h_{qk} \mtS_q^k \vcx = \mtH(\{ \mtS_q\}_{q=1}^Q)\vcx,
\end{equation}
where $\{h_{qk}\}$ is the parameter applied to the $k$th signal shift with respect to the $q$th GSO. The multiple GSOs now act as inductive biases about the graph and / or data to aid modeling. In contrast to e.g., learning the GSOs, this approach reduces the filter parameters to $Q(K+1)$ and the computational cost to $\fnO(QK|\stE|)$. In addition, because these filters are linear in the parameters, their data-driven design reduces to solving a least squares problem, similar to convolutional filtering.

\section{Graph Filter Banks and Wavelets} \label{sec:wavelets}


{In many instances, a single graph filter suffices to smooth data, identify discontinuities, or classify a signal. However, the outputs of multiple filters (a \emph{filter bank}) can also be combined to generate more nuanced representations of the data. The combined filter coefficients can serve as feature vectors in machine learning tasks or be leveraged in regularization problems when one has \emph{a priori} modeling information that the graph signal of interest belongs to a class of signals whose filter bank coefficients exhibit specific structural patterns (e.g., they are sparse).

Throughout this section, unless stated otherwise, we assume the underlying graph is undirected and the graph shift operator is Hermitian (including real symmetric).}

%
%
\begin{figure*}[t]
\centering
\includegraphics[width=7in]{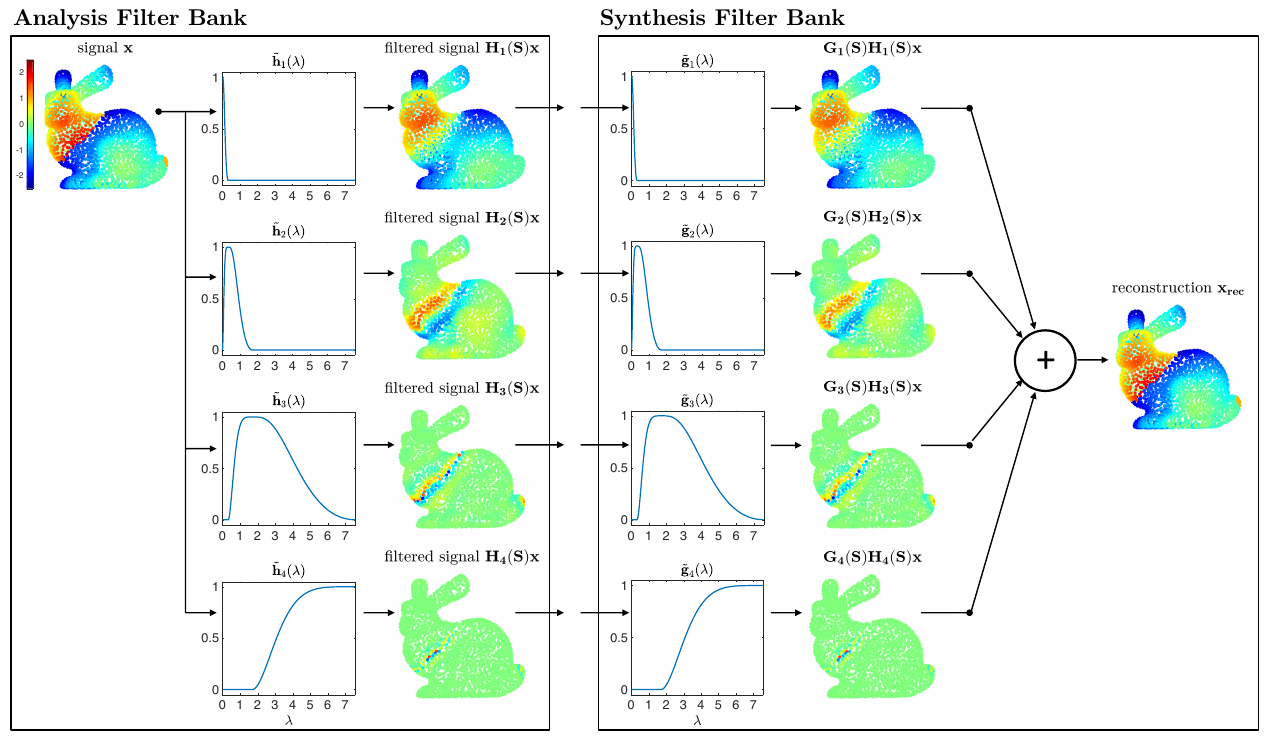}
\caption{An undecimated single-level four-channel graph filter bank. 
The signal is piecewise smooth with respect to the Stanford bunny graph \cite{bunny}. As a result, the non-zero coefficients of the bandpass and highpass channels ($m=2,3,4$) cluster around the two discontinuities at the midsection and tail of the bunny. The synthesis filters $\{\fntg_m\}$ used here are the same as the analysis filters $\{\fnth_m\}$, although they do not need to be in general. The filters $\{\fnth_m\}$ chosen for this example with the design method of \cite{shuman2013spectrum} satisfy the tight Parseval frame condition \eqref{Eq:Parseval}, leading to perfect reconstruction.}
\label{Fig:filter_bank}
\end{figure*}


\subsection{Undecimated Single-Level $M$-Channel Graph Filter Banks}
A single-level graph filter bank without any downsampling (\emph{undecimated}) applies $M$ different filters to a signal $\vcx$ and concatenates the outputs into a single vector of length $MN$: 
\begin{eqnarray*}
\vcalpha := \left[\mtH_1(\mtS)\vcx;~\mtH_2(\mtS)\vcx;~\ldots;~\mtH_M(\mtS)\vcx \right].
\end{eqnarray*}
An example shown in Fig. \ref{Fig:filter_bank}.
When the filters, often called the \emph{analysis filters}, are linear, the graph filter bank constitutes a linear transform from $\fdX^{\stV}$ (the graph signal) to $\fdX_{MN}$ (the filtered signals). Equivalently, we can interpret each of the $MN$ output coefficients as the inner product between the graph signal $\vcx$ and a \emph{dictionary atom} of the form 
$\vcvphi_{im}:=\mtH_m(\mtS)\vcdelta_i,$
where $[\vcdelta_i]_j =1$ if $j=i$ and 0 otherwise. Each atom $\vcvphi_{im}$ can be viewed as a pattern defined through the filter $\vcth_m$ in the spectral domain and then centered at vertex $i$ (cf. \cite[Fig. 1]{shuman2020Localized}).

The most common method to reconstruct the signal from the output coefficients is through a synthesis filter bank. For an undecimated single-level $M$-channel graph filter bank, shown in Fig. \ref{Fig:filter_bank}, the reconstructed signal is given by
\begin{eqnarray}\label{Eq:bank_rec}
\vcx_{\mathbf{rec}}=\sum_{m=1}^M \mtG_m(\mtS) \mtH_m(\mtS)\vcx,
\end{eqnarray}
where $\{\mtG_m(\mtS)\}$ are the \emph{synthesis filters}.

\smallskip
\noindent \textbf{Parseval frames.} 
{
The dictionary atoms  $\left\{\vcvphi_{im}\right\}_{i=1,2,\ldots,N;m=1,2,\ldots,M}$ form a \emph{tight Parseval frame} if $\sum_{m=1}^M \sum_{i=1}^N |\langle \vcx,\vcvphi_{im}\rangle|^2 =||\vcx||_2^2$.
A sufficient condition for these atoms to form a tight Parseval frame is that}
\begin{eqnarray} \label{Eq:Parseval}
\sum_{m=1}^M [\fnth_m(\lambda_i)]^2=1,\hbox{ for each }i=1,2,\ldots,N; 
\end{eqnarray}
{that is, the chosen filters cover the entire spectrum evenly in the sense that the sums of their squared values are the same at every eigenvalue.}
Benefits of designing the filters to meet this condition include (i) $||\vcx||_2 = ||\vcalpha||_2$; i.e., the filter bank preserves the energy of the signal, which also helps avoid numerical instabilities; and (ii) using the same filters for the synthesis filter bank as the analysis filter bank (i.e., $\mtG_m(\mtS)=\mtH_m(\mtS)$ for all $m$) results in perfect reconstruction of the signal, because
\begin{eqnarray*}
\sum_{m=1}^M \mtH_m(\mtS) \mtH_m(\mtS)\vcx =  \mtV \left[ \sum_{m=1}^M [\diag(\vcth_m)]^2 \right]\mtV^{\fnH}\vcx = \vcx.
\end{eqnarray*} 
Examples of such tight spectral graph filter frames include those constructed and investigated in \cite{leonardi_multislice,shuman2013spectrum,dong2017sparse,gobel2018construction}.

As discussed in Sec. \ref{sec:filterDesign}, using polynomial filters circumvents the need to exactly compute the eigenvectors of $\mtS$ and also enables local processing. However, \cite{tay2017almost} shows that it is not possible to design a filter bank comprised of polynomial filters that satisfies $\sum_{m=1} [\fnth_m(\lambda)]^2=1$ for all $\lambda \in [\lambda_{\min},\lambda_{\max}]$ (the desired condition for a graph-independent tight frame guarantee since the idea is to not compute all of the eigenvalues). References \cite{tay2017almost,sakiyama2016spectral,fan_algorithms} explore different methods to design polynomial filter banks that approximately satisfy the tight frame condition, while \cite{jiang2019nonsubsampled} allows the polynomial synthesis filters $\{\fntg_m\}$ to be different from the polynomial analysis filters $\{\fnth_m\}$, and outlines a method to design the filters to satisfy $\sum_{m=1}^M \fnth_m(\lambda)\fntg_m(\lambda)=1$, guaranteeing perfect reconstruction.
\smallskip

\noindent \textbf{Spectral graph wavelets.} A seminal example of undecimated graph filter banks are \emph{spectral graph wavelets}, introduced in \cite{hammond2011wavelets} 
and later extended to tight frames \cite{leonardi_multislice,shuman2013spectrum,dong2017sparse,gobel2018construction}. Analogous to wavelet filter banks for discrete-time signals (see, 
e.g., \cite{strang1996wavelets}), choosing the filters to be dilated versions of each 
other with wider support in the highpass filters at the upper end of the spectrum yields atoms that are increasingly (as the filters become more dilated) localized in the vertex domain. As a result, 
the spectral graph wavelet filter bank coefficients are sparse for signals that are smooth or piecewise smooth with respect to the underlying graph \cite{hammond2011wavelets,ricaud_sparsity_SPIE_2013}. This phenomenon is illustrated in Fig. \ref{Fig:filter_bank}, where the coefficients in the bandpass and highpass filters (channels $m=2,3,4$) are  (i) close to 0 except around the discontinuities in the piecewise smooth signal, and (ii) increasingly sparse at higher scales (larger $m$). 
Spectral graph wavelets have been applied in community mining \cite{tremblay2014graph}, mobility pattern analysis \cite{dong2013inference}, semi-supervised learning \cite{shuman_SSL_SAMPTA_2011}, 3D action recognition from depth cameras \cite{kerola2014spectral}, fMRI data analysis \cite{behjat2015anatomically}, and network topology analysis \cite{donnat2018learning}.



%

\subsection{Downsampling and Critically-Sampled Graph Filter Banks}


Without any downsampling, the $M$-channel graph filter bank is a redundant transform. In many applications, this is just fine and the fact that the output coefficients are sparse for specific classes of signals can be leveraged in regularization and machine learning problems. In some applications, it is desirable to subsample the output coefficients, keeping only those associated with the vertices in the set $\stV_m$ at the $m$th channel, reducing the overall storage cost. When $\sum_{m=1}^M |\stV_m| =N$, the filter bank is said to be \emph{critically sampled} \cite{narang_icip}. 
With a typical synthesis filter bank comprised of upsampling the output coefficients from each channel, filtering, and summing, the reconstructed signal is given by (cf. \eqref{Eq:bank_rec} for the effect of the downsampling and upsampling):
\begin{eqnarray}\label{Eq:rec_cs}
\vcx_{\mathbf{rec}}=\sum_{m=1}^M \mtG_m(\mtS) \mtM_{\stV_m}^{\top} \mtM_{\stV_m} \mtH_m(\mtS)\vcx,
\end{eqnarray}
where $\mtM_{\stV_m}$ is a $|\stV_m| \times N$ selection matrix with $\left[\mtM_{\stV_m}\right]_{k,i}=1$ if vertex $i$ is the $k$th element of $\stV_m$ and 0 otherwise{, and $\mtM_{\stV_m}^{\top}$ is the corresponding upsampling operator.} 
\smallskip

\noindent \textbf{Perfect reconstruction.}  While \cite{anis2017critical} investigates how to select the sampling sets $\{\stV_m\}_{m}$ to minimize the reconstruction error $||\vcx_{\mathbf{rec}}-\vcx||_2$ for a fixed choice of filters, a broader question is whether it is possible to jointly select the filters $\{\mtH_m(\mtS)\}_{m}$ and $\{\mtG_m(\mtS)\}_{m}$ and the sampling sets $\{\stV_m\}_{m}$ to recover the original signal $\vcx$ exactly from the subsampled outputs $\left\{ \mtM_{\stV_m} \mtH_m(\mtS)\vcx \right\}_{m}$.

\begin{figure*}[t]
\centering
\includegraphics[width=7.1in]{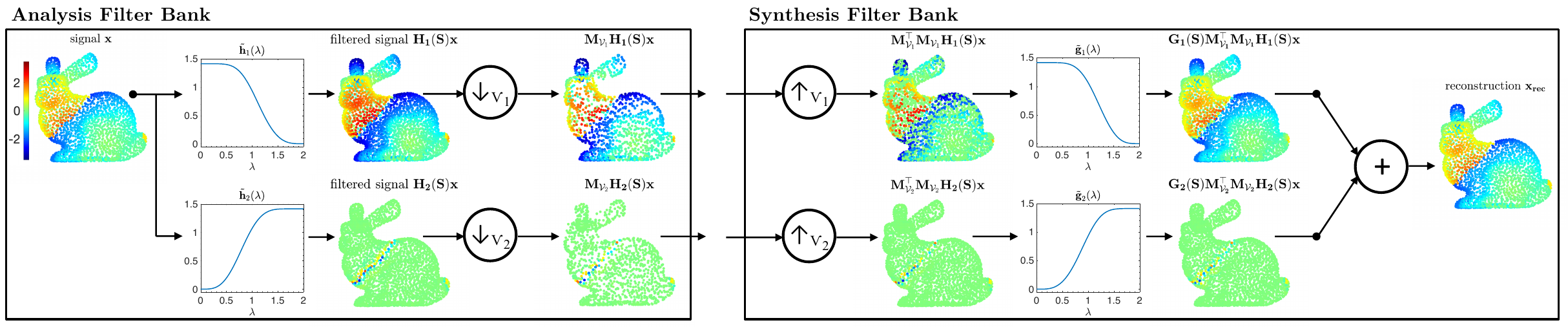}
\caption{A single-level critically-sampled two-channel generalized graph filter bank for the same signal shown in Fig. \ref{Fig:filter_bank}. This filter bank combines the perfect reconstruction biorthogonal filters of \cite[Ex. 3]{tay2015techniques} with the generalized filter bank approach of \cite{pavez2022two} for arbitrary graphs. The graph is partitioned into two approximately equal-sized complementary sets of vertices, $\stV_1$ and $\stV_2$. Here, $\mtbS=\vcL$,
$\vcQ=
\begin{bmatrix}
[\vcL]_{\stV_1,\stV_1} & \vcZeros \\
\vcZeros & [\vcL]_{\stV_2,\stV_2}
\end{bmatrix}$,
 $\vcS=\vcQ^{-1}\vcL$ {(not Hermitian in general), and $\vcH_i(\vcS)=\vcV \fnh_i(\vcLambda)\vcV^{-1}=\vcV \fnh_i(\vcLambda)\vcV^H \vcQ$.} Although they look similar in shape, the synthesis filters are not the same as the analysis filters.}\label{Fig:filter_bank2}
\end{figure*}

Indeed, when the underlying graph has special structural properties, it is possible to guarantee perfect reconstruction. For example, 
when the graph is bipartite and $\mtS=\mtL_{\text{n}}$, the spectrum of normalized Laplacian eigenvalues (contained in $[0,2]$) is symmetric around $\lambda=1$ and the eigenvectors associated with eigenvalues $\lambda$ and $2-\lambda$ are closely related, leading to a \emph{spectral folding} effect analogous to aliasing in one-dimensional signal processing. 
Specifically, with $M=2$ and the downsampling sets selected according to the bipartition $\{\stV_1,\stV_2 \}$, for each eigenvalue and $m=1,2$, we have
\begin{eqnarray}\label{Eq:spec_folding}
\vcGamma_{\lambda}\left(\mtM_{\stV_m}^{\top} \mtM_{\stV_m}\vcx\right)=\frac{1}{2}\left[\vcGamma_{\lambda}(\vcx) +\vcJ_{\stV_m}\vcGamma_{2-\lambda}(\vcx) \right],
\end{eqnarray}
where  $\vcGamma_{\lambda}$ performs an orthogonal projection of a vector onto the eigenspace associated with eigenvalue $\lambda$,
and $\vcJ_{\stV_m}=2\mtM_{\stV_m}^{\top} \mtM_{\stV_m}-\vcI_N$. A key takeaway from \eqref{Eq:spec_folding} is that the portion of the downsampled and upsampled signal in the eigenspace associated with $\lambda$ only depends on the portions of the original signal in the eigenspaces associated with $\lambda$ and $2-\lambda$.
Ref. \cite{narang2012perfect} shows that for this case, the following two conditions are necessary and sufficient for perfect reconstruction
\begin{eqnarray*}
\fntg_1(\lambda) \fnth_1(\lambda)+\fntg_2(\lambda)\fnth_2(\lambda)&=&2, \\
\fntg_1(\lambda) \fnth_1(2-\lambda)-\fntg_2(\lambda)\fnth_2(2-\lambda)&=&0.
\end{eqnarray*}
Leveraging these conditions, \cite{narang2012perfect,narang_bior_filters,zeng2017bipartite,tay2015techniques,tay2017bipartite,tay2017critically} design two-channel critically-sampled perfect reconstruction graph filter banks. 

Other special types of graph with structural properties that can be leveraged to generate critically-sampled perfect reconstruction graph filter banks include shift-invariant graphs that have a circulant graph Laplacian \cite{ekambaram_icip,ekambaram2013globalsip,kotzagiannidis2017splines} and $M$-block cyclic graphs \cite{teke2016,teke2017ii,tay2019m}. 

The generalized critically-sampled filter banks of \cite{pavez2022two} extend the spectral folding idea from bipartite graphs to arbitrary graphs. They do this by taking
the 
filtering basis vectors $\{\bar{\vcv}_i\}_{i=1,2,\ldots,N}$  to be the solutions to the \emph{generalized} eigenvalue problem
 \begin{equation}\label{Eq:gen_eig}
 \mtbS \bar{\vcv}_i=\bar{\lambda}_i \vcQ \bar{\vcv}_i,
 \end{equation}
 so that {$\bar{\vcv}_i^{\fnH}\vcQ \bar{\vcv}_j = 0$} for $i \neq j$; i.e., the filtering basis is {orthonormal} with respect to the inner product {$\langle \bar{\vcv}_i, \bar{\vcv}_j \rangle_{\vcQ}:=\bar{\vcv}_j^{\fnH}\vcQ \bar{\vcv}_i$} instead of the standard dot product. In \eqref{Eq:gen_eig},
 $\mtbS$ is  a {Hermitian (including real symmetric)} positive semi-definite matrix with {off-diagonal} sparsity pattern matching the adjacency matrix (e.g., a Laplacian). If, for any partition $\{\stV_1,\stV_2\}$ of the vertices, $\vcQ$ is selected to be equal to $\left[
\begin{array}{cc}
[\mtbS]_{\stV_1,\stV_1} & \vcZeros \\
\vcZeros & [\mtbS]_{\stV_2,\stV_2}
\end{array}
\right]$, then a spectral folding property analogous to the one for bipartite graphs holds (but this time for arbitrary graphs), leading to perfect reconstruction. Fig.~\ref{Fig:filter_bank2} shows an example of such a critically-sampled two-channel generalized graph filter bank. 
\smallskip

\noindent \textbf{Orthogonality and biorthogonality.}  
A critically-sampled filter bank is said \emph{orthogonal} if $\sum_{m=1}^M \mtH_m(\mtS) \mtM_{\stV_m}^{\top} \mtM_{\stV_m} \mtH_m(\mtS)=\mtI_N$, in which case selecting the synthesis filters to be the same as the analysis filters leads to perfect reconstruction (cf. \eqref{Eq:rec_cs}), and is said to be \emph{biorthogonal} if $\sum_{m=1}^M \mtG_m(\mtS) \mtM_{\stV_m}^{\top} \mtM_{\stV_m} \mtH_m(\mtS)=\mtI_N$, again guaranteeing perfect reconstruction. 
Refs.  \cite{narang2012perfect,narang_bior_filters,tay2015techniques,tay2014design,zhang2016design,tay2016near,tay2017bipartite} examine orthogonal, near orthogonal, and biorthogonal filter designs for critically-sampled filter banks on bipartite graphs. The primary motivation for using biorthogonal filters with bipartite graphs is that it is impossible to choose polynomial filters that yield an orthogonal filter bank \cite{narang_bior_filters}. 

\subsection{Alternative Structures for Arbitrary Graphs}
A number of alternative structures for perfect signal reconstruction on arbitrary graphs have also been proposed:
\begin{enumerate}
\item graph extensions of lifting transforms \cite{jansen,narang_lifting_graphs,tay2018cascade,jiang2020design}, pyramid transforms \cite{shuman_TSP_multiscale}, and oversampled filter banks \cite{tanaka2014m,sakiyama2014oversampled};
\item subgraph-based filter banks for graph signals \cite{tremblay2016subgraph} where the downsampling is performed by partitioning the graph into connected subsets of vertices and representing each subset by a single supernode;
\item filter banks where the synthesis portion (upsampling and filtering) is replaced with a different interpolation operator  \cite{chen2015discrete,li_mcsfb_2018};
\item filter banks where the downsampling is performed in the graph spectral domain instead of in the vertex domain \cite{sakiyama2019two,sakiyama2019m,miraki2021spectral};
\item filter banks that first replace the arbitrary underlying graph by a maximum spanning tree \cite{nguyen2014downsampling,zheng2019framework};
\item multi-dimensional separable filter banks that first decompose an arbitrary graph into sums of bipartite graphs \cite{narang2012perfect,narang_bior_filters,zeng2017bipartite};
\item filter banks that first decompose an arbitrary graph into sums of circulant graphs \cite{ekambaram_icip,ekambaram2013globalsip,kotzagiannidis2017splines};
\item filter banks that work with a similarity-transformed adjacency matrix \cite{teke2017ii}. 
\end{enumerate}


\subsection{Multi-Level Graph Filter Banks} 
In classical multi-level filter banks for time series data or images, multiple levels of filtering and downsampling are applied. For example, in the classical logarithmic wavelet filter bank, at each level, another filter bank is applied to the downsampled output of the lowpass channel from the previous level \cite{strang1996wavelets}. Numerous works have investigated extensions to multi-level filter banks, lifting transforms, and pyramids for graph signals (e.g., \cite{narang2012perfect,ekambaram_icip,ekambaram2013globalsip,shuman_TSP_multiscale,tremblay2016subgraph,sakiyama2019two,pavez2022two}). In classical time series analysis or image processing, the structure of the underlying domain enables regular sampling (e.g., every other time sample) that preserves the notion of frequency entailed by filtering at each level of the multi-level filter bank. 
One main difference and significant challenge in the graph setting is that -- unless the graph is highly symmetric -- it is not obvious how to define a coarser graph at each subsequent level of the filter bank in a way that maintains a clear correspondence between the eigenvectors of the shift operator that are used for graph filtering at one level, and the eigenvectors of the shift operator on the coarsened graph that are used for filtering the downsampled signal on that coarsened graph (c.f.,\cite{loukas2018spectrally,loukas2019graph,jin2020graph}).

%


\subsection{Data-Adapted Transforms / Dictionary Learning}
All of the design elements discussed so far -- the graph(s), filters, and downsampling sets -- can be adapted either to the specific graph signal being analyzed or to an additional set of representative training signals. For example, \cite{thanou2014learning} presents a method to learn polynomial filters that yield sparse representations of the training signals and \cite{behjat2016signal} presents a method to learn filters that yield a tight frame with each resulting filter subband capturing the same amount of energy on average across the training signals. This approach is particularly beneficial when the energy of a typical signal from the class of interest is concentrated on a small region of the spectrum, which the authors show is the case for brain fMRI data \cite{behjat2016signal}. Meanwhile, \cite{narang2013critically} is just one of many examples of constructing the underlying graph from the signal, in this case for the purpose of image coding.

\section{Graph Neural Networks} \label{sec:GNN}

Graph neural networks (GNNs) are nonlinear layered architectures, in which each layer comprises a bank of graph filters (Sec.~\ref{sec:wavelets}) and an activation function that is (typically) pointwise and nonlinear \cite{Scarselli2009-GNN, bronstein2021geometric, gama2020graphs}. {This nonlinear nature allow us to capture more complex relationships than the linear graph filters, and their compositional form allows for a sequential extraction of features, typically enhancing representational capabilities over simpler nonlinear graph filters.} 

The basic building block of GNNs is the \emph{graph perceptron}, which is a straightforward extension of graph filters \cite{gama2020graphs}.

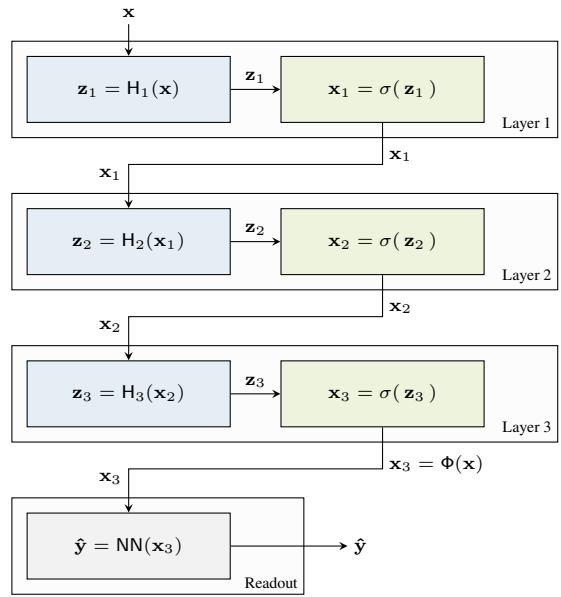
\begin{figure}[!t]
\vspace{-.1in}
    \centering
    {\def \myfactor {0.675}

\def \unit  {\myfactor cm}

\tikzstyle{block} = [ rectangle,
                      minimum width = \unit,
                      minimum height = \unit,
                      fill = figBaseColor!10,
                      draw = black,
                      text = black]

\tikzstyle{filter} = [ block,
                      minimum width  = 4.0*\unit,
                      minimum height = 1.3*\unit]

\tikzstyle{nonlinearity} = [ filter,
                             minimum width  = 4.0*\unit,
                             fill = figThirdColor!20]

\def \deltainput     {( 0.0,-1.5)}
\def \deltaoutput    {( 0.0,-0.9)}
\def \deltalayer     {3.0}
\def \deltaconnector {0.95}
\def \deltasigma     {( 5.0, 0.0)}
\def \deltafeature   {1.5}

\def \one   {$\displaystyle{\vcz_{1}  = \fnH_{1}(\vcx)    }$}
\def \two   {$\displaystyle{\vcz_{2}  = \fnH_{2}(\vcx_{1}) }$}
\def \three {$\displaystyle{\vcz_{3}  = \fnH_{3}(\vcx_{2})}$}
\def \readout{$\displaystyle{\vchy = \mathsf{NN}(\vcx_{3})}$}
\def \sigmaone   {$\displaystyle{\vcx_{1} = {\sigma} (\, \vcz_1 \, )}$}
\def \sigmatwo   {$\displaystyle{\vcx_{2} = {\sigma} (\, \vcz_2 \, )}$}
\def \sigmathree {$\displaystyle{\vcx_{3} = {\sigma} (\, \vcz_3 \, )}$}

%
{\fontsize{7}{7}\selectfont\begin{tikzpicture}[scale = \myfactor]

  \pgfdeclarelayer{bg}     
  \pgfsetlayers{bg,main}   

  \node (input) [rectangle, minimum width = 0.1*\unit] {$\vcx$};
  \path (input)      ++ \deltainput node [filter]       (L1 Filter1) {\one};
  \path (L1 Filter1) ++ \deltasigma node [nonlinearity] (L1 F1)      {\sigmaone};
  \path[draw, -stealth] (L1 Filter1.east) -- node [above] {$\vcz_1$} (L1 F1.west);

  \path (L1 Filter1) ++ (0,-\deltalayer) node [filter]       (L2 Filter1) {\two};
  \path (L2 Filter1) ++ \deltasigma      node [nonlinearity] (L2 F1)      {\sigmatwo};
  \path[draw, -stealth] (L2 Filter1.east) --  node [above] {$\vcz_2$} (L2 F1.west);

  \path (L2 Filter1) ++ (0,-\deltalayer) node [filter]       (L3 Filter1) {\three};
  \path (L3 Filter1) ++ \deltasigma      node [nonlinearity] (L3 F1)      {\sigmathree};
  \path[draw, -stealth] (L3 Filter1.east) --  node [above] {$\vcz_3$} (L3 F1.west);
  
  \path (L3 Filter1) ++ (0,-\deltalayer) node [filter, fill = black!5]       (Readout) {\readout};

  \path[draw, -stealth] (input.south) -- (L1 Filter1.north);
  \path (L1 F1.south) ++ (0,-\deltaconnector) node [] (aux1) {};
  \path[draw, -stealth] (L1 F1.south) -- node [below right] {$\vcx_1$} (aux1.north)
                                      --                         (aux1.north -| L2 Filter1.north)
                                      -- node [above left]  {$\vcx_1$} (L2 Filter1.north);
  \path (L2 F1.south) ++ (0,-\deltaconnector) node [] (aux1) {};
  \path[draw, -stealth] (L2 F1.south) -- node [below right] {$\vcx_2$} (aux1.north)
                                      --                         (aux1.north -| L2 Filter1.north)
                                      -- node [above left]  {$\vcx_2$} (L3 Filter1.north);
  \path (L3 F1.south) ++ (0,-\deltaconnector) node [] (aux1) {};
  \path[draw, -stealth] (L3 F1.south) -- node [below right] {$\vcx_3 = \fnPhi(\vcx)$} (aux1.north)
  --                         (aux1.north -| Readout.north)
  -- node [above left]  {$\vcx_3$} (Readout.north);
  \path[draw, -stealth] (Readout.east) -- ++ (2.0, 0) -- ++ (0.3, 0)
                        node [right]{$\vchy$};

  \begin{pgfonlayer}{bg}
      \path (L1 Filter1.west |- L1 F1.south) ++ (-0.3,-0.3)
           node [filter, anchor = south west,
                 fill = black!1,
                 minimum width  = 10.75*\unit,
                 minimum height = 1.9*\unit,]
        (layer)
        {};
      \path (layer.south east) ++ (0.0,0.0) node [above left] {\fontsize{6}{6}\selectfont Layer 1};
      \path (L1 Filter1.west |- L2 F1.south) ++ (-0.3,-0.3)
           node [filter, anchor = south west,
                 fill = black!1,
                 minimum width  = 10.75*\unit,
                 minimum height = 1.9*\unit,]
        (layer)
        {};
      \path (layer.south east) ++ (0.0,0.0) node [above left] {\fontsize{6}{6}\selectfont Layer 2};
      \path (L1 Filter1.west |- L3 F1.south) ++ (-0.3,-0.3)
           node [filter, anchor = south west,
                 fill = black!1,
                 minimum width  = 10.75*\unit,
                 minimum height = 1.9*\unit,]
        (layer)
        {};
      \path (layer.south east) ++ (0.0,0.0) node [above left] {\fontsize{6}{6}\selectfont Layer 3};
      \path (L2 Filter1.west |- Readout.south) ++ (-0.3,-0.3)
      node [filter, anchor = south west,
      fill = black!1,
      minimum width  = 5.75*\unit,
      minimum height = 1.9*\unit,]
      (layer)
      {};
      \path (layer.south east) ++ (0.0,0.0) node [above left] {\fontsize{6}{6}\selectfont Readout};

  \end{pgfonlayer}

\end{tikzpicture}}}
    \caption{Schematic for a GNN with $3$ layers. The input $\vcx$ is processed by a graph filter $\fnH_{1}$ and then an activation function $\fnsigma$. The output of this block -- a graph perceptron -- is fed into another graph perceptron corresponding to layer $2$. The output of the GNN is the output of the third, cascaded graph perceptron. 
    Each layer has a different filter whose coefficients are learned from data. If required for the problem, the output of the last layer of the GNN, in this case $\fnPhi(\vcx) = \vcx_{3}$ can be fed into a readout layer to finally compute the target value $\vchy$. 
    Depending on the nature of this readout layer, the distributed nature of the GNN may be violated. 
    See paragraph on `Readout Layer' for more details.}
    \label{fig:GNN}
 \vspace{-.2in}
\end{figure}

\begin{defBox}{Graph perceptron} A graph perceptron is a nonlinear mapping comprising a linear graph filter $\fnH(\vcx)$ nested into an activation function (a pointwise nonlinear function) $\fnsigma:\fdR \to \fdR$, i.e.,
\begin{equation}
\vcy = \fnsigma(\fnH(\vcx))
\end{equation}
where $\fnsigma(\vcx)$ signifies $[\fnsigma(\vcx)]_{i} = \fnsigma([\vcx]_{i})$.
\end{defBox}
Graph perceptrons can be built using any of the filters reported in Table~\ref{tab:filters} and the activation function can take different forms e.g., $\sigma(x) = \text{ReLU}(x) = \max\{0, x\}$ or hyperbolic tangent $\sigma(x) = \tanh(x)$.
{If we let $\fnH(\vcx)$ be a convolutional filter of the form $\fnH(\vcx) = h_{1} \mtS \vcx$, the graph perceptron becomes the usual expression of GCNs given by $\vcx_{1} = \fnsigma(h_{1} \mtS \vcx_{0})$, where $h_{1}$ is the learnable coefficient. Extension to multi-featured graph signals comes in \eqref{eq:GCNN}.}
Cascading graph perceptrons gives rise to a graph neural network (GNN) \cite{gama2020graphs}.
Formally, a GNN $\fnPhi:\fdX^{\stV} \to \fdX^{\stV}$ comprising $L$ layers is given by
\begin{equation} \label{eq:GNN}
    \fnPhi(\vcx) = \vcx_{L} \ \text{where} \ \vcx_{\ell} = \fnsigma \big( \fnH_{\ell}(\vcx_{\ell-1}) \big),~\ell=1,\ldots,L
\end{equation}
with $\vcx_{0} = \vcx$.
That is, the input to the GNN is a graph signal that is processed by a graph perceptron to form the output of layer $\ell = 1$, i.e., $\vcx_{1} = \fnsigma ( \fnH_{1}(\vcx_{0}))$. Signal $\vcx_{1}$ is the input of the next layer and it is processed by another graph perceptron to output $\vcx_{2} = \fnsigma ( \fnH_{2}(\vcx_{1}))$.
This procedure is repeated for all layers, 
and the GNN output is that of the last layer, $\vcx_{L}$; see Fig.~\ref{fig:GNN}.

The graph filters incorporate the topology of the data structure, and these filters are dependent on the parameters at each layer. 
Grouping all filter parameters in the set $\stH$, we estimate $\stH$ in a data-driven fashion from a training set $\stT = \{\vcx_{i}\}_{i=1}^{|\stT|}$ by minimizing a task-dependent cost function $\fnJ: \fdR^{N} \to \fdR$: 
\begin{equation} \label{eq:ERM}
    \min_{\stH} \sum_{\vcx_{i} \in \stT} \fnJ \big( \fnPhi(\vcx_{i}) \big).
\end{equation}

In general, it is assumed the samples in $\stT$ are independent, identically distributed, and thus \eqref{eq:ERM} becomes an empirical risk minimization problem \cite{Vapnik91-ERM}.
For a supervised learning setting, we have output samples $\vcy_{i}$ for the training data $\stT = \{(\vcx_{i},\vcy_{i})\}_{i=1}^{|\stT|}$ and thus the objective function in \eqref{eq:ERM} becomes $\fnJ(\fnPhi(\vcx_{i}),\vcy_{i})$.
For semi-supervised learning -- e.g., node classification, where we have $\stT = \{\vcx,\vcby\}$ with inputs $\vcx$ typically available for all nodes and output $\vcby$ available only at a subset of nodes -- the i.i.d. assumption on the samples does not hold.
The
objective is not necessarily to find the filter parameters $\stH$ that minimize $\fnJ(\cdot)$, but rather to take gradient descent steps that would improve the generalization performance on unseen data; see \cite[Ch. 8]{goodfellow2016deep} for more details on training neural networks. {GNNs have taken over as a very powerful and  promising tool in machine learning, with notable applications in recommender systems \cite{ying2018graph}, drug discovery \cite{stokes2020deep}, biology \cite{senior2020improved,gainza2020deciphering}, and time of arrival prediction \cite{derrow2021eta}.}

Choosing the form of filters $\fnH_{\ell}$ determines the overall GNN characteristics \cite{isufi2021edgenets}.
In the following, we discuss the convolutional filters (Sec.~\ref{subsec:GCNN}) and the non-convolutional filters (Sec.~\ref{subsec:GNN}).
We close with a brief overview of other uses of graph filters in GNN-style architectures (Sec.~\ref{subsec:otherGNN}).

\titledParagraph{Multiple features.}
The representation power of the GNN in \eqref{eq:GNN} can be increased by utilizing a bank of filters (see Sec.~\ref{sec:wavelets}) instead of a single filter \cite{isufi2021edgenets}.
To see this, consider that the input graph signal $\vcx = \vcx_{0}$ gets processed by $F_{1}$ different graph filters $\{\fnH_{1}^{f}\}_{f=1}^{F_{1}}$, creating a set of $F_{1}$ output graph signals $\stX_{1}^{F_{1}} = \{\vcx_{1}^{1},\ldots,\vcx_{1}^{F_{1}}\}$ after applying the activation function to each of them, i.e. $\vcx_{1}^{f} = \sigma(\fnH_{1}^{f}(\vcx))$.
At the next layer, we have a set of $F_{1}$ input graph signals, instead of just a single one.
If we want to use a bank of filters again, the most general linear operation would be to use a distinct bank for each of the input graph signals.
That is, if we want $F_{2}$ output graph signals, we need $F_{2}$ filters $\{\fnH_{2}^{1f},\ldots,\fnH_{2}^{F_{2}f}\} = \{\fnH_{2}^{gf}\}_{g=1}^{F_{2}}$ for each input graph signal $f \in \{1,\ldots,F_{1}\}$.
Doing so creates a set of $F_{2}F_{1}$ graph signals, each one obtained as $\vcx_{2}^{gf} = \fnsigma(\fnH_{2}^{gf}(\vcx_{1}^{f}))$.
To prevent the number of signals from growing exponentially, a summary is created by adding up the signals resulting from each of the $g$th filters, i.e. $\vcx_{2}^{g} = \sum_{f=1}^{F_{1}}\vcx_{1}^{f}$.
In this way, we can think of the layer as taking $F_{1}$ input signals, and giving $F_{2}$ output signals.
Repeating this for every layer, we can think of taking $F_{\ell-1}$ input signals, and giving $F_{\ell}$ output signals, which leads to a generic description of GNNs as follows
\begin{equation} \label{eq:GNNmulti}
    \fnPhi(\stX^{F}) = \stX_{L}^{F_{L}} \ \text{where} \ \vcx_{\ell}^{g} = \fnsigma \Big( \sum_{f=1}^{F_{\ell-1}} \fnH_{\ell}^{gf}(\vcx_{\ell-1}^{f}) \Big),
\end{equation}
for $g=1,\ldots,F_{\ell}$ at every layer $\ell=1,\ldots,L$, and where $\stX_{\ell}^{F_{\ell}} = \{\vcx_{\ell}^{1},\ldots,\vcx_{\ell}^{F_{\ell}}\}$ is the set of $F_{\ell}$ features at layer $\ell$.
The resulting filter bank can be interpreted as an undecimated, analysis filter bank (Sec.~\ref{sec:wavelets}), where the filter coefficients are learned from data instead of being designed.
Understanding each layer as a learnable filter bank may allow us to impose certain characteristics, such as Parseval tight frames (cf. \eqref{Eq:Parseval}), during design or training.
The values of $\{F_{\ell}\}$ and $L$ are hyperparameters, and are often used as a proxy for representational capability (see \cite[Ch. 5]{goodfellow2016deep} for the relationship between capacity, width, and generalization).

\titledParagraph{Readout layer.} The GNN output $\fnPhi(\stX^{F}) = \stX_{L}^{F_{L}}$ (cf. \eqref{eq:GNNmulti}) at each vertex is a vector of dimension $F_L$. 
Thus, the dimensions of the GNN output
may not match the dimensions of the target output $\vcy$. A readout layer is therefore used to match the dimensions and decode the GNN encoded embeddings into the final output, {see \cite[Ch. 9]{goodfellow2016deep}}. Depending on whether we use the GNN for centralized or distributed processing, the readout layer has different forms. In a centralized processing, all node features are usually concatenated into a vector of size $NF_L$ $\vcx_{\text{GNN}} = [(\vcx_{L}^{1})^{\Tr},\ldots,(\vcx_{L}^{F_{L}})^{\Tr}]^{\Tr}$ and then mapped to the output dimension as per e.g., the linear transform $\vcu = \mttheta\vcx_{\text{GNN}}$, where the matrix of parameters $\mttheta$ matches the output dimensions.
Instead, in distributed processing,\footnote{GNNs are distributed architectures if the graph filters are distributable.}
the readout layer must also be local. One conventional case is to consider a readout layer operating on a single node.
That is, let vector $\boldsymbol{\chi}_i = [[\vcx_{L}^{1}]_{i}, \ldots, [\vcx_{L}^{F_{L}}]_{i}]^\Tr \in \fdR^{F_L}$ be the vector of GNN output features at node $i$ and suppose the target output is a real scalar. Then, the readout layer at node $i$ is of the form $u_i = \vctheta^\top\boldsymbol{\chi}_i $, where vector $\vctheta \in \fdR^{F_L}$ is common for all nodes.
The 
 readout layers can also be nonlinear multi-layer perceptron layers.
In either case, they are 
 trainable parameters and are used to minimize cost \eqref{eq:ERM} or alternative objective functions. 

\titledParagraph{Pooling.}
Pooling is included in regular CNNs to construct regional summaries of information.
This mainly serves two objectives: 
(i) control the computational cost by trading spatial information with feature information (i.e., reducing the size of the images while increasing the number of features); (ii)
aggregate global information in the deeper CNN layers.
%
Pooling approaches have also been developed for GNNs and can be interleaved with the graph perceptron layers \cite{moura2020pooling}. These also follow two different lines: (i)
use some sort of multiscale hierarchical clustering algorithm \cite{defferrard2016chebnets}, creating ever smaller graph supports at each layer; or 
(ii) use graph sampling methods \cite{Gama2019-Archit} that leave the graph topology unaltered.
The former 
is typically of more interest in abstract networks where the graph supports can be manipulated.
The latter 
is typically of more interest for physical networks and distributed processing where we want to use the original topology to process signals in deeper layers.
However, in many applications such as those involving physical graphs (robotic, sensor, or power grid networks), nodes have computational power and thus the cost of computing the GNN output is naturally distributed among these nodes.
In these cases, pooling is less crucial and may not be needed.


\subsection{Graph Convolutional Neural Networks} \label{subsec:GCNN}

The most popular GNN architectures are those that use a GCF 
at each layer; i.e., substitute $\fnH_{\ell}$ in \eqref{eq:GNN} with \eqref{eq:graphConv}.
This leads to graph convolutional neural networks (GCNN) \cite{Bruna2014-SpectralGNN, defferrard2016chebnets, Gama2019-Archit}.
The GCNN can 
 be compactly written as \cite{ruiz2021graph}
%
\begin{equation} \label{eq:GCNN}
    \fnPhi(\mtX) = \mtX_{L} \ \text{where} \ \mtX_{\ell} = \fnsigma \Big( \sum_{k=1}^{K} \mtS^{k} \mtX_{\ell-1} \mtH_{\ell k}\Big),
\end{equation}
where $\mtX_{\ell} \in \fdR^{N \times F_{\ell}}$ collects the $F_{\ell}$ graph signal features $\vcx_{\ell}^{g}$ obtained at the output of layer $\ell$, and $\mtH_{\ell k} \in \fdR^{F_{\ell} \times F_{\ell-1}}$ contains the $k$th filter parameters of the $F_{\ell}F_{\ell-1}$ filters involved in \eqref{eq:GNNmulti}; i.e. $[\mtH_{\ell k}]_{gf} = h_{\ell k}^{gf}$ for $f=1,\ldots,F_{\ell-1}$ and $g=1,\ldots,F_{\ell}$.
%
The multiplications $\mtS^{k}$ on the left of $\mtX_{\ell-1}$ shift the different signals locally over the graph up to $k$ hops away, 
whereas the multiplications on the right carry out a linear combination of values contained in the same node via the filter bank coefficients, and as such, can be arbitrary (which is the case when $\mtH_{\ell k}$ is learned from data). 
%

This structure, coupled with the pointwise nature of the activation function, makes the GCNN a local architecture that respects Properties~\ref{prop:perm_equiv}-\ref{prop:compCost} and Property~\ref{prop:LipCont} of the graph convolutional filter \cite{gama2020stability}. 
GCNNs are also Lipschitz continuous to changes in the underlying graph support (cf. \eqref{eq:stabilityGraphFilter}), albeit with a slightly modified constant.
They can, however, process information located in large GSO eigenvalues in a stable manner, a feat that cannot be achieved by linear graph convolutions (see \cite{gama2020stability}).
This makes GCNNs better suited for problems in which information located at large GSO eigenvalues is important.


\titledParagraph{Implementations.}
While it is perfectly feasible to implement GCNNs via \eqref{eq:GCNN}, different (sub-)implementations, often derived from a different stating point, have become popular. These include:

\begin{enumerate}
\item GCNNs with orthogonal polynomials such as Chebyshev \cite{defferrard2016chebnets,he2022convolutional}, Berstein \cite{he2021bernnet}, and Jacobi \cite{wang2022powerful}.
\item The GCN of \cite{Kipf2017-GCN} uses in \eqref{eq:GCNN} $\mtS = \mtD^{-1/2}(\mtI+ \mtA) \mtD^{-1/2}$, $K=1$ and, more crucially, $\mtH_{\ell 0} = \mtZeros$ for all $\ell$. This forces all the learned filters to be low-pass filters leading to the \emph{oversmoothing} problem so thoroughly discussed in the GCNN literature \cite{nt2019revisiting,Chen2020-Oversmoothing}.
\item A Simplifying Graph Convolutional (SGC) Network \cite{Weinberger2019-SGC} further exacerbates this problem by setting $\mtH_{\ell k} = \mtZeros$ for all $k < K$ for some order $K$.
\item A Graph Isomorphism Network (GIN) \cite{Jegelka2019-GIN} can be obtained from \eqref{eq:GCNN} by setting $\mtS$ to be the binary adjacency matrix, $K=1$, and $\mtH_{\ell 0} = (1+\sceps_{\ell}) \mtH_{\ell 1}$ for some pre-defined $\sceps_{\ell}$. GINs further propose to use $K=0$ for some intermediate layers to mimic a node-based multi-layer perceptron (MLP).
\item GraphSAGE \cite{Hamilton2017-GraphSAGE} with a linear `aggregate' function is obtained from \eqref{eq:GCNN} by setting $K=1$ and then normalizing feature-wise the resulting graph signal. While GraphSAGE does not suffer from oversmoothing, forcing $K=1$ precludes sharp transitions.
\item A Jumping Knowledge Network (JKNet) \cite{xu2018representation} with a summation aggregation can be seen as computing the GCNN operation \eqref{eq:GCNN} where residual connections are used to account for multi-hop neighbors.
\end{enumerate}

The design of GNN architectures is evolving at a fast pace and thus many newer architectures become readily available each month.
While we have only mentioned the most popular ones, we would like to encourage readers to identify whether these new architectures are convolutional in nature, and thus fit the description in \eqref{eq:GCNN}, as the ones above, or they are essentially non-convolutional and may be equivalently described by leveraging other filter structures as we discuss in the next section. 

\subsection{Non-convolutional GNNs} \label{subsec:GNN}

In principle, we can build a different GNN architecture by exchanging the 
filters in \eqref{eq:GNNmulti} with any of the types discussed in Sec.~\ref{sec:other}. These GNNs will exploit different aspects of the data structure, closely following the properties that the chosen graph filters themselves exhibit.


Rational graph filters (Sec.~\ref{subsec_ratGF}) lead to GNNs that are convolutional in practice, but capable of achieving much sharper frequency transitions with fewer learnable parameters.
Cayleynets \cite{levie2018cayleynets} and ARMANets \cite{isufi2021edgenets} are two such examples.

Node-varying graph filters \eqref{eq.NVfilt} lead to non-convolutional GNNs \cite{gama2021node} as a means of learning frequency content creation {(refer to \cite{gama2020stability} for a thorough discussion on the effects of having new frequencies at the otuput)}.
Edge-varying graph filters \eqref{eq.EVfilt} also lead to non-convolutional GNNs \cite{isufi2021edgenets}, with graph attention transformers (GATs, \cite{Velickovic2018-GAT}) and natural graph convolutions \cite{pim2021natural} being two of the most popular exponents. The edge varying filter has been used here to propose a broad family of GNN solutions as a way to show benefits and limitations of the different architectures and how they trade parameter sharing with permutation equivariance.
In fact, neither the node varying nor the edge varying GNNs are permutation equivariant architectures. 
This is particularly useful in applications where nodes are distinguishable. For example, in power grids, some nodes represent generators while others represent consumers, and thus it may be useful that they learn different behaviors. 

\subsection{Other Uses of Filtering in GNNs} \label{subsec:otherGNN}

Graph filters also play other key roles in GNNs.
For instance, graph wavelets (Sec.~\ref{sec:wavelets}) can be used in lieu of filters in \eqref{eq:GNNmulti} to avoid training procedures (cf. \eqref{eq:ERM}) and to gain interpretability.
The resulting architectures are known as graph scattering transforms \cite{Gama2019-Scattering, Lerman2020-Scattering} and have been used successfully in biological applications \cite{Wolf2019-GeometricScattering} and 3D point clouds \cite{ioannidis20scattering}, where data is scarce or the training process is computationally intensive.

Nonlinear graph filters, such as max or median filters \eqref{eq.Med_filt}, can be used as local activation functions (instead of pointwise ones).
They preserve permutation equivariance (Property~\ref{prop:perm_equiv}), while achieving a higher expressive power.
Using these filters also implies that the activation functions are learnable \cite{ruiz2019invariance, iancu2020graph}.

\section{Applications in Signal Processing}\label{sec:SP}



Graph filters have found widespread use in several signal processing application areas. These include the standard problems of graph signal reconstruction from partial and noisy observations (Sec.~\ref{subsec_denInterp}), anomaly detection over networks (Sec.~\ref{subsec_AnomDet}), and network topology inference (Sec.~\ref{subsec_gti}).
Graph filters are also key components of many recently developed graph-based image processing methods (Sec.~\ref{subsec_imProc}).
Lastly, due to their local implementation, graph filters have been used for distributed signal processing tasks (Sec.~\ref{subsec_distSP}).

\subsection{Signal Reconstruction}\label{subsec_denInterp}

This task consists of reconstructing graph signals from one or more noisy (and possibly partial) observations. 
%
Filtering by regularization [Sec.~\ref{subsec_regFilt}] has been extensively used for this task, and different regularizers have been developed to match signal priors in different settings. A second strategy is to fit the observed signals with a graph filter and use this filter to reconstruct the missing values. This strategy is first discussed in \cite{sandryhaila2013discrete} and subsequently extended to the various graph filters discussed in Sec.~\ref{sec:other}. These techniques are applied in sensor networks \cite{jablonski2017graph} and speech enhancement \cite{wang2021speech}, among others.
A third strategy to reconstruct graph signals is to represent them as sparse linear combinations of atoms of an overcomplete graph-based dictionary \cite{thanou2014learning}; that is, write a signal as $\vcx = \mtD_{\stG}\vcs$, where $\mtD_{\stG} \in \fdR^{N \times NS}$ is the graph-based dictionary and $\vcs \in \fdR^{NS}$ is a sparse vector. Graph filters are used to define the atoms of the dictionary, as $\mtD_{\stG} = [\mtH_1(\mtS), \ldots, \mtH_S(\mtS)]$, where each $\mtH_s(\mtS)$ is a graph convolutional filter \cite{thanou2014learning}. The advantage over graph-agnostic dictionaries is that the filter order dictates both the atoms' vertex locality and the number of trainable parameters. Differently, 
\cite{zheng2010graph,ramamurthy2012learning,yankelevsky2016dual} learn dictionaries where the columns in $\mtD_{\stG}$ behave as smooth graph signals. 
Using the Tikhonov regularizer, this boils down to solving versions of
\begin{align}\label{eq.dicLearn}
 \underset{\mtD_{\stG}, \mtX}{\text{min}}~&\|\mtY - \mtD_{\stG}\mtX	\|_F^2 + \gamma_1\trace(\mtD_{\stG}^\top\mtL\mtD_{\stG})+ \gamma_2\trace(\mtX^\top\mtL_c\mtX) \nonumber \\
&\text{s.t.} \,\,  \|\vcx_i\|_0 \le T~\forall~i,
\end{align}
where $\mtL$ and $\mtL_c$ are Laplacians of two graphs capturing the manifold structure of the dictionary atoms $\mtD_{\stG}$ and of the data $\mtX$, respectively. The constraint \eqref{eq.dicLearn} imposes a maximum sparsity $T$ on each column $\vcx_i$ of $\mtX$ and $\gamma_1, \gamma_2>0$ control the respective trade-offs. Further, 
\cite{yankelevsky2019finding} augments problem \eqref{eq.dicLearn} with graph wavelets to learn multi-scale atoms that facilitate scalability to large graphs. Finally, \cite{thanou2018learning} considers quantization effects on the learned atoms when such dictionaries are used for distributed signal processing tasks.

%

\subsection{Anomaly Detection}\label{subsec_AnomDet}

Many graph signals -- including, e.g., voltage measurements in power grids \cite{ramakrishna2021grid} or brain imaging recordings in healthy patients \cite{hu2016matched} -- exhibit bandlimited (cf. \eqref{eq:bandlimited}) and/or low-pass behavior \cite{ortega2018graph,ramakrishna2020user}.
When an anomaly occurs, these signals contain unexpected components in their high-pass spectrum. We can leverage graph filters to localize such components associated, e.g., with corrupted signals \cite{ramakrishna2021grid} or non-healthy patients \cite{hu2016matched}.
%
The idea is to design a graph filter $\mtH(\mtS)$ and form a hypothesis test on the filtered signal $\vcy = \mtH(\mtS)\vcx$ of the form
\begin{align}\label{eq.hypTest}
\begin{split}
&\stH_0:~~\fnf(\vcy) \le \gamma\\
&\stH_1:~~\fnf(\vcy) > \gamma\\
\end{split},
\end{align}
where $\fnf(\vcy)$ is a transformation of the filtered output (e.g., $\fnf(\vcy) = \|\vcy\|_2$) and $\gamma$ is a threshold; that is, the filtered signal shows different characteristics under the null hypothesis $\stH_0$ and the alternative hypothesis $\stH_1$.

The work in \cite{sandryhaila2014discrete} considers such a setting to detect anomalous sensors. The input signal is filtered with a high-pass convolutional filter and the signal is classified as anomalous if one or more GFT coefficients exceed a threshold. References 
\cite{xiao2020anomalous,xiao2020nonlinear,ferrer2022volterra} consider nonlinear filters  (Sec.~\ref{subsec_nlinFilt}) to reconstruct the data under normal behavior, and the low-pass signal components are used to detect and localize anomalous sensors. This idea is extended in \cite{francisquini2022community} to identify a cluster of abnormal nodes. The work in \cite{egilmez2014spectral} proposes an unsupervised setting for the scenario when we do not have knowledge of how normal and/or anomalous graph signals behave. Under the assumption that normal data are more frequent than abnormal ones, the authors use two complementary ideal step graph filters -- one low-pass and one high-pass -- with the same cut-off frequency, and optimize the cut-off frequency to minimize the cluster size.

In the context of brain imaging, 
\cite{hu2016matched} uses 
similar principles to detect early stage Alzheimer's disease. 
Two band-pass filters are built (one per type of patient) to localize signal components not belonging to that class. Subsequently, an energy-based Neyman-Pearson detector is derived from the filtered outputs. Reference \cite{isufi2018blind} generalizes this to the setting where information about the alternative hypothesis $\stH_1$ is unavailable, leading to a Neyman-Pearson detector only with respect to hypothesis $\stH_0$.

\subsection{Network Topology Inference}\label{subsec_gti}

Often the graph $\graph$ is unavailable and, accordingly, network topology inference from a set of (graph signal) measurements is a prominent yet challenging problem. Early foundational contributions can be traced back to the statistical literature of graphical model selection~\cite{dempster_cov_selec, meinshausen06}. Recently, the fresh signal representation perspectives offered by GSP through graph filters have sparked renewed interest~\cite{mateos2019connecting}.
At its core, network topology inference assumes some relation between the set of observed graph signals $\mtX = [\vcx_1, \vcx_2, \ldots, \vcx_m]$ and the GSO $\mtS$ to be recovered.
This relation can be modeled as each $\vcx_i$ being the output of a graph (convolutional) filter defined on the unknown $\mtS$.
Intuitively, this means that the observations $\vcx_i$ were generated through (linear) local interactions in the unknown graph, so that the topologocal information of $\mtS$ is contained in $\mtX$. 


Different assumptions on the filter type generating $\mtX$ lead to different formulations of the topology inference problem.
We can state a generic version of this inverse problem as
\begin{equation}\label{eq:network_topology_inference}
	\min_{\mtS \in \stS} \,\, \fnf(\mtX, \mtH(\mtS)) + \fnr(\mtS),
\end{equation}
where the fitting loss $\fnf(\cdot)$ quantifies how well $\mtX$ can be modeled as the output of a filter $\mtH(\mtS)$, the regularizer $\fnr(\cdot)$ promotes desirable properties on $\mtS$ such as sparsity, and the feasibility set $\stS$ encodes the type of GSO that we are looking for (e.g., Laplacian or adjacency matrix). 

\smallskip
\noindent\textbf{Smoothness.} 
A first type of (convolutional) graph filter considered in the literature is the low-pass graph filter.
This is a reasonable modeling assumption in averaging dynamics such as opinion formation.
This model leads to signals $\vcx_i$ being slow varying, which when $\mtS$ is the graph Laplacian implies smoothness of $\vcx_i$ on the unknown graph (cf.~\eqref{eq:quad_und}).
Two early proponents of this model are~\cite{DongLaplacianLearning} and~\cite{Kalofolias2016inference_smoothAISTATS16}.
Although their regularizers $\fnr(\mtS)$ are different, in both cases the fitting term is $\fnf(\mtX, \mtS) = \mathrm{Tr}(\mtX^\top \mtL \mtX)$ so that it penalizes graphs not leading to a smooth representation of the observed signals.

\smallskip
\noindent\textbf{Stationarity.} 
Another approach is  to not assume any specific form (low-pass, band-pass, high-pass) for the graph filter generating $\mtX$~\cite{segarra2017topoidTSIPN}.
Under certain statistical assumptions on the inputs to the filter, this setting leads to the signals $\vcx_i$ being graph stationary on the unknown $\mtS$~\cite{marques2017stationary, perraudin2017stationary, girault2015stationary}; (cf. Sec.~\ref{subsec_regFilt}, Wiener filter).
In short, this implies that the covariance matrix $\mtsigma_x$ of the observed signals shares the eigenvectors with $\mtS$, or, equivalently and more practically, that $\mtsigma_x$ and $\mtS$ commute.
Ref. ~\cite{segarra2017topoidTSIPN} uses a two-step procedure to first estimate the eigenvectors of $\mtS$ and then restate~\eqref{eq:network_topology_inference}, where only the eigenvalues of $\mtS$ are unknown.
In contrast, the commutativity property can be imposed through a fitting term of the form $\fnf(\mtX, \mtS) = \| \hat{\mtsigma}_x \mtS - \mtS \hat{\mtsigma}_x \|_\mathrm{F}$, where $\hat{\mtsigma}_x$ is an estimate of the covariance matrix~\cite{shafipourdirectedTopoID2018}.
Other assumptions on the filter include modeling the signal as the superposition of several heat-diffusion filters~\cite{thanou17,coutino2020state}, or as a 
consensus-like process \cite{zhuconsensusinference2020, egilmezTopoID2019}. 


\subsection{Image Processing}\label{subsec_imProc}

Graph-based image processing complements conventional image processing approaches with new insights and techniques for tasks such as image reconstruction and filtering \cite{lezoray2012image, cheung2018graph}. Images have a natural grid structure that can be represented as a graph, where each node is a pixel, an edge connects two pixels, and the graph signal is the pixel intensity.
The edge weights can be set via the Gaussian kernels as
\begin{equation}\label{eq.pix_graph}
w_{ij} = \text{exp}\bigg(-\frac{\|\vcf_i - \vcf_j	\|_2^2}{\sigma_l^2}	\bigg)\text{exp}\bigg(-\frac{(x_i - x_j)^2}{\sigma_x^2}	\bigg),
\end{equation}
where $\vcf_i$ is the location (feature) of pixel $i$, $x_i$ its intensity, and $\sigma_l, \sigma_x$ are two parameters. Such edge weighs are a combination of the geometric distance (pixels' locations) and photometric distance (signal intensities $x_i$), where a larger distance implies a smaller weight. 
Graph-based image processing works mainly with undirected graphs and low-pass filtering since connected pixel nodes have a stronger edge weight if they are close (either geometrically or photometrically). Graph filters are used for image reconstruction (e.g., denoising, debluring) and edge-preserving filtering (i.e., preserve edges appearing in an image, not graph edges).


\smallskip
\noindent\textbf{Image reconstruction.} This task consists of reconstructing an image signal $\vcx$ from a degraded version, which can be noisy, blurred, or have missing pixels. These are all ill-posed inverse problems and regularization is typically used. In the GSP language, this is a graph signal reconstruction task and regularized filtering [Sec.~\ref{subsec_regFilt}] is often used to impose low-pass behavior. The Tikhonov regularized filter \eqref{eq.Tikproblem} is leveraged for image denoising in \cite{liu2014progressive}. Reference \cite{pang2017graph} explores the connection with manifold regularization and provides an explanation why low-pass filtering is particularly useful for denoising depth images. Reference \cite{elmoataz2008nonlocal} uses a form of the total variation regularizer \eqref{eq.Variation2} to denoise the image. Reference \cite{yaugan2016spectral} approaches the problem from a Wiener filtering perspective \eqref{eq.WeinFilt}-\eqref{eq.Wienresp}. Since regularized filters are particular forms of rational graph filtering,  \cite{tian2014chebyshev} proposes a non-parametric rational filter to denoise the image. Finally, \cite{zhang2008graph} considers a smoothing graph filter of the form $\mtH(\mtS) = e^{-\gamma\mtL} = \sum_{k = 0}^{\infty}\frac{\gamma^k}{k!}(-\mtL)^k$ (a.k.a. the heat kernel) to perform low-pass graph filtering. This can be seen as a convolutional filter of order $K\to\infty$ with frequency response $\fnth(\lambda) = e^{-\gamma\lambda}$, which for $\gamma>0$ acts as a low-pass filter.

\smallskip
\noindent\textbf{Edge-preserving filtering.} Some conventional image filters that preserve image edges can also be interpreted from a GSP perspctive.
Ref. \cite{gadde2013bilateral,girault2018irregularity} study the bilateral image filter and show that it is an order $K = 1$ low-pass 
GCF (compare to \eqref{eq.pix_graph}) that smooths the image. To boost smoothing,  \cite{onuki2016graph} develops the trilateral filter, which has a rational graph frequency response, explaining its improved performance. 
GCFs are also used for guided image filtering in \cite{knyazev2015accelerated, huang2019fast}. 

Simple forms of 
GCFs are also used for smoothing and edge enhancement with low computational cost.
A convolutional filter of small order (e.g., two) is used to smooth the image, and a successive filter of the form $\mtH(\mtL) = \mtI + h_1\mtL$ is used to sharpen the edges \cite{sadreazami2017data_a,sadreazami2017data}. Ref. \cite{lu2022dct} uses 
GCFs to efficiently implement the sparse low-pass discrete cosine transform in the vertex domain. Lastly, median graph filters \eqref{eq.Med_filt} are used in \cite{salembier2018ship} to detect ships in image data.
%

\subsection{Distributed Signal Processing}\label{subsec_distSP}

Because of their local implementation (Property~\ref{prop:loc}), graph filters are readily distributable, where the graph captures both the signal structure and the distributed communication pattern. For example, we may want to denoise sensor measurements (the graph signal) over a network where each node can exchange information only locally. The research on \emph{distributed graph filtering} has evolved in three main directions: (i) using graph filters to approximate a desired operation and apply it in a distributed fashion; 
(ii) analyzing the filtering performance when facing distributed communication challenges such as interference, asynchronous implementation, and quantization; and, (iii) estimating the filter parameters distributively.



\smallskip
\noindent\textbf{Distributed tasks.} Using graph filters for distributed processing implies first matching a desired operator [Sec.~\ref{Ss:design_opMatch}] and then deploying the filter. Here, we first discuss distributed average consensus and then other general operators. 


\smallskip
\emph{1) Consensus:} Distributed average consensus is a cornerstone method underpinning myriad distributed estimation and detection tasks \cite{olfati2005consensus}. Given a graph $\stG = (\stV, \stE)$ representing connectivities $\stE$ between different agents $\stV$, we want the agents to estimate the average value of their signals $\vcx$ by only exchanging information with their local neighbors. 
Let $\bar{x}:=1/N \sum_{n = 1}^N x_n$ be the true average, $\vcy := \bar{x}\mathbf{1}$ the vector of averages,  and $\mtB := \frac{1}{N} \mathbf{1} \mathbf{1}^\top$ the consensus operator. Then $\vcy = \mtB \vcx$.
%
%
%
%
Graph convolutional filters can be used to reach \emph{exact} and \emph{finite-time} consensus as long as we design appropriately their coefficients, as stated by the following proposition. 
%
\begin{proposition}[Finite-time consensus \cite{sandryhaila2014finite}] Average consensus can be computed exactly in finite-time by a graph convolutional filter of appropriate order if the Laplacian eigenvalue $\lambda_1 = 0$ is of multiplicity one, i.e., the graph is connected.
\end{proposition}
Since the constant vector $\vcv_1 = 1/\sqrt{N} \mathbf{1}$ is an eigenvector of the Laplacian  (i.e., $\mtL\mathbf{1} = 0$), we can find the filter coefficients to achieve the frequency response%
\begin{align}\label{eq.consFresp}
\fnth(\lambda_n)=\left\{
\begin{array}{ll}
1&\text{for}~\lambda_n = 0~(n = 1) \\
0 &\text{for}~\lambda_n > 0~(n = 2, \ldots, N)
\end{array}.
\right.
\end{align}
%
%
References \cite{sandryhaila2014finite,safavi2014revisiting} provide closed-form solutions for $\{h_k\}$, whereas \cite{coutino2018limits} discusses theoretical limits on the minimum filter order for which consensus can be achieved.
%

Exact finite-time consensus can be challenging due 
 to numerical issues related to computing close-by eigenvalues. 
Approximate consensus is analyzed in \cite{sandryhaila2014finite} for the convolutional filter (cf. \eqref{eq:graphConv}), in \cite{Segarra2017-GraphFilterDesign} for the node varying filter \eqref{eq.NVfilt}, and in \cite{Coutino2019-EdgeVariant} for the edge varying filter \eqref{eq.EVfilt}. The common observation is that filters with higher orders approximate better the consensus operator; however, instabilities during the design phase arise and more advanced design strategies are needed \cite{coutinominguez2021cascaded}.
When the underlying graph comes from a random distribution, reaching consensus via graph filtering can be improved by accounting for the distribution of the eigenvalues \cite{Kruzick2018-filter}.

Recent literature also discusses the links between consensus via graph filtering and control theory. Specifically,  \cite{yi2019average} discusses both finite-time and asymptotic consensus, and derives conditions when they can be achieved even over uncertain graphs. References \cite{apers2016accelerating,ran2021fast} focus on graph filters of order two to accelerate consensus. By linking the eigenvalues of the respective graph Laplacian with the graph properties, \cite{apers2016accelerating} provides optimal filter design for finite-time consensus and characterizes the convergence rate. On the other hand, \cite{ran2021fast} shows that for some graphs it is impossible to accelerate consensus. Finite-time consensus over directed graphs is discussed in \cite{charalambous2018laplacian,li2021fast}, where, as for the undirected case, the multiplicity of the eigenvalues influences the number of steps. Reference \cite{li2021fast} discusses asymptotic consensus for unknown directed graphs, and \cite{kruzick2018optimal} considers finite-time consensus over random graphs. Lastly, \cite{ran2020group} focuses on group consensus via graph filtering, i.e., that nodes within a group achieve average consensus, but different groups can have different values. 


\smallskip
\emph{2) General operator:} In Sec.~\ref{Ss:design_opMatch}, we discussed filter design strategies to match any general operator $\mtB$ via graph convolutional  filters. Exploiting the filter locality, it is then possible to implement $\mtB$ (or an approximation) distributively over the graph. Distributed operator matching via graph filters is investigated for the convolutional filter in \cite{chen2014signal,safavi2014revisiting,shuman2018distributed}, rational filter in \cite{Isufi2017-ARMA,emirov2020polynomial}, node varying filter in \cite{Segarra2017-GraphFilterDesign}, edge varying in \cite{Coutino2019-EdgeVariant,coutinominguez2021cascaded}, and for other modifications of these filters in \cite{emirov2020polynomial}. The common theme is to approximate $\mtB$ with a low filter order, so as to limit the communication costs. Reference \cite{romero2020fast} details this challenge and designs the minimum order convolutional filter to either match or approximate the operator.

\smallskip
\noindent\textbf{Distributed challenges.} In distributed graph filtering, we must also account for the communication challenges, including:


%
\smallskip
\emph{1) Interference:} In distributed processing, the communication edge weights $\mthS$ may differ from the nominal ones $\mtS$ used to design the filter. Property~\ref{prop:LipCont} characterizes the impact of small differences in the GSO on the output of graph convolutional filters, which are Lipschitz. 
However, it focuses on small relative perturbations, while we often encounter larger perturbations such as link losses. The effect of link losses on graph filters is discussed in \cite{isufi2017filtering,gao2021stability}. The following result generalizes Property~\ref{prop:LipCont} to this stochastic setting.
\begin{proposition}[\cite{gao2021stability}] If the edges in the graph realization $\mthS \subseteq \mtS$ are preserved independently with a probability $p$ and the filter is Lipschitz (Property~\ref{prop:LipCont}) with constant $C$, the expected squared deviation of the filter output is bounded as
    \begin{equation}
    \fdE \big[ \| \mtH(\mthS)\vcx - \mtH(\mtS)\vcx 	\|_2^2	\big] \le \alpha NC^2(1-p)\|\vcx\|_2^2 + \fnO((1-p)^2),
    \end{equation}
    where $\alpha$ is either 2 or the maximal node degree, depending of the choice of shift operator.
\end{proposition}
Similar results are developed in \cite[Proposition 1]{gao2021stochastic} for convolutional filters and in \cite[Thm. 3]{isufi2017filtering} for distributed rational filters. Ref. \cite{saad2020accurate} considers the setting where the link preservation probabilities are different for each edge, and characterizes the statistical output of both convolutional 
and node varying 
filters \cite[Proposition 1]{saad2020accurate}. The authors then consider such a statistical deviation to design robust filters and propose a cross-layer protocol 
to run graph filters over an asymmetric wireless sensor network. Robust data-driven learning of graph filters in stochastic settings is also investigated for 
GNNs: 
 \cite{gao2021stochastic}
shows that by learning the parameters on a perturbed graph, we can achieve robust transference; 
and \cite{gao2022learning}
proposes a constraint-learning framework where the parameters are optimized in the expectation while bounding the output variance.

\smallskip
\emph{2) Asynchronous implementation:} Another challenge in distributed filtering is that nodes cannot always communicate in a synchronous manner. {Asynchronous communication enables scalability \cite{baudet1978asynchronous}, as it avoids the need for global synchronization; however, in general, it compromises the guarantee that the filter output converges.}
%
The work in \cite{teke2020iir} provides sufficient conditions for an asynchronous implementation to converge to the designed output in a mean-squared error sense. Similar results are derived for filter banks in \cite{teke2020node} and for the edge varying filter in \cite{coutino2019asynchronous}.

\smallskip
\emph{3) Signal quantization:} In a distributed setting, we may also need to account for the low communication capacity between sensors. In these cases, the exchanged signal shifts $\vcx^{(k)} = \mtS^k\vcx$ need to be quantized prior to transmission. The quantized signal can be written as $\vctx^{(k)} = \vcx^{(k)} + \vcn_{\text{q}}^{(k)}$, where $\vcn_{\text{q}}^{(k)}$ is the quantization error. In turn, the quantized filter output becomes 
\begin{equation}
\vcy_{\text{q}} = \sum_{k = 0}^Kh_k\mtS^k\vcx + \sum_{k = 1}^Kh_k\sum_{\kappa = 0}^{k-1}\mtS^{k-\kappa}\vcn_{\text{q}} := \mtH(\mtS)\vcx + \vceps_{\text{q}},
\end{equation}
where is $\vceps_{\text{q}}$ the accumulated quantization error. This quantization error distorts the filter output and needs to be accounted for during the filter design phase. If $\tilde\fnbeta(\lambda)$ is the desired frequency response and $\text{MSE}_{Q}(\vch)$ is the mean squared quantization error, the robust filter design problem looks like
\begin{equation}
\begin{aligned}
& \underset{\vch}{\text{minimize}}
& & \int_\lambda \bigg| \sum_{k = 0}^Kh_k\lambda^k - \tilde\fnbeta(\lambda)	\bigg|^2d\lambda \\
& \text{subject to}
& &\text{MSE}_{Q}(\vch) \le \gamma
\end{aligned},
\end{equation}
where $\gamma$ controls the distortion and needs to be set in accordance with the quantization step size \cite{chamon2017finite,saad2021quantization}. Ref. \cite{nobre2019optimized} further discusses optimal quantization schemes and links them with the graph topology, while \cite{saad2021quantization} discusses robust quantization in the presence of link losses. Ref. \cite{thanou2018learning} further discusses the impact of quantization errors when learning localized dictionaries (cf.~\eqref{eq.dicLearn}) via graph filters, while \cite{li2021task} discusses a joint design of signal sampling and recovery under quantization.



{\smallskip
\noindent\textbf{Filter estimation.} The above works consider filters  that are designed centrally and implemented distributively. A recent stream of works consider the task of estimating the filter coefficients distributively when data is available. Formally, consider a series of input-output pairs $\{\vcx^{(t)}, \vcy^{(t)}\}$, where for each $t$, model \eqref{eqn_lms_model} holds. We can reformulate the latter as
\begin{equation}\label{eqn_lms_model_2}
	\vcy^{(t)} = \mtZ^{(t)} \vch + \vcnu^{(t)},
\end{equation}
where $\mtZ^{(t)} = [\vcx^{(t)}, \mtS \vcx^{(t)}, \ldots, \mtS^K \vcx^{(t)}]$. Under standard statistical assumptions, we can find the filter parameters $\vch$ that minimize $ \fdE \| \vcy^{(t)} - \mtZ^{(t)} \vch  \|^2_2$ by using classical centralized linear regression techniques~\cite{hua2020online}.
More interestingly, we can decompose this objective among the $N$ nodes as
\begin{equation}\label{eq_filt_adaptive}
\vch^* = \argmin_{\vch}  \sum_{i=1}^N  \fdE | y^{(t)}_i - \vcz^{(t)\top}_i \vch  |^2,
\end{equation}
where $\vcz^{(t)\top}_i$ is the $i$-th row of $\mtZ^{(t)}$.
The reformulation in~\eqref{eq_filt_adaptive} leads to a decentralized solution.
In particular, diffusion strategies are attractive since they are scalable, robust, and enable continuous learning and adaptation. A distributed adapt-then-combine diffusion least mean squares (LMS) algorithm takes the following form at every node $i$
\begin{subequations}
	\label{eq_diffusion_atc}
	\begin{align}
		\vcpsi_i^{(t+1)}  & = \vch_i^{(t)} + \mu_i \vcz_i^{(t)} \left( y_i^{(t)} - \vcz^{(t)\top}_i \vch_i^{(t)}  \right), 	\label{eq_diffusion_a} \\
		\vch_i^{(t+1)} & =  c_{ii} 	\vcpsi_i^{(t+1)}  + \sum_{j \in \stN_{i}} c_{j i} \vcpsi_j^{(t+1)},     \label{eq_diffusion_c} 
	\end{align}
\end{subequations}
where $\mu_i > 0$ is a local step size and $\{c_{j i}\}$ are non-negative combination parameters satisfying 
$c_{j i} = 0$ if $j \not \in \stN_{i}$, and $\sum_{j=1}^N c_{j i} = 1$~\cite{hua2020online}.
In the adaptation step~\eqref{eq_diffusion_a}, each node $i$ updates its local parameter estimate $\vch_i^{(t)}$ to an auxiliary intermediate vector $\vcpsi_i^{(t+1)}$.
In the combination step~\eqref{eq_diffusion_c}, node $i$ aggregates its own intermediate vector $\vcpsi_i^{(t+1)}$ and those from its neighbors to update its estimate $\vch_i^{(t+1)}$. We run $T$ iterations of this diffusion algorithm to estimate the filter parameters at each node. Assuming that vectors $\vcz_i^{(t)}$ are drawn from a zero-mean random process that is white over the temporal dimension $t$, the following result holds.
\begin{proposition}[\cite{hua2020online}]\label{prop_data_driven_filter_design}
	For any initial condition, the iterative algorithm in~\eqref{eq_diffusion_atc} converges asymptotically in the mean towards the optimal vector $\vch^*$ (i.e., the expected value of the error goes to zero) if the step sizes $\mu_i$ are small enough.
\end{proposition}
Several extensions of this basic formulation have been proposed. First, a state space formulation is discussed in \cite{ding2021minimum}, which allows also to find the minimum filter order. Second,  model~\eqref{eqn_lms_model} assumes instantaneous diffusion, where node $i$ processes $\vcy^{(t)}_i$ at each time instant $t$ by collecting samples of $\vcx^{(t)}$ that are up to $K$ hops away. Since this limits the practical implementation,~\cite{hua2020online} also considers a modified model where the successive shifts in the filter are applied to different time samples of the input.
Third, the convergence rate of LMS is notoriously slow. To alleviate this problem, (i) \cite{hua2020online} presents a modified adaptation step~\eqref{eq_diffusion_a} based on Newton's method where Hessian information is considered but at an increased (per iteration) computational cost; and (ii) \cite{alinaghi2021graph} considers recursive least squares adaptive estimators. Lastly, 
\cite{elias2020adaptive} extends these techniques to nonlinear filters, and \cite{scardapane2020distributed} discusses the distributed parameter estimation of GNNs.}


\section{Applications in Machine Learning} \label{sec:ML}



In machine learning, graph filters act as a parameterized mapping between input-output data pairs and use the graph structure as an inductive bias. Particular properties of interest include the limited number of parameters, permutation equivariance, and the linear computation cost. Hence, graph filters have been useful in the standard tasks of semi-supervised learning on graphs [Sec.~\ref{subsec:ssl}] and unsupervised learning, especially in clustering-like algorithms [Sec.~\ref{subsec:unsuperv}]. Graph filters 
 have also been used for graph-based matrix completion  [Sec.~\ref{subsec:matComp}] and Gaussian processes [Sec.~\ref{subsec:gausProc}]. Lastly, we review some applications in computer graphics and 
 computer vision [Sec.~\ref{subsec:compvis}].

\subsection{Semi-Supervised Learning}\label{subsec:ssl}

Semi-supervised learning on graphs classifies unlabelled nodes given labels on some other nodes. Graph filters can be used to weigh and propagate the label information of multi-hop neighbors to the unknown nodes. Mathematically, consider the label matrix $\mtX \in \fdR^{N\times C}$ such that row $n$ represents the label 
of node $n$ among the $C$ classes, i.e., entry $[\mtX]_{nc} = 1$ if node $n$ belongs to class $c \in [C]$ and zero otherwise. We consider that only $M \ll N$ nodes are labeled, treat each column of $\mtX$ as a graph signal, and infer the labels as
\begin{equation}
\mtY = \mtH(\mtS)\mtX,
\end{equation}
where the unlabeled node $m$ is assigned to class $c$ for which entry $[\mtY]_{mc}$ is highest. The filter parameters are estimated as
\begin{equation}\label{eq.filtDesigCfilt}
\underset{\stH}{\text{minimize}}~\big\| \mtM \big(\mtH(\mtS)\mtX - \mtX\big)	\big\|_\mathrm{F} + \gamma \fnr(\stH, \mtY),
\end{equation}
where $\mtM = \diag(\vcm)$ and $\vcm  \in \{0,1\}^{N}$ is a masking matrix to compute the error only on the labeled nodes. Instead, $\fnr(\stH, \mtY)$ is a regularizer on the filter parameters $\stH$ (e.g., norm two) or on the output $\mtY$ (e.g., smooth label variation, cf. \eqref{eq:quad_und}). 
Refs. \cite{sandryhaila2013discrete,chen2014semi} consider the convolutional filter  \eqref{eq:graphConv} for binary and multi-class classification of blog networks and indirect bridge monitoring, respectively. 
Ref. \cite{girault2014semi} considers the Wiener graph filter \eqref{eq.Wienresp} and shows improvement upon conventional label propagation algorithms. To further improve the expressivity of the mapping, \cite{fan2022graph} uses a bank of filters with multiple GSOs (cf. \eqref{eq:multiGSOfilt}), where each GSO represents a different similarity graph built from node features. Finally, \cite{berberidis2018adaptive} considers a bank of convolutional filters, where each filter is fitted to a particular class. Then, the unlabelled nodes are assigned to the class with the highest filter output. These works, however, solve the classification problem via regression-like cost functions (e.g., Frobeius norm $\|\cdot\|_\mathrm{F}$ in \eqref{eq.filtDesigCfilt}) which may lead to a suboptimal performance despite the efficient and convex implementation properties. {GNNs are also a valid alternative, given their state-of-the-art performance in this task \cite{Kipf2017-GCN,song2022graph}
}

\subsection{Unsupervised Learning}\label{subsec:unsuperv}

The canonical task in unsupervised learning on graphs consists of grouping nodes in the absence of labels into different clusters such that nodes are tightly connected within clusters and loosely connected between them.
In this context, graph filters have been used to tackle the scalability issues of different variants of spectral clustering, a conventional unsupervised learning technique. Graph filters have also been used as a signal model to detect clusters in networks when no topology information is available. 

\smallskip
\noindent\textbf{Spectral clustering.} 
This is a family of algorithms that compute spectral embeddings of data points based on the eigenvectors of a graph Laplacian matrix; Alg.~\ref{alg:SpecCls} shows the steps of a spectral clustering algorithm \cite{von2007tutorial}. 
The spectral embeddings (step 4) are built from the $k$ eigenvectors associated with the lower variation on the graph (cf. \eqref{eq.Eigvar}); hence, behaving as an ideal low-pass graph filter. 
This step and the $k$-means clustering in step 7 are the computational bottlenecks of spectral clustering and limit its scalability. 
Thus, approximate solutions are often preferred to trade accuracy with scalability \cite{tremblay2020approximating}. 
The scalability of spectral clustering is enhanced via graph filtering in \cite{tremblay2016compressive}. 
The ideal graph filter is approximated via convolutional (cf. \eqref{eq.ChebCoeff}) or rational \cite{rimleanscaia2020rational} Chebyshev fitting; $k$-means is run only on a sampled number of nodes; and the cluster labels on the remaining nodes are obtained by solving a smooth regularized problem (cf. \eqref{eq.Tikproblem}). 
The filtering operations adopted in compressive spectral clustering are also implemented via the power method in \cite{boutsidis2015spectral} and via an asynchronous implementation in \cite{teke2020nodeSC}.


%
\smallskip
\noindent\textbf{{Blind} community detection.}
As discussed in Section~\ref{subsec_gti}, graph filters can serve as generative models for nodal observations, inspiring a range of network inference methods. 
Inferring the entire graph structure is often only the first step of a longer pipeline where the ultimate goal is to obtain interpretable information from graph signals. 
To this end, a feature that is often sought in network science is the community
structure that offers a coarse description of graphs.
For this task, applying conventional methods necessitates a two-step procedure comprising graph learning and community detection. 
{An alternative line of work, called blind community detection, recovers the communities directly from the observed signals bypassing the intermediate network inference step~\cite{wai_blind_2020}.} 
More precisely, under the assumption that the observed signals $\vcx$ are obtained by passing white noise through a low-pass filter, it follows that the leading eigenvectors of the covariance $\mtSigma_x$ coincide with the $k$ lowest eigenvalues of $\mtL$ (see step 4 in Algorithm~\ref{alg:SpecCls}); see~\cite{wai_blind_2020} for theoretical guarantees.
Once this information is attained, the same steps as spectral clustering can be followed to reveal the community structure.
The benefit of this direct 
approach stems from the fact that fewer observations are needed to recover the coarse community features compared to the detailed graph structure.
Blind recovery of network features has been extended to community detection in dynamic graphs~\cite{roddenberry_blind_2020}, node centrality estimation~\cite{roddenberry_blind_centrality_2021, he_blind_centrality_2022}, and topology change-point detection~\cite{kaushik_change_point_2021, shaked_change_point_2021, marenco_change_point_2022}.

\begin{algorithm}[!t]
\caption{Spectral clustering blueprint.}\label{alg:SpecCls}
\begin{algorithmic}[1]
\State \textbf{Input:} A set of $N$ $d-$dimensional data points $\vcf_1, \ldots, \vcf_N$ and the number of clusters $k$;
\State Build an undirected similarity sparse graph $\stG$ (cf. \eqref{eq.graph}) with each node a data point (e.g., a $K$ nearest neighbor graph of $N$ nodes);
\State Set $\mtS = \mtL \in \fdR^{N \times N}$ to be a graph Laplacian of $\stG$;
\State Take the $k$ eigenvectors $\mtU_k \in \fdR^{N \times k}$ of $\mtL$ associated with the $k$ lowest eigenvalues (smoothest, cf. \eqref{eq.Eigvar}); 
\State Normalize $\mtU_k$ row-wise {(unit norm)} to have $\mttU_k \in \fdR^{N \times k}$;
\State Treat each node $n$ as a data point in $\fdR^k$ and define its feature vector $\vctf_n \in \fdR^k$ as the $n$th row of $\mttU_k$
\begin{equation}
\vctf_n = \mttU_k^\top\vcdelta_n,
\end{equation}
where $\vcdelta_n \in \fdR^N$ is a Dirac vector with $[\vcdelta_n]_m = 1$ if $m = n$ and zero otherwise;
\State Obtain the $k$ clusters via $k-$means with the Euclidean distance $d_{nm} = \|\vctf_n-\vctf_m\|_2$.
\end{algorithmic}
\end{algorithm}


\subsection{Matrix Completion and Collaborative Filtering}\label{subsec:matComp}

Matrix completion comprises filling the missing entries of a partially observed matrix. While its staple application is in recommender systems \cite{chen2022review} it is also used in bioinformatics \cite{natarajan2014inductive}, signal processing \cite{weng2012low}, and chemistry \cite{lee2001survey}, to name a few. Graphs have been used to capture the structural side information among the rows and columns of this matrix and the entries are treated as signals over these graphs. Then, graph filters have been used to interpolate the missing values in a form akin to the signal reconstruction task seen in Sec.~\ref{subsec_denInterp}.

Specifically, consider matrix $\mtR \in \mathbb{R}^{R \times C}$ capturing interactions between $RC$ entities, e.g., $R$ users interacting with $C$ items in a recommender system. We observe only a portion of $\mtR$, which we represent with the masked version $\mtM \odot \mtR$ where $\mtM \in \{0,1\}^{R\times C}$ is the masking matrix with $[\mtM]_{ij} = 1$ if $[\mtR]_{ij}$ is observed and zero otherwise. Then, the canonical matrix completion problem consists of solving
\begin{equation}\label{eq.MatCompl}
\begin{aligned}
\underset{\mtX}{\text{minimize}}~\|\mtM \odot (\mtX - \mtR)\|_F^2 + \gamma \fnr(\mtX)
\end{aligned}
\end{equation}
which looks for a matrix $\mtX \in \mathbb{R}^{R \times C}$ that is close to $\mtR$ on the observed entries while at the same time having a particular structure, e.g., low rank via the nuclear norm $\fnr(\mtX) := \|\mtX\|_*$. Such solutions suffer when an entire row/column of $\mtR$ is not observed or when the low-rank structure in $\mtR$ does not hold. In these cases, one can exploit side information, including, e.g., user features (age, gender, geolocation) for $R$ or social interaction within them; or item features (category, co-purchase) for $C$. Such side information can be used to build two graphs $\mtS_R \in \mathbb{R}^{R \times R}$ and $\mtS_C \in \mathbb{R}^{C \times C}$ and treat each row $\vcr^r$ and column $\vcr_c$ of $\mtR$ as signals on these graphs, respectively. If the side information is unavailable, the graph can be built based on a similarity distance by using the available values in $\mtR$ \cite{gao2021graph}.

\smallskip
\noindent\textbf{Regularized filtering.} Under the assumption 
that connected entities have similar preferences (e.g., similar users tend to like similar products, or co-purchased products tend to be liked similarly), regularized filtering  (Sec.~\ref{subsec_regFilt}) is used to smooth the available values into the adjacent nodes by imposing $\fnr(\mtX) = \text{Tr}(\mtX^\top\mtS_R\mtX) + \text{Tr}(\mtX\mtS_C\mtX^\top)$ as a regularizer in \eqref{eq.MatCompl}, with $\mtS_R$ and $\mtS_C$ being some Laplacian form \cite{gu2010collaborative,du2011user,kalofolias2014matrix,rao2015collaborative}. Such regularizers impose a low-pass filtering behavior on the two graphs. 

\smallskip
\noindent\textbf{Collaborative filtering.} The above graph-based regularizer may be suboptimal because the low-pass filtering leads to interpolated values that are similar in strongly connected nodes. {While this issue could be tackled by choosing a different graph regularizer, going down this path often leads to a trial and error process of choosing regularizer kernels. 
Another approach is to learn the parameters of a graph convolutional filter, in order to gather multi-hop neighbor information.} 
The filter parameters are designed as
\begin{equation}\label{eq.filtDesigCfilt}
\underset{\stH}{\text{minimize}}~\sum_{(r,c)\in \stT}\bigg|\big[	\mtH(\mtS_R)\vcx^r\big]_c - [\mtR]_{rc}	\bigg|^2 + \gamma \fnr(\stH),
\end{equation}
which fits to the available interactions while regularizing them (e.g., norm two). Reference \cite{huang2018rating} shows that a learned graph convolutional filter in this setting behaves as bandstop, in which the low-pass component smoothes the available values while the high-pass component improves diversity. Furthermore, the vanilla nearest neighbor collaborative filter is the particular case of an order one graph convolutional filter. 
Ref. \cite{isufi2021accuracy} uses a filter bank of convolutional filters to balance recommendation accuracy with diversity, 
while \cite{nikolakopoulos2019personalized} follows 
a graph convolutional approach over an item-item graph with the shift operator being a random walk Laplacian. Differently, \cite{chen2020revisiting,he2020lightgcn} treat matrix $\mtR$ as the interactions of a bipartite graph and build a convolutional filter 
on this graph. Reference \cite{shen2021powerful} discusses further the details of these techniques with regularizer filtering for recommender systems. {Extensions to GNNs could be found in \cite{gao2021graph}.}

\subsection{{Supervised Learning with Gaussian Processes}}\label{subsec:gausProc}

{
Consider the common scenario where we are given input-output data pairs $\{\vcx_n, \vcy_n\}$, with each input  $\vcx_n \in \fdR^{K}$ and each output $\vcy_n \in \fdR^{N}$. We wish to learn a model of the form
\begin{eqnarray}\label{Eq:gpmodel}
\vcy_n = \fnf(\vcx_n) + \vceps_n, 
\end{eqnarray}
where $\{\vceps_n\}$ are white Gaussian noise vectors and  $\fnf: \fdR^{K} \rightarrow \fdR^{N}$ is an unknown, multi-output function. 
In Gaussian process (GP) regression \cite{schulz2018tutorial,wang2020intuitive},
$\fnf(\vcx_n)$ is modeled to be distributed as a GP
\begin{equation}
\fnf(\vcx_n) \sim \stG\stP (\fnm(\vcx_n), \fnK(\vcx_n, \vcx_m)),
\end{equation}
which is a distribution over functions characterized by a mean function $\fnm(\vcx) = \mathbb{E}[\fnf(\vcx)]$ (i.e., the weighted average of the evaluations at $\vcx$ of all functions in the distribution) and a covariance kernel function 
\begin{eqnarray*}
\fnK(\vcx_n, \vcx_m)= \mathbb{E}\left[\bigl(\fnf(\vcx_n)-\fnm(\vcx_n)\bigr)\bigl(\fnf(\vcx_m)-\fnm(\vcx_m)\bigr)\right]
\end{eqnarray*}
that models the dependence between function values at two inputs $\vcx_n$ and $\vcx_m$.
}
%
%


{
When the outputs $\vcy_n \in \fdR^{N}$ are graph signals, an alternative model to \eqref{Eq:gpmodel} is 
\begin{equation} \label{Eq:gpmodel_graph}
\vcy_n = \mtH(\mtS)\fnf(\vcx_n) + \vceps_n,
\end{equation}
where again $\fnf(\vcx_n)$ is assumed to be a Gaussian process with covariance kernel  $\fnK(\vcx_n, \vcx_m)$.}
%
Consequently, the covariance matrix between the respective outputs is $\text{Cov}(\vcy_n, \vcy_m) = \fnK(\vcx_n, \vcx_m) \mtH(\mtS)\mtH(\mtS)^\top$ \cite{venkitaraman2020gaussian, zhi2020gaussian}. 

The advantage of {incorporating the graph filters into the regression model \eqref{Eq:gpmodel_graph}} is that we can now impose particular signal behavior properties. We mention three examples. 
First, \cite{venkitaraman2020gaussian} considers $\mtH(\mtS)$ to be a low-pass rational filter of order one \eqref{eq.Tiksol}, {so that the model \eqref{Eq:gpmodel_graph} outputs graph signals that are smooth with respect to the underlying graph}. Second, 
 \cite{borovitskiy2021matern} generalizes the Mat\'ern kernel to the graph setting, yielding a graph filter of the form $\mtH(\mtS) = \big(	\frac{2\alpha}{\beta^2}\mtI_N +	\mtS\big)^{\frac{\alpha}{2}}$, where $\alpha, \beta >0$ and $\mtS$ is some form of the Laplacian. Third, to further enhance the kernel flexibility, \cite{zhi2020gaussian} considers a graph convolutional filter, where the parameters are estimated from the data to ensure a valid kernel. Because of the multi-hop locality of graph filters, such a parametric approach weighs accordingly the information of multi-hop neighbors and has shown a better performance compared with regularized-filtering kernels.

\subsection{Computer Graphics and Computer Vision}\label{subsec:compvis}

In computer graphics and in computer vision -- including subdomains such as virtual reality, geographic information systems, and autonomous driving -- two types of sensing data have become increasingly prevalent 
\cite{lozes2015pde,chen20203d,hu2021graph,jiao2022graph}. First, using light detection and ranging (LiDAR) sensing, the external surfaces of objects are often represented with 3D point clouds and their physical coordinates (and possibly color information). Second, using depth cameras such as Microsoft Kinect, depth maps can be associated with the pixels of 2D images. For both types of data, graph filters have proved useful in common tasks such as object classification \cite{khasanova2017graph}, object tracking \cite{cui2018spectral,gao2019graph,weng2020gnn3dmot}, motion estimation and forecasting \cite{thanou2016graph,liang2020learning}, facial recognition \cite{wang2019linkage}, visual localization \cite{lassance2021improved}, segmentation \cite{wang2019dynamic,giraldo2021emerging}, pose estimation \cite{cai2019exploiting,choi2020pose2mesh}, pose transfer \cite{liu20213d}, compression \cite{thanou2016graph,chen2017fast}, registration \cite{chen2017fast}, surface smoothing \cite{taubin1995signal,taubin1996optimal,chen2017fast,dinesh2020point}, edge detection \cite{bayram2017spectral}, inpainting \cite{akyazi2018graph}, deblurring \cite{yamamoto2016deblurring}, and color denoising \cite{dinesh20193d}. We highlight a few examples.

\smallskip
\noindent\textbf{Surface smoothing.} In one of the earliest examples of using the eigenvectors of a discrete Laplacian to perform graph filtering (1995), Taubin \cite{taubin1995signal,taubin1996optimal} smooths polyhedral surfaces (also called \emph{surface fairing}) by (i) creating a graph by connecting each pair of vertices that share a face in the polyhedral surface, and (ii) updating the vertex locations by applying a lowpass polynomial filter of the random walk Laplacian $\mtL_{\text{rw}}$ to each vector of coordinates; e.g., 
$\vcx_{\text{updated}}=\mtH(\mtL_{\text{rw}}) \vcx$, $\vcy_{\text{updated}}=\mtH(\mtL_{\text{rw}}) \vcy$, and $\vcz_{\text{updated}}=\mtH(\mtL_{\text{rw}}) \vcz$.

\smallskip
\noindent\textbf{Point cloud compression.} To compress a single 3D point cloud in a manner that enhances application-dependent features such as edges, key points, or flatness, \cite{chen2017fast} suggests to resample the point cloud with a resampling distribution that is proportional to the norms of filtered attributes. That is, the probability of resampling vertex $i$ is proportional to $||\vcdelta_i^{\top} \mtH(\mtS)\mtX ||_2$, where $\mtX$ is an $N \times K$ matrix of attributes, with the $i$th row corresponding to the selected attributes (e.g., 3D coordinates, RGB colors, textures) of the $i$th vertex. For example, when $\mtX$ is just the $N \times 3$ matrix of the coordinates, $\mtS=\mtL_{\text{rw}}$, and $\mtH(\mtS)$ is a highpass graph filter, this resampling strategy leads to choosing relatively more points along the contours (e.g., corner points, edges, end points) of the 3D point cloud. The result can be beneficial for contour-based registration to align point clouds.

Given a sequence of 3D point clouds, \cite{thanou2016graph}
(i) uses graph wavelet coefficients as feature vectors to compute point-to-point correspondences between a sparse set of points from point clouds at each successive time; (ii) uses those sparse point-to-point correspondences to estimate motion over time; (iii) interpolates the motion to get a complete point-to-point correspondence mapping over the sequence of point clouds; and, (iv) leverages that motion map to compress and efficiently code the entire sequence of point clouds.

\smallskip
\noindent\textbf{Object tracking}.
This problem consists of identifying an object in a sequence of images, and following its movement through time. 
This can typically be done by graph matching of the object through the sequence of images.
An alternative approach \cite{cui2018spectral}, is to consider the object of interest as a grid graph and designing a graph filter tailored to identifying the object (see Sec.~\ref{sec:filterDesign}).
In particular, \cite{cui2018spectral} learns a graph convolutional filter via least-squares (see Sec.~\ref{sec:wavelets}).
Subsequently, \cite{gao2019graph} considers popular solutions in the space of spatio-temporal Siamese networks, and suggests to replace these networks by graph convolutional networks (see Sec.~\ref{sec:GNN}).

A more challenging problem is that of multi-object tracking, where many different objects have to be tracked simultaneously. 
The typical approach consists of first learning discriminative features for each object, and then tracking the temporal evolution of those features. 
In \cite{weng2020gnn3dmot}, a feature extraction mechanism based on GNNs is proposed. 
The main idea is to exploit the relationship between the objects to learn features that are more discriminative, and thus, easier to distinguish during tracking.
This can be achieved by learning graph filters (either by themselves, or included within a GNN) to highlight high-frequency features -- that is, the ones that are more different across the elements of the graph (see Sec.~\ref{sec:wavelets}).

\section{Where to Start} \label{sec:Start}



Despite the extensive analysis of the different filtering forms, we purposely did not address in detail questions about \emph{when} to use a particular filter type or \emph{when} to choose filter banks or GNNs. While Secs.~\ref{sec:other}, \ref{sec:wavelets}, \ref{sec:GNN} and Table~\ref{tab:filters} discuss the advantages and limitations for each method, we believe there is no single recipe about what solution to use when, and this depends largely on the task at hand. That said, our general recommendation is to start simple, checking if the task can be accomplished with a single graph filter before moving to a filter bank, and seeing if the filter bank provides sufficient representations before proceeding to graph neural networks. Within the class of single filters, we recommend to start with    
the convolutional form (specifically a polynomial filter), and consider the more involved node or edge varying filters when a non-spectral operator is provided or a low (distributed) computation cost is a priority. Within the class of filter banks, the least complicated starting place is probably a single-level tight frame filter bank like the one shown in Fig. \ref{Fig:filter_bank}, as it avoids many extra choices about which vertices to downsample or how to reconnect the downsampled vertices in a coarser graph. We recommend moving to nonlinear filters, filter banks, or GNNs when interested in learning a nonlinear mapping from data. However, the expressive power of the latter (filter order, number of features, and layers) does not have to be too large, as widely suggested by the literature in computer science. 

{For hands-on practitioners, entry points from a GSP perspective are the toolboxes \cite{perraudin2014gspbox,girault2017grasp}, whereas from a machine learning perspective (especially GNNs), we suggest PyTorch Geometric \cite{fey2019fast} and the toolbox available at \url{https://github.com/alelab-upenn/graph-neural-networks}, which contains several of the filtering solutions discussed in this overview.
}

\section{A Look Ahead} \label{sec:Future}


We have identified the following main promising directions regarding fundamental research on graph filters.
\begin{enumerate}
\item The computation cost of graph filters is at best linear in the number of edges in the graph. While this may allow scalability to tens of thousands of nodes, it becomes a challenge for web-scale graphs containing billions of nodes and edges. In these cases, sparsifying techniques on the filter implementation are needed but at the same time the implications of these solutions into the output become more challenging to address. 
\item Some recent works have shown that for particular classes of graphs we can exploit the graph frequency density distribution to improve the filter design. However, it is still unaddressed how to use properties of particular graph families to aid learning and to understand better how the statistical topological properties affect the filter frequency response.
\item We focused on the role of graph filters over static and idealistic graphs. However, real networks are dynamic, noisy, and the respective signals are also time varying. Therefore, one of the biggest challenges is to extend graph filters to this dynamic setting in a principled manner by accounting for the variability in the graph signals and in the number of nodes and edges \cite{isufi20162,grassi2017time,das2022graph}.
\item In several nonlinear tasks (e.g., classification) graph filters are often designed via suboptimal losses to prioritize convex and mathematically tractable solutions. 
Further improvement can be achieved by using non-convex losses and 
iterative algorithms to find the filter parameters.
\item Federated learning tackles the problem of training a model when the data and/or the parameters are located on several different machines \cite{li2020federated}. The central tenet of federated learning involves exchanging messages among these machines in order to train models, while satisfying security, privacy, and communication constraints. This exchange of messages can be interpreted as the implementation of one or more graph filters, and thus, graph filtering has the potential to be a useful framework for analyzing and synthesizing federated learning methods.
\item Regarding applications, graph filters and respective extensions have potential in power and water networks \cite{ramakrishna2021grid,zhou2022bridging}, Internet of Things \cite{jablonski2017graph}, and finance \cite{cardoso2020algorithms}.
\end{enumerate}

Finally, we remark that graphs represent only pairwise relationships between data points but complex networks and data may often be better represented by higher-order network structures \cite{bick2021higher} such as multi-relational graphs \cite{stanley2020multiway}, cell or simplicial complexes \cite{barbarossa2020topological,schaub2021signal,yang2022simplicial,roddenberry_icml_21}, and hypergraphs \cite{barbarossa2016introduction,zhang2019introducing,leus2021topological}. Developing and analysing filters in these settings is an interesting avenue with large potential in both signal processing and machine learning.


%
%
%
%




\linespread{0.9}

\begin{table*}[t]
\centering
\caption{Summary of the different graph filtering forms and their properties (P). "?" means that property has not been proven to hold or not.}
\begin{tabular}{p{2cm} | p{.1cm}p{.1cm}p{.1cm}p{.1cm}p{.1cm}p{.1cm}p{.1cm}p{.1cm}|p{11.5cm}}
\hline
\hline
Filter / Properties & P1 & P2 & P3& P4& P5& P6& P7& P8&Discussion \& Recommendation\\
\hline
Convolutional \eqref{eq:graphConv} %
& \checkmark & \checkmark & \checkmark& \checkmark& \checkmark& \checkmark& \checkmark& \checkmark & Extends naturally from the conventional convolutional filters in DSP and respects also the convolution theorem in \eqref{eq:freqResponse}. May require high-orders $K$ and suffers numerical instabilities for large powers $\mtS^k$. Recommendation is to use them as the baseline solution but often with a normalized GSO.\\
\hline
Rational \eqref{eq.ratOut} & \checkmark & \checkmark & \checkmark & \checkmark & \checkmark & \checkmark & \checkmark & \checkmark & Requires lower orders to approximate a given frequency response. Design is more challenging and requires solving a non-convex constrained problem (cf. \eqref{eq.ratFitt}). Obtaining the output implies approximating an inverse problem via iterative methods (cf. \eqref{eq.ratFilter}). Recommendation is to use them when the design could be centralized and the implementation distributed to reduce the communication cost of higher-oder convolutional filters.\\
\hline
Node var. \eqref{eq.NVfilt} & \checkmark & $\stX$ & $\stX$ & $\stX$ & \checkmark & \checkmark & \checkmark & \checkmark & Can approximate a broader family of operators than convolutional/rational while maintaining local implementation. It is not permutation equivariant thus cannot be transferred across graphs. Hence, recommendation is to consider them for approximating a desired operator over a fixed graph.\\
\hline
Edge var. \eqref{eq.EVfilt} & \checkmark & $\stX$ & $\stX$ & $\stX$ & \checkmark & \checkmark & ? & ? & Increases further the DoFs w.r.t. the node varying filter while maintaining the local implementation. The design problem to fit it into a defined operator is a least squares problem but with higher dimensions compared with the node varying and convolutional filter. As the node varying filter, it cannot be transferred across graphs and the high DoFs need to be reduced when used in a data-driven fashion (regularize design problem or share parameters). Recommendation is to consider for approximating complex tasks on a fixed graph or when a large amount of data is available. \\
\hline
Volterra \eqref{eq.polyGF} & $\stX$ & $\stX$ & $?$ & $\checkmark$ & \checkmark & \checkmark & ? & ? & It is more flexible than the convolutional filter but shares parameters among nodes and enjoys a local implementation. Spectral design is more challenging. It is more appropriate for data fitting compared with the node and edge varying filters because of the low number of parameters and permutation equivariance. It can be a good alternative to the convolutional filtering when the spectral interpretation is not needed and to the node domain filters when parameters are estimated from data. Can still run into overfitting and numerical instabilities, thus, orthogonal polynomials are recommended. \\
\hline
Median \eqref{eq.Med_filt} & $\stX$ & $\stX$ & $\checkmark$ & $\checkmark$ & \checkmark & \checkmark & $?$ & ? & Allows tackling outliers in graph signals propagation via a median operation of locally shifted inputs. Enjoys a local implementation and parameter sharing but the design is feasible only in a data driven fashion. Its application domain is more restricted than the convolutional filter but for denoising in anomalous signals it can be a viable tool.\\
\hline
Tikhonov \eqref{eq.Tiksol} & \checkmark & \checkmark & \checkmark & \checkmark & \checkmark & \checkmark & \checkmark & \checkmark & Particular form of rational filtering of order one for undirected graphs. Typically used to smooth the observed signals. \\
\hline
Sobolev \eqref{eq.Sobsol} & \checkmark & \checkmark & \checkmark & \checkmark & \checkmark & \checkmark & \checkmark & \checkmark & Particular form of rational filtering which can achieve arbitrary order for undirected graphs. It generalizes the Tikhonov filter to smooth observed signals.\\
\hline
Quadratic shift variation \eqref{eq.totVsol} & \checkmark & \checkmark & \checkmark & \checkmark & $\stX$ & \checkmark & ? & $?$ & Inverse smooth filtering on directed graphs. Differently from the undirected counterpart it penalizes sharp shifted signal transitions and obtaining a local implementation with iterative solvers is challenging.\\
\hline
Trend filtering \eqref{eq.GTF1} & $\stX$ & $\stX$ & $?$ & \checkmark & $?$ & $?$ & ? & $?$ & Performs sparse filtering on undirected graphs by penalizing sharp transitions that happen only at a few nodes. It is more appropriate to use where the signal has similar values in group of nodes but arbitrary values in different groups. Proving what properties this type of filter satisfies is more challenging because it lacks a closed-form solution.\\
\hline
Total variation \eqref{eq.TV1} & $\stX$ & $\stX$ & $?$ & \checkmark & $?$ & $?$ & ? & $?$ & Performs sparse filtering on directed graphs by penalizing sharp shifts at a few nodes. It complements the smoothness total variation counterpart \eqref{eq.totVsol}. As for the graph trend filter, it lacks a closed-form solution and can be solved only with iterative methods.\\
\hline
Wiener filtring \eqref{eq.Wienresp} & $\checkmark$ & $\checkmark$ & $\checkmark$ & $\stX$ & $\stX$ & $\stX$ & $\stX$ & $?$ & Performs optimal statistical filtering for stationary graph signals. In the vanilla form, it is a rational filter that does not respect the graph sparsity, hence, many of the properties do not hold. But if we approximate its frequency response either with polynomial or rational filters, we could implement an approximation where all the properties hold. \\
\hline
Multi-GSO \eqref{eq:multiGSOfilt} & $\checkmark$ & $\stX$ & $\checkmark$ & $\checkmark$ & $\checkmark$ & $\checkmark$ & $?$ & $?$ & Performs convolutional filtering over multiple GSOs to represent input-output relations. It inherits several properties of the convolutional form but has a higher descriptive power. Yet, differently from node domain and nonlinear filters, it has less chances to overfit the data. The challenge remains to build multiple GSOs that can aid the problem at hand. \\
\hline\hline
\end{tabular}\\
\label{tab:filters}
\end{table*}

\balance
\bibliographystyle{bibFiles/IEEEtranD}
\bibliography{bibFiles/myIEEEabrv,bibFiles/biblioFilters}

\begin{thebibliography}{100}
\providecommand{\url}[1]{#1}
\csname url@samestyle\endcsname
\providecommand{\newblock}{\relax}
\providecommand{\bibinfo}[2]{#2}
\providecommand{\BIBentrySTDinterwordspacing}{\spaceskip=0pt\relax}
\providecommand{\BIBentryALTinterwordstretchfactor}{4}
\providecommand{\BIBentryALTinterwordspacing}{\spaceskip=\fontdimen2\font plus
\BIBentryALTinterwordstretchfactor\fontdimen3\font minus
  \fontdimen4\font\relax}
\providecommand{\BIBforeignlanguage}[2]{{%
\expandafter\ifx\csname l@#1\endcsname\relax
\typeout{** WARNING: IEEEtran.bst: No hyphenation pattern has been}%
\typeout{** loaded for the language `#1'. Using the pattern for}%
\typeout{** the default language instead.}%
\else
\language=\csname l@#1\endcsname
\fi
#2}}
\providecommand{\BIBdecl}{\relax}
\BIBdecl

\bibitem{oppenheim2001discrete}
A.~V. Oppenheim and R.~W. Schafer, \emph{Discrete-Time Signal Processing},
  3rd~ed.\hskip 1em plus 0.5em minus 0.4em\relax Upper Saddle River, NJ:
  Pearson, 2010.

\bibitem{goodfellow2016deep}
I.~Goodfellow, Y.~Bengio, and A.~Courville, \emph{Deep Learning}, ser. Adaptive
  Comput. Mach. Learning.\hskip 1em plus 0.5em minus 0.4em\relax Cambridge, MA:
  The {MIT} Press, 2016.

\bibitem{marques2017stationary}
A.~G.~Marques, S.~Segarra, G.~Leus, and A.~Ribeiro, ``Stationary graph
  processes and spectral estimation,'' \emph{{IEEE} Trans. Signal Process.},
  vol.~65, no.~22, pp. 5911--5926, 2017.

\bibitem{bronstein2021geometric}
\BIBentryALTinterwordspacing
M.~M. Bronstein, J.~Bruna, T.~Cohen, and P.~Veli{\v{c}}kovi{\'c}, ``Geometric
  deep learning: Grids, groups, graphs, geodesics, and gauges,''
  \emph{arXiv:2104.13478v2}, 2022. [Online]. Available:
  \url{http://arxiv.org/abs/2104.13478}
\BIBentrySTDinterwordspacing

\bibitem{ortega2018graph}
A.~Ortega, P.~Frossard, J.~Kova{\v{c}}evi{\'{c}}, J.~M.~F. Moura, and
  P.~Vandergheynst, ``Graph signal processing: Overview, challenges and
  applications,'' \emph{Proc. {IEEE}}, vol. 106, no.~5, pp. 808--828, 2018.

\bibitem{dong2020graph}
X.~Dong, D.~Thanou, L.~Toni, M.~Bronstein, and P.~Frossard, ``Graph signal
  processing for machine learning: A review and new perspectives,''
  \emph{{IEEE} Signal Process. Mag.}, vol.~37, no.~6, pp. 117--127, 2020.

\bibitem{chung1997spectral}
F.~R.~K. Chung, \emph{Spectral Graph Theory}, ser. Regional Conf. Ser.
  Math.\hskip 1em plus 0.5em minus 0.4em\relax Providence, RI: Amer. Math.
  Soc., 1997, no.~92.

\bibitem{Segarra2017-GraphFilterDesign}
S.~Segarra, A.~G.~Marques, and A.~Ribeiro, ``Optimal graph-filter design and
  applications to distributed linear networks operators,'' \emph{{IEEE} Trans.
  Signal Process.}, vol.~65, no.~15, pp. 4117--4131, 2017.

\bibitem{shuman2018distributed}
{D. I Shuman}, P.~Vandergheynst, D.~Kressner, and P.~Frossard, ``Distributed
  signal processing via {C}hebyshev polynomial approximation,'' \emph{{IEEE}
  Trans. Signal Inform. Process. Networks}, vol.~4, no.~4, pp. 736--751, 2018.

\bibitem{Coutino2019-EdgeVariant}
M.~Coutino, E.~Isufi, and G.~Leus, ``Advances in distributed graph filtering,''
  \emph{{IEEE} Trans. Signal Process.}, vol.~67, no.~9, pp. 2320--2333, 2019.

\bibitem{taubin1995signal}
G.~Taubin, ``A signal processing approach to fair surface design,'' in
  \emph{ACM Conf. Comput. Graph. Interactive Techn.}, 1995, pp. 351--358.

\bibitem{taubin1996optimal}
G.~Taubin, T.~Zhang, and G.~Golub, ``Optimal surface smoothing as filter
  design,'' in \emph{Eur. Conf. Comput. Vision}, 1996, pp. 283--292.

\bibitem{smola2003kernels}
A.~J. Smola and R.~Kondor, ``Kernels and regularization on graphs,'' in
  \emph{Conf. Comput. Learning Theory}, 2003, pp. 144--158.

\bibitem{zhou2004regularization}
D.~Zhou and B.~Sch{\"{o}}lkopf, ``A regularization framework for learning from
  graph data,'' in \emph{Int. Conf. Mach. Learning}, 2004.

\bibitem{bougleux2007discrete}
S.~Bougleux, A.~Elmoataz, and M.~Melkemi, ``Discrete regularization on weighted
  graphs for image and mesh filtering,'' in \emph{Int. Conf. Scale Space
  Variational Methods Comput. Vision}, vol. 4485, 2007, pp. 128--139.

\bibitem{zhang2008graph}
F.~Zhang and E.~R. Hancock, ``Graph spectral image smoothing using the heat
  kernel,'' \emph{Pattern Recognition}, vol.~41, no.~11, pp. 3328--3342, 2008.

\bibitem{crovella2003graph}
M.~Crovella and E.~Kolaczyk, ``Graph wavelets for spatial traffic analysis,''
  in \emph{Joint Conf. {IEEE} Comput. Commun. Soc.}, 2003, pp. 1848--1857.

\bibitem{coifman2006diffusion}
R.~R. Coifman and M.~Maggioni, ``Diffusion wavelets,'' \emph{Appl. Comput.
  Harmonic Anal.}, vol.~21, no.~1, pp. 53--94, 2006.

\bibitem{shuman2013emerging}
{D. I Shuman}, S.~K. Narang, P.~Frossard, A.~Ortega, and P.~Vandergheynst,
  ``The emerging field of signal processing on graphs: Extending
  high-dimensional data analysis to networks and other irregular domains,''
  \emph{{IEEE} Signal Process. Mag.}, vol.~30, no.~3, pp. 83--98, 2013.

\bibitem{puschel2008algebraic}
M.~P{\"{u}}schel and J.~M.~F. Moura, ``Algebraic signal processing: Foundation
  and {$1$-D} time,'' \emph{{IEEE} Trans. Signal Process.}, vol.~56, no.~8, pp.
  3572--3585, 2008.

\bibitem{sandryhaila2013discrete}
A.~Sandryhaila and J.~M.~F. Moura, ``Discrete signal processing on graphs,''
  \emph{{IEEE} Trans. Signal Process.}, vol.~61, no.~7, pp. 1644--1656, 2013.

\bibitem{sandryhaila2014discrete}
A.~Sandyhaila and J.~M.~F. Moura, ``Discrete signal processing on graphs:
  Frequency analysis,'' \emph{{IEEE} Trans. Signal Process.}, vol.~62, no.~12,
  pp. 3042--3054, 2014.

\bibitem{gama2020graphs}
F.~Gama, E.~Isufi, G.~Leus, and A.~Ribeiro, ``Graphs, convolutions, and neural
  networks: From graph filters to graph neural networks,'' \emph{{IEEE} Signal
  Process. Mag.}, vol.~37, no.~6, pp. 128--138, 2020.

\bibitem{ma2021unified}
Y.~Ma, X.~Liu, T.~Zhao, Y.~Liu, J.~Tang, and N.~Shah, ``A unified view on graph
  neural networks as graph signal denoising,'' in \emph{{ACM} Int. Conf. Inf.
  Knowl. Manag.}, 2021, pp. 1202--1211.

\bibitem{zhu2021interpreting}
M.~Zhu, X.~Wang, C.~Shi, H.~Ji, and P.~Cui, ``Interpreting and unifying graph
  neural networks with an optimization framework,'' in \emph{{ACM} Web Conf.},
  2021, pp. 1215--1226.

\bibitem{ruiz2021graph}
L.~Ruiz, F.~Gama, and A.~Ribeiro, ``Graph neural networks: Architectures,
  stability and transferability,'' \emph{Proc. {IEEE}}, vol. 109, no.~5, pp.
  660--682, 2021.

\bibitem{isufi2021edgenets}
E.~Isufi, F.~Gama, and A.~Ribeiro, ``{EdgeNets}: Edge varying graph neural
  networks,'' \emph{{IEEE} Trans. Pattern Anal. Mach. Intell.}, vol.~44,
  no.~11, pp. 3862--3877, 2021.

\bibitem{ortega2022introduction}
A.~Ortega, \emph{Introduction to Graph Signal Processing}.\hskip 1em plus 0.5em
  minus 0.4em\relax Cambridge, UK: Cambridge University Press, 2022.

\bibitem{tremblay2018design}
N.~Tremblay, P.~Gon{\c{c}}alves, and P.~Borgnat, ``Design of graph filters and
  filterbanks,'' in \emph{Cooperative and Graph Signal Processing: Principles
  and Applications}.\hskip 1em plus 0.5em minus 0.4em\relax Academic Press,
  2018, ch.~11, pp. 299--324.

\bibitem{shuman2020Localized}
{D. I Shuman}, ``Localized spectral graph filter frames: A unifying framework,
  survey of design considerations, and numerical comparison,'' \emph{{IEEE}
  Signal Process. Mag.}, vol.~37, no.~6, pp. 43--63, 2020.

\bibitem{ramakrishna2020user}
R.~Ramakrishna, H.-T. Wai, and A.~Scaglione, ``A user guide to low-pass graph
  signal processing and its applications: Tools and applications,''
  \emph{{IEEE} Signal Process. Mag.}, vol.~37, no.~6, pp. 74--85, 2020.

\bibitem{jablonski2017graph}
I.~Jab{\l}o{\'{n}}ski, ``Graph signal processing in applications to sensor
  networks, smart grids, and smart cities,'' \emph{{IEEE} Sensors J.}, vol.~17,
  no.~23, pp. 7659--7666, 2017.

\bibitem{gama2022controlGNN}
F.~Gama, Q.~Li, E.~Tolstaya, A.~Prorok, and A.~Ribeiro, ``Synthesizing
  decentralized controllers with graph neural networks and imitation
  learning,'' \emph{{IEEE} Trans. Signal Process.}, vol.~70, pp. 1932--1946,
  2022.

\bibitem{ramakrishna2021grid}
R.~Ramakrishna and A.~Scaglione, ``Grid-graph signal processing (grid-{GSP}): A
  graph signal processing framework for the power grid,'' \emph{{IEEE} Trans.
  Signal Process.}, vol.~69, pp. 2725--2739, 2021.

\bibitem{eisen2020wirelessEGNN}
M.~Eisen and A.~Ribeiro, ``Optimal wireless resource allocation with random
  edge graph neural networks,'' \emph{{IEEE} Trans. Signal Process.}, vol.~68,
  pp. 2977--2991, 2020.

\bibitem{chowdhury2021uwmmse}
A.~Chowdhury, G.~Verma, C.~Rao, A.~Swami, and S.~Segarra, ``Unfolding {WMMSE}
  using graph neural networks for efficient power allocation,'' \emph{{IEEE}
  Trans. Wireless Commun.}, vol.~20, no.~9, pp. 6004--6017, 2021.

\bibitem{candelieri2014graph}
A.~Candelieri, D.~Conti, and F.~Archetti, ``A graph based analysis of leak
  localization in urban water networks,'' \emph{Procedia Eng.}, vol.~70, pp.
  228--237, 2014.

\bibitem{jain2014big}
R.~K. Jain, J.~M. Moura, and C.~E. Kontokosta, ``Big data + big cities: Graph
  signals of urban air pollution,'' \emph{{IEEE} Signal Process. Mag.},
  vol.~31, no.~5, pp. 130--136, 2014.

\bibitem{von2007tutorial}
U.~Von~Luxburg, ``A tutorial on spectral clustering,'' \emph{Stat. Comput.},
  vol.~17, no.~4, pp. 395--416, 2007.

\bibitem{giannakis2018topology}
G.~B. Giannakis, Y.~Shen, and G.~V. Karanikolas, ``Topology identification and
  learning over graphs: Accounting for nonlinearities and dynamics,''
  \emph{Proc. IEEE}, vol. 106, no.~5, pp. 787--807, 2018.

\bibitem{mateos2019connecting}
G.~Mateos, S.~Segarra, A.~G. Marques, and A.~Ribeiro, ``Connecting the dots:
  Identifying network structure via graph signal processing,'' \emph{{IEEE}
  Signal Process. Mag.}, vol.~36, no.~3, pp. 16--43, 2019.

\bibitem{dong2019learning}
X.~Dong, D.~Thanou, M.~Rabbat, and P.~Frossard, ``Learning graphs from data: A
  signal representation perspective,'' \emph{{IEEE} Signal Process. Mag.},
  vol.~36, no.~3, pp. 44--63, 2019.

\bibitem{huang2018rating}
W.~Huang, A.~G. Marques, and A.~R. Ribeiro, ``Rating prediction via graph
  signal processing,'' \emph{{IEEE} Trans. Signal Process.}, vol.~66, no.~19,
  pp. 5066--5081, 2018.

\bibitem{sporns2016networks}
O.~Sporns, \emph{Networks of the Brain}.\hskip 1em plus 0.5em minus 0.4em\relax
  MIT Press, 2016.

\bibitem{jackson2010social}
M.~O. Jackson, \emph{Social and Economic Networks}.\hskip 1em plus 0.5em minus
  0.4em\relax Princeton University Press, 2010.

\bibitem{huang2016graph}
W.~Huang, L.~Goldsberry, N.~F. Wymbs, S.~T. Grafton, D.~S. Bassett, and
  A.~Ribeiro, ``Graph frequency analysis of brain signals,'' \emph{IEEE J. Sel.
  Topics Signal Process.}, vol.~10, no.~7, pp. 1189--1203, 2016.

\bibitem{xing2022graph}
L.~Xing and L.~Sela, ``Graph neural networks for state estimation in water
  distribution systems: Application of supervised and semisupervised
  learning,'' \emph{J. Water Resour. Plan. Manag.}, vol. 148, no.~5, p.
  04022018, 2022.

\bibitem{wu2020comprehensive}
Z.~Wu, S.~Pan, F.~Chen, G.~Long, C.~Zhang, and S.~Y. Philip, ``A comprehensive
  survey on graph neural networks,'' \emph{IEEE Trans. Neural Netw. Learn.
  Syst}, vol.~32, no.~1, pp. 4--24, 2020.

\bibitem{lu2011link}
L.~L{\"u} and T.~Zhou, ``Link prediction in complex networks: A survey,''
  \emph{Phys. A: Stat. Mech. Appl.}, vol. 390, no.~6, pp. 1150--1170, 2011.

\bibitem{mallat2012scattering}
S.~Mallat, ``Group invariant scattering,'' \emph{Commun. Pure, Appl. Math.},
  vol.~65, no.~10, pp. 1331--1398, 2012.

\bibitem{willsky1997signals}
A.~V. Oppenheim, A.~S. Willsky, and S.~H. Nawab, \emph{Signals and Systems},
  2nd~ed., ser. Prentice Hall Signal Process.\hskip 1em plus 0.5em minus
  0.4em\relax Prentice Hall, 1997.

\bibitem{schafer2010discreteSP}
A.~V. Oppenheim and R.~W. Schafer, \emph{Discrete-Time Signal Processing},
  3rd~ed.\hskip 1em plus 0.5em minus 0.4em\relax Pearson, 2010.

\bibitem{sardellitti2017graph}
S.~Sardellitti, S.~Barbarossa, and P.~Di~Lorenzo, ``On the graph {F}ourier
  transform for directed graphs,'' \emph{IEEE J. Sel. Topics Signal Process.},
  vol.~11, no.~6, pp. 796--811, 2017.

\bibitem{shafipour2018directed}
R.~Shafipour, A.~Khodabakhsh, G.~Mateos, and E.~Nikolova, ``A directed graph
  {F}ourier transform with spread frequency components,'' \emph{{IEEE} Trans.
  Signal Process.}, vol.~67, no.~4, pp. 946--960, 2018.

\bibitem{girault2018irregularity}
B.~Girault, A.~Ortega, and S.~S. Narayanan, ``Irregularity-aware graph fourier
  transforms,'' \emph{IEEE Transactions on Signal Processing}, vol.~66, no.~21,
  pp. 5746--5761, 2018.

\bibitem{gama2020stability}
F.~Gama, J.~Bruna, and A.~Ribeiro, ``Stability properties of graph neural
  networks,'' \emph{{IEEE} Trans. Signal Process.}, vol.~68, pp. 5680--5695,
  2020.

\bibitem{Isufi2017-ARMA}
E.~Isufi, A.~Loukas, A.~Simonetto, and G.~Leus, ``Autoregressive moving average
  graph filtering,'' \emph{{IEEE} Trans. Signal Process.}, vol.~65, no.~2, pp.
  274--288, 2016.

\bibitem{Kruzick2018-filter}
S.~Kruzick and J.~M.~F. Moura, ``Optimal filter design for signal processing on
  random graphs: Accelerated consensus,'' \emph{{IEEE} Trans. Signal Process.},
  vol.~66, no.~5, pp. 1258--1272, 2018.

\bibitem{druskin1989two}
V.~L. Druskin and L.~A. Knizhnerman, ``Two polynomial methods of calculating
  functions of symmetric matrices,'' \emph{USSR Comput. Math. Math. Phys.},
  vol.~29, no.~6, pp. 112--121, 1989.

\bibitem{tseng2021minimax}
C.-C. Tseng and S.-L. Lee, ``Minimax design of graph filter using {C}hebyshev
  polynomial approximation,'' \emph{IEEE Trans. Circuits Syst. II: Express
  Br.}, vol.~68, no.~5, pp. 1630--1634, 2021.

\bibitem{pakiyarajah2022minimax}
D.~Pakiyarajah and C.~U. Edussooriya, ``Minimax design of
  computationally-efficient {FIR} graph filters using semidefinite
  programming,'' \emph{IEEE Trans. Circuits Syst. II: Express Br.}, 2022.

\bibitem{trefethen2019approximation}
L.~N. Trefethen, \emph{Approximation Theory and Approximation Practice,
  Extended Edition}.\hskip 1em plus 0.5em minus 0.4em\relax SIAM, 2019.

\bibitem{coutino2020fast}
M.~Coutino, S.~P. Chepuri, T.~Maehara, and G.~Leus, ``Fast spectral
  approximation of structured graphs with applications to graph filtering,''
  \emph{Algorithms}, vol.~13, no.~9, p. 214, 2020.

\bibitem{Fan2020-Spectrum}
T.~Fan, {D. I Shuman}, S.~Ubaru, and Y.~Saad, ``Spectrum-adapted polynomial
  approximation for matrix functions with applications in graph signal
  processing,'' \emph{Algorithms}, vol.~13, no.~11, p. 295, 2020.

\bibitem{Kruzick2017-filter}
S.~Kruzick and J.~M.~F. Moura, ``Graph signal processing: Filter design and
  spectral statistics,'' in \emph{{IEEE} Int. Workshop Comput. Advances
  Multi-Sensor Adaptive Process.}, 2017, pp. 1--5.

\bibitem{liu2018filter}
J.~Liu, E.~Isufi, and G.~Leus, ``Filter design for autoregressive moving
  average graph filters,'' \emph{{IEEE} Trans. Signal Inform. Process.
  Networks}, vol.~5, no.~1, pp. 47--60, 2018.

\bibitem{Sakiyama2017-polynomial}
A.~Sakiyama, T.~Namiki, and Y.~Tanaka, ``Design of polynomial approximated
  filters for signals on directed graphs,'' in \emph{{IEEE} Global Conf. Signal
  and Inform. Process.}, 2017, pp. 633--637.

\bibitem{Segarra2017-Blind}
S.~Segarra, G.~Mateos, A.~G. Marques, and A.~Ribeiro, ``Blind identification of
  graph filters,'' \emph{{IEEE} Trans. Signal Process.}, vol.~65, no.~5, pp.
  1146--1159, 2017.

\bibitem{Ramirez2021-Blind}
D.~Ram{\'\i}rez, A.~G. Marques, and S.~Segarra, ``Graph-signal reconstruction
  and blind deconvolution for structured inputs,'' \emph{Signal Process.}, vol.
  188, p. 108180, 2021.

\bibitem{Natali2020-Topology}
A.~Natali, M.~Coutino, and G.~Leus, ``Topology-aware joint graph filter and
  edge weight identification for network processes,'' in \emph{{IEEE} Int.
  Workshop Mach. Learning Signal Process.}, 2020, pp. 1--6.

\bibitem{rey2022robust}
S.~Rey, V.~M. Tenorio, and A.~G. Marques, ``Robust graph filter identification
  and graph denoising from signal observations,'' \emph{arXiv preprint
  arXiv:2210.08488}, 2022.

\bibitem{Rey2021-Filter}
S.~Rey and A.~G. Marques, ``Robust graph-filter identification with graph
  denoising regularization,'' in \emph{{IEEE} Int. Conf. Acoust., Speech and
  Signal Process.}, 2021, pp. 5300--5304.

\bibitem{Ahmed2014-Deconvolution}
A.~Ahmed, B.~Recht, and J.~Romberg, ``Blind deconvolution using convex
  programming,'' \emph{{IEEE} Trans. Inf. Theory}, vol.~60, no.~3, pp.
  1711--1732, 2014.

\bibitem{Iwata2020-Dconvolution}
K.~Iwata, K.~Yamada, and Y.~Tanaka, ``Graph blind deconvolution with sparseness
  constraint,'' \emph{arXiv preprint arXiv:2010.14002}, 2020.

\bibitem{Yu2020-Blind}
Y.~Zhu, F.~J.~I. Garcia, A.~G. Marques, and S.~Segarra, ``Estimating network
  processes via blind identification of multiple graph filters,'' \emph{{IEEE}
  Trans. Signal Process.}, vol.~68, pp. 3049--3063, 2020.

\bibitem{Iglesias2020-Demixing}
F.~J. Iglesias, S.~Segarra, and A.~G. Marques, ``Blind demixing of diffused
  graph signals,'' \emph{arXiv preprint arXiv:2012.13301}, 2020.

\bibitem{shi2015infinite}
X.~Shi, H.~Feng, M.~Zhai, T.~Yang, and B.~Hu, ``Infinite impulse response graph
  filters in wireless sensor networks,'' \emph{{IEEE} Signal Process. Lett.},
  vol.~22, no.~8, pp. 1113--1117, 2015.

\bibitem{levie2018cayleynets}
R.~Levie, F.~Monti, X.~Bresson, and M.~M. Bronstein, ``Cayleynets: Graph
  convolutional neural networks with complex rational spectral filters,''
  \emph{{IEEE} Trans. Signal Process.}, vol.~67, no.~1, pp. 97--109, 2018.

\bibitem{jiang2019decentralised}
J.~Jiang and D.~B. Tay, ``Decentralised signal processing on graphs via matrix
  inverse approximation,'' \emph{Signal Process.}, vol. 165, pp. 292--302,
  2019.

\bibitem{cheng2020preconditioned}
C.~Cheng, N.~Emirov, and Q.~Sun, ``Preconditioned gradient descent algorithm
  for inverse filtering on spatially distributed networks,'' \emph{{IEEE}
  Signal Process. Lett.}, vol.~27, pp. 1834--1838, 2020.

\bibitem{loukas2015distributed}
A.~Loukas, A.~Simonetto, and G.~Leus, ``Distributed autoregressive moving
  average graph filters,'' \emph{{IEEE} Signal Process. Lett.}, vol.~22,
  no.~11, pp. 1931--1935, 2015.

\bibitem{isufi2017autoregressive}
E.~Isufi, A.~Loukas, and G.~Leus, ``Autoregressive moving average graph
  filters: {A} stable distributed implementation,'' in \emph{{IEEE} Int. Conf.
  Acoust., Speech and Signal Process.}, 2017, pp. 4119--4123.

\bibitem{hayes2009statistical}
M.~H. Hayes, \emph{Statistical Digital Signal Processing and Modeling}.\hskip
  1em plus 0.5em minus 0.4em\relax John Wiley \& Sons, 2009.

\bibitem{aittomaki2019graph}
T.~Aittom{\"a}ki and G.~Leus, ``Graph filter design using sum-of-squares
  representation,'' in \emph{Eur. Signal Process. Conf.}, 2019, pp. 1--5.

\bibitem{jiang2019stable}
A.~Jiang, B.~Ni, J.~Wan, and H.~K. Kwan, ``Stable {ARMA} graph filter design
  via partial second-order factorization,'' in \emph{IEEE Int. Symp. Circuits
  Syst.}, 2019, pp. 1--5.

\bibitem{pakiyarajah2021wls}
D.~Pakiyarajah and C.~U. Edussooriya, ``{WLS} design of {ARMA} graph filters
  using iterative second-order cone programming,'' in \emph{{IEEE} Int. Conf.
  Acoust., Speech and Signal Process.}, 2022, pp. 5937--5941.

\bibitem{desbrun1999implicit}
M.~Desbrun, M.~Meyer, P.~Schr{\"o}der, and A.~H. Barr, ``Implicit fairing of
  irregular meshes using diffusion and curvature flow,'' in \emph{ACM Conf.
  Comput. Graph. Interactive Techn.}, 1999, pp. 317--324.

\bibitem{rimleanscaia2020rational}
O.~Rimleanscaia and E.~Isufi, ``Rational {Chebyshev} graph filters,'' in
  \emph{Asilomar Conf. Signals, Systems and Comput.}, 2020, pp. 736--740.

\bibitem{tseng2020rational}
C.-C. Tseng, ``Rational graph filter design using spectral transformation and
  {IIR} digital filter,'' in \emph{IEEE Region 10 Conf.}, 2020, pp. 247--250.

\bibitem{cheng2019iterative}
C.~Cheng, J.~Jiang, N.~Emirov, and Q.~Sun, ``Iterative {Chebyshev} polynomial
  algorithm for signal denoising on graphs,'' in \emph{Int. Conf. Samp. Theory
  and Appl.}, 2019, pp. 1--5.

\bibitem{hua2020online}
F.~Hua, R.~Nassif, C.~Richard, H.~Wang, and A.~H. Sayed, ``Online distributed
  learning over graphs with multitask graph-filter models,'' \emph{{IEEE}
  Trans. Signal Inform. Process. Networks}, vol.~6, pp. 63--77, 2020.

\bibitem{alinaghi2021graph}
A.~Alinaghi, S.~Weiss, V.~Stankovic, and I.~Proudler, ``Graph filter design for
  distributed network processing: a comparison between adaptive algorithms,''
  in \emph{2021 Sensor Signal Processing for Defence Conference (SSPD)}, 2021,
  pp. 1--5.

\bibitem{coutinominguez2021cascaded}
M.~A. Coutino~Minguez and G.~Leus, ``A cascaded structure for generalized graph
  filters,'' \emph{{IEEE} Trans. Signal Process.}, 2021.

\bibitem{gama2021node}
F.~Gama, B.~G. Anderson, and S.~Sojoudi, ``Node-variant graph filters in graph
  neural networks,'' in \emph{{IEEE} Data Sci. Learning Workshop}, 2022, pp.
  1--6.

\bibitem{xiao2021distributed}
Z.~Xiao, H.~Fang, and X.~Wang, ``Distributed nonlinear polynomial graph filter
  and its output graph spectrum: Filter analysis and design,'' \emph{{IEEE}
  Trans. Signal Process.}, vol.~69, pp. 1--15, 2021.

\bibitem{segarra2016center}
S.~Segarra, A.~G. Marques, G.~R. Arce, and A.~Ribeiro, ``Center-weighted median
  graph filters,'' in \emph{{IEEE} Global Conf. Signal and Inform. Process.},
  2016, pp. 336--340.

\bibitem{segarra2017design}
S.~Segarra, A.~G. Marques, G.~R. Arce, and A.~Ribeiro, ``Design of weighted
  median graph filters,'' in \emph{IEEE Int. Workshop Comp. Adv. Multi-Sensor
  Adaptive Process.}, 2017, pp. 1--5.

\bibitem{ruiz2019invariance}
L.~Ruiz, F.~Gama, A.~G. Marques, and A.~Ribeiro, ``Invariance-preserving
  localized activation functions for graph neural networks,'' \emph{{IEEE}
  Trans. Signal Process.}, vol.~68, pp. 127--141, 2019.

\bibitem{iancu2020graph}
B.~Iancu, L.~Ruiz, A.~Ribeiro, and E.~Isufi, ``Graph-adaptive activation
  functions for graph neural networks,'' in \emph{IEEE Int. Workshop Mach.
  Learn. Signal Process.}, 2020, pp. 1--6.

\bibitem{giraldo2022reconstruction}
J.~H. Giraldo, A.~Mahmood, B.~Garcia-Garcia, D.~Thanou, and T.~Bouwmans,
  ``Reconstruction of time-varying graph signals via {Sobolev} smoothness,''
  \emph{{IEEE} Trans. Signal Inform. Process. Networks}, vol.~8, pp. 201--214,
  2022.

\bibitem{chen2017bias}
P.-Y. Chen and S.~Liu, ``Bias-variance tradeoff of graph {Laplacian}
  regularizer,'' \emph{{IEEE} Signal Process. Lett.}, vol.~24, no.~8, pp.
  1118--1122, 2017.

\bibitem{romero2016kernel}
D.~Romero, M.~Ma, and G.~B. Giannakis, ``Kernel-based reconstruction of graph
  signals,'' \emph{{IEEE} Trans. Signal Process.}, vol.~65, no.~3, pp.
  764--778, 2016.

\bibitem{yang2021node}
M.~Yang, M.~Coutino, G.~Leus, and E.~Isufi, ``Node-adaptive regularization for
  graph signal reconstruction,'' \emph{IEEE Open J. Signal Process.}, vol.~2,
  pp. 85--98, 2021.

\bibitem{wang2015trend}
Y.-X. Wang, J.~Sharpnack, A.~Smola, and R.~Tibshirani, ``Trend filtering on
  graphs,'' in \emph{Int. Conf. Artificial Intell., Statist.}, 2015, pp.
  1042--1050.

\bibitem{varma2019vector}
R.~Varma, H.~Lee, J.~Kova{\v{c}}evi{\'c}, and Y.~Chi, ``Vector-valued graph
  trend filtering with non-convex penalties,'' \emph{{IEEE} Trans. Signal
  Inform. Process. Networks}, vol.~6, pp. 48--62, 2019.

\bibitem{girault2015stationary}
B.~Girault, ``Stationary graph signals using an isometric graph translation,''
  in \emph{Eur. Signal Process. Conf.}, 2015, pp. 1516--1520.

\bibitem{perraudin2017stationary}
N.~Perraudin and P.~Vandergheynst, ``Stationary signal processing on graphs,''
  \emph{{IEEE} Trans. Signal Process.}, vol.~65, no.~13, pp. 3462--3477, 2017.

\bibitem{zhang2015graph}
C.~Zhang, D.~Flor{\^e}ncio, and P.~A. Chou, ``Graph signal processing-a
  probabilistic framework,'' \emph{Microsoft Res., Redmond, WA, USA, Tech. Rep.
  MSR-TR-2015-31}, 2015.

\bibitem{girault2014semi}
B.~Girault, P.~Gon{\c{c}}alves, E.~Fleury, and A.~S. Mor, ``Semi-supervised
  learning for graph to signal mapping: A graph signal {Wiener} filter
  interpretation,'' in \emph{{IEEE} Int. Conf. Acoust., Speech and Signal
  Process.}, 2014, pp. 1115--1119.

\bibitem{isufi2018distributed}
E.~Isufi, P.~Di~Lorenzo, P.~Banelli, and G.~Leus, ``Distributed {Wiener}-based
  reconstruction of graph signals,'' in \emph{IEEE Stat. Signal Process.
  Workshop}, 2018, pp. 21--25.

\bibitem{zheng2022wiener}
C.~Zheng, C.~Cheng, and Q.~Sun, ``Wiener filters on graphs and distributed
  polynomial approximation algorithms,'' \emph{arXiv preprint
  arXiv:2205.04019}, 2022.

\bibitem{sevi2018harmonic}
H.~Sevi, G.~Rilling, and P.~Borgnat, ``Harmonic analysis on directed graphs and
  applications: {From Fourier} analysis to wavelets,'' \emph{Appl. Comput.
  Harmon. Anal.}, vol.~62, pp. 390--440, 2023.

\bibitem{Hua2019-combination}
F.~Hua, C.~Richard, J.~Chen, H.~Wang, P.~Borgnat, and P.~Gon{\c c}alves,
  ``Learning combination of graph filters for graph signal modeling,''
  \emph{{IEEE} Signal Process. Lett.}, vol.~26, no.~12, pp. 1912--1916, 2019.

\bibitem{fan2019global}
J.~Fan, C.~Tepedelenlioglu, and A.~Spanias, ``Global optimization of graph
  filters with multiple shift matrices,'' in \emph{Asilomar Conf. Signals,
  Systems and Comput.}, 2019, pp. 2082--2086.

\bibitem{emirov2022polynomial}
N.~Emirov, C.~Cheng, J.~Jiang, and Q.~Sun, ``Polynomial graph filters of
  multiple shifts and distributed implementation of inverse filtering,''
  \emph{Sampl. Theory Signal Process. Data Anal.}, vol.~20, no.~1, pp. 1--39,
  2022.

\bibitem{bunny}
{Stanford University Computer Graphics Laboratory}, ``{The Stanford 3D Scanning
  Repository},'' http://graphics.stanford.edu/data/3Dscanrep/.

\bibitem{shuman2013spectrum}
{D. I Shuman}, C.~Wiesmeyr, N.~Holighaus, and P.~Vandergheynst,
  ``Spectrum-adapted tight graph wavelet and vertex-frequency frames,''
  \emph{IEEE Trans. Signal Process.}, vol.~63, no.~16, pp. 4223--4235, 2015.

\bibitem{leonardi_multislice}
N.~Leonardi and D.~{Van De Ville}, ``Tight wavelet frames on multislice
  graphs,'' \emph{IEEE Trans. Signal Process.}, vol.~61, no.~13, pp.
  3357--3367, 2013.

\bibitem{dong2017sparse}
B.~Dong, ``Sparse representation on graphs by tight wavelet frames and
  applications,'' \emph{Appl. Comput. Harmon. Anal.}, vol.~42, no.~3, pp.
  452--479, 2017.

\bibitem{gobel2018construction}
F.~G{\"o}bel, G.~Blanchard, and U.~von Luxburg, ``Construction of tight frames
  on graphs and application to denoising,'' in \emph{Handbook of Big Data
  Analytics}.\hskip 1em plus 0.5em minus 0.4em\relax Springer, 2018, pp.
  503--522.

\bibitem{tay2017almost}
D.~B. Tay, Y.~Tanaka, and A.~Sakiyama, ``Almost tight spectral graph wavelets
  with polynomial filters,'' \emph{IEEE J. Sel. Topics Signal Process.},
  vol.~11, no.~6, pp. 812--824, 2017.

\bibitem{sakiyama2016spectral}
A.~Sakiyama, K.~Watanabe, and Y.~Tanaka, ``Spectral graph wavelets and filter
  banks with low approximation error,'' \emph{IEEE Trans. Signal Inf. Process.
  Netw.}, vol.~2, no.~3, pp. 230--245, 2016.

\bibitem{fan_algorithms}
T.~Fan, {D. I Shuman}, S.~Ubaru, and Y.~Saad, ``Spectrum-adapted polynomial
  approximation for matrix functions with applications in graph signal
  processing,'' \emph{Algorithms}, vol.~13, no. 11, 295, pp. 1--22, 2020.

\bibitem{jiang2019nonsubsampled}
J.~Jiang, C.~Cheng, and Q.~Sun, ``Nonsubsampled graph filter banks: {T}heory
  and distributed algorithms,'' \emph{IEEE Trans. Signal Process.}, vol.~67,
  no.~15, pp. 3938--3953, 2019.

\bibitem{hammond2011wavelets}
D.~K. Hammond, P.~Vandergheynst, and R.~Gribonval, ``Wavelets on graphs via
  spectral graph theory,'' \emph{Appl. Comput. Harmonic Anal.}, vol.~30, no.~2,
  pp. 129--150, 2011.

\bibitem{strang1996wavelets}
G.~Strang and T.~Nguyen, \emph{Wavelets and Filter Banks}.\hskip 1em plus 0.5em
  minus 0.4em\relax {SIAM}, 1996.

\bibitem{ricaud_sparsity_SPIE_2013}
B.~Ricaud, {D. I Shuman}, and P.~Vandergheynst, ``On the sparsity of wavelet
  coefficients for signals on graphs,'' in \emph{{SPIE} Wavelets and Sparsity},
  2013.

\bibitem{tremblay2014graph}
N.~Tremblay and P.~Borgnat, ``Graph wavelets for multiscale community mining,''
  \emph{IEEE Trans. Signal Process.}, vol.~62, pp. 5227--5239, 2014.

\bibitem{dong2013inference}
X.~Dong, A.~Ortega, P.~Frossard, and P.~Vandergheynst, ``Inference of mobility
  patterns via spectral graph wavelets,'' in \emph{{IEEE} Int. Conf. Acoust.,
  Speech and Signal Process.}, 2013, pp. 3118--3122.

\bibitem{shuman_SSL_SAMPTA_2011}
{D. I Shuman}, M.~J. Faraji, and P.~Vandergheynst, ``Semi-supervised learning
  with spectral graph wavelets,'' in \emph{Int. Conf. Samp. Theory and Appl.},
  2011.

\bibitem{kerola2014spectral}
T.~Kerola, N.~Inoue, and K.~Shinoda, ``Spectral graph skeletons for {3D} action
  recognition,'' in \emph{Asian Conf. Comp. Vision}, 2014, pp. 417--432.

\bibitem{behjat2015anatomically}
H.~Behjat, N.~Leonardi, L.~S{\"o}rnmo, and D.~Van De~Ville,
  ``Anatomically-adapted graph wavelets for improved group-level {fMRI}
  activation mapping,'' \emph{NeuroImage}, vol. 123, pp. 185--199, 2015.

\bibitem{donnat2018learning}
C.~Donnat, M.~Zitnik, D.~Hallac, and J.~Leskovec, ``Learning structural node
  embeddings via diffusion wavelets,'' in \emph{ACM Int. Conf. Knowl. Discov.
  Data Min.}, 2018, pp. 1320--1329.

\bibitem{narang_icip}
S.~K. Narang and A.~Ortega, ``Local two-channel critically sampled filter-banks
  on graphs,'' in \emph{IEEE Int. Conf. Image Process.}, 2010, pp. 333--336.

\bibitem{anis2017critical}
A.~Anis and A.~Ortega, ``Critical sampling for wavelet filterbanks on arbitrary
  graphs,'' in \emph{{IEEE} Int. Conf. Acoust., Speech and Signal Process.},
  2017, pp. 3889--3893.

\bibitem{tay2015techniques}
D.~B. Tay and J.~Zhang, ``Techniques for constructing biorthogonal bipartite
  graph filter banks,'' \emph{{IEEE} Trans. Signal Process.}, vol.~63, no.~21,
  pp. 5772--5783, 2015.

\bibitem{pavez2022two}
E.~Pavez, B.~Girault, A.~Ortega, and P.~A. Chou, ``Two channel filter banks on
  arbitrary graphs with positive semi definite variation operators,''
  \emph{{IEEE} Trans. Signal Process.}, vol.~71, pp. 917--932, 2023.

\bibitem{narang2012perfect}
S.~K. Narang and A.~Ortega, ``Perfect reconstruction two-channel wavelet
  filter-banks for graph structured data,'' \emph{IEEE. Trans. Signal
  Process.}, vol.~60, no.~6, pp. 2786--2799, 2012.

\bibitem{narang_bior_filters}
S.~K. Narang and A.~Ortega, ``Compact support biorthogonal wavelet filterbanks
  for arbitrary undirected graphs,'' \emph{IEEE Trans. Signal Process.},
  vol.~61, no.~19, pp. 4673--4685, 2013.

\bibitem{zeng2017bipartite}
J.~Zeng, G.~Cheung, and A.~Ortega, ``Bipartite approximation for graph wavelet
  signal decomposition,'' \emph{{IEEE} Trans. Signal Process.}, vol.~65,
  no.~20, pp. 5466--5480, 2017.

\bibitem{tay2017bipartite}
D.~B. Tay and A.~Ortega, ``Bipartite graph filter banks: {P}olyphase analysis
  and generalization,'' \emph{{IEEE} Trans. Signal Process.}, vol.~65, no.~18,
  pp. 4833--4846, 2017.

\bibitem{tay2017critically}
D.~B. Tay, Y.~Tanaka, and A.~Sakiyama, ``Critically sampled graph filter banks
  with polynomial filters from regular domain filter banks,'' \emph{Signal
  Process.}, vol. 131, pp. 66--72, 2017.

\bibitem{ekambaram_icip}
V.~N. Ekambaram, G.~C. Fanti, B.~Ayazifar, and K.~Ramchandran, ``Circulant
  structures and graph signal processing,'' in \emph{{IEEE} Int. Conf. Image
  Process.}, 2013.

\bibitem{ekambaram2013globalsip}
V.~N. Ekambaram, G.~Fanti, B.~Ayazifar, and K.~Ramchandran,
  ``Critically-sampled perfect-reconstruction spline-wavelet filterbanks for
  graph signals,'' in \emph{{IEEE} Glob. Conf. Signal and Inform. Process.},
  2013, pp. 475--478.

\bibitem{kotzagiannidis2017splines}
M.~S. Kotzagiannidis and P.~L. Dragotti, ``Splines and wavelets on circulant
  graphs,'' \emph{Appl. Comput. Harmon. Anal.}, vol.~47, no.~2, pp. 481--515,
  2019.

\bibitem{teke2016}
O.~Teke and P.~P. Vaidyanathan, ``Graph filter banks with {M-channels}, maximal
  decimation, and perfect reconstruction,'' in \emph{{IEEE} Int. Conf. Acc.,
  Speech, and Signal Process.}, 2016, pp. 4089--4093.

\bibitem{teke2017ii}
O.~Teke and P.~P. Vaidyanathan, ``Extending classical multirate signal
  processing theory to graphs -- {Part II: M}-channel filter banks,''
  \emph{IEEE Trans. Signal Process.}, vol.~65, no.~2, pp. 423--437, 2017.

\bibitem{tay2019m}
D.~B. Tay and A.~Ortega, ``M-channel graph filter banks: {P}olyphase analysis
  and structures,'' \emph{{IEEE} Signal Process. Lett.}, vol.~26, no.~5, pp.
  730--734, 2019.

\bibitem{tay2014design}
D.~B. Tay and Z.~Lin, ``Design of near orthogonal graph filter banks,''
  \emph{{IEEE} Signal Process. Lett.}, vol.~22, no.~6, pp. 701--704, 2014.

\bibitem{zhang2016design}
X.~Zhang, ``Design of orthogonal graph wavelet filter banks,'' in \emph{Conf.
  {IEEE} Ind. Electron. Soc.}, 2016, pp. 889--894.

\bibitem{tay2016near}
D.~B. Tay, Y.~Tanaka, and A.~Sakiyama, ``Near orthogonal oversampled graph
  filter banks,'' \emph{{IEEE} Signal Process. Lett.}, vol.~23, no.~2, pp.
  277--281, 2016.

\bibitem{jansen}
M.~Jansen, G.~P. Nason, and B.~W. Silverman, ``Multiscale methods for data on
  graphs and irregular multidimensional situations,'' \emph{J. R. Stat. Soc.
  Ser. B Stat. Methodol.}, vol.~71, no.~1, pp. 97--125, 2009.

\bibitem{narang_lifting_graphs}
S.~K. Narang and A.~Ortega, ``Lifting based wavelet transforms on graphs,'' in
  \emph{{APSIPA ASC}}, 2009, pp. 441--444.

\bibitem{tay2018cascade}
D.~B. Tay, A.~Ortega, and A.~Anis, ``Cascade and lifting structures in the
  spectral domain for bipartite graph filter banks,'' in \emph{{APSIPA ASC}},
  2018, pp. 1141--1147.

\bibitem{jiang2020design}
J.~Jiang, D.~B. Tay, Q.~Sun, and S.~Ouyang, ``Design of nonsubsampled graph
  filter banks via lifting schemes,'' \emph{{IEEE} Signal Process. Lett.},
  vol.~27, pp. 441--445, 2020.

\bibitem{shuman_TSP_multiscale}
{D. I Shuman}, M.~Faraji, and P.~Vandergheynst, ``A multiscale pyramid
  transform for graph signals,'' \emph{{IEEE} Trans. Signal Process.}, vol.~64,
  no.~8, pp. 2119--2134, 2016.

\bibitem{tanaka2014m}
Y.~Tanaka and A.~Sakiyama, ``{$M$}-channel oversampled graph filter banks,''
  \emph{{IEEE} Trans. Signal Process.}, vol.~62, no.~14, pp. 3578--3590, 2014.

\bibitem{sakiyama2014oversampled}
A.~Sakiyama and Y.~Tanaka, ``Oversampled graph {L}aplacian matrix for graph
  filter banks,'' \emph{{IEEE} Trans. Signal Process.}, vol.~62, no.~24, pp.
  6425--6437, 2014.

\bibitem{tremblay2016subgraph}
N.~Tremblay and P.~Borgnat, ``Subgraph-based filterbanks for graph signals,''
  \emph{IEEE Trans. Signal Process.}, vol.~64, no.~15, pp. 3827--3840, 2016.

\bibitem{chen2015discrete}
S.~Chen, R.~Varma, A.~Sandryhaila, and J.~Kova\v{c}evi{\'{c}}, ``Discrete
  signal processing on graphs: {S}ampling theory,'' \emph{IEEE Trans. Signal
  Process.}, vol.~63, no.~24, pp. 6510--6523, 2015.

\bibitem{li_mcsfb_2018}
S.~Li, Y.~Jin, and {D. I Shuman}, ``Scalable {$M$}-channel critically sampled
  filter banks for graph signals,'' \emph{IEEE Trans. Signal Process.},
  vol.~67, no.~15, pp. 3954--3969, 2019.

\bibitem{sakiyama2019two}
A.~Sakiyama, K.~Watanabe, Y.~Tanaka, and A.~Ortega, ``Two-channel critically
  sampled graph filter banks with spectral domain sampling,'' \emph{{IEEE}
  Trans. Signal Process.}, vol.~67, no.~6, pp. 1447--1460, 2019.

\bibitem{sakiyama2019m}
A.~Sakiyama, K.~Watanabe, and Y.~Tanaka, ``M-channel critically sampled
  spectral graph filter banks with symmetric structure,'' \emph{{IEEE} Signal
  Process. Lett.}, vol.~26, no.~5, pp. 665--669, 2019.

\bibitem{miraki2021spectral}
A.~Miraki, H.~Saeedi-Sourck, N.~Marchetti, and A.~Farhang, ``Spectral domain
  spline graph filter bank,'' \emph{{IEEE} Signal Process. Lett.}, vol.~28, pp.
  469--473, 2021.

\bibitem{nguyen2014downsampling}
H.~Q. Nguyen and M.~N. Do, ``Downsampling of signals on graphs via maximum
  spanning trees,'' \emph{{IEEE} Trans. Signal Process.}, vol.~63, no.~1, pp.
  182--191, 2014.

\bibitem{zheng2019framework}
X.~Zheng, Y.~Y. Tang, and J.~Zhou, ``A framework of adaptive multiscale wavelet
  decomposition for signals on undirected graphs,'' \emph{{IEEE} Trans. Signal
  Process.}, vol.~67, no.~7, pp. 1696--1711, 2019.

\bibitem{loukas2018spectrally}
A.~Loukas and P.~Vandergheynst, ``Spectrally approximating large graphs with
  smaller graphs,'' in \emph{Int. Conf. Mach. Learn.}, 2018, pp. 3237--3246.

\bibitem{loukas2019graph}
A.~Loukas, ``Graph reduction with spectral and cut guarantees,'' \emph{J. Mach.
  Learn. Res.}, vol.~20, no. 116, pp. 1--42, 2019.

\bibitem{jin2020graph}
Y.~Jin, A.~Loukas, and J.~JaJa, ``Graph coarsening with preserved spectral
  properties,'' in \emph{Int. Conf. Artificial Intell., Statist.}, 2020, pp.
  4452--4462.

\bibitem{thanou2014learning}
D.~Thanou, {D. I Shuman}, and P.~Frossard, ``Learning parametric dictionaries
  for signals on graphs,'' \emph{{IEEE} Trans. Signal Process.}, vol.~62,
  no.~15, pp. 3849--3862, 2014.

\bibitem{behjat2016signal}
H.~Behjat, U.~Richter, D.~Van De~Ville, and L.~S{\"o}rnmo, ``Signal-adapted
  tight frames on graphs,'' \emph{IEEE Trans. Signal Process.}, vol.~64,
  no.~22, pp. 6017--6029, 2016.

\bibitem{narang2013critically}
S.~K. Narang, Y.-H. Chao, and A.~Ortega, ``Critically sampled graph-based
  wavelet transforms for image coding,'' in \emph{{APSIPA} ASC}, 2013, pp.
  1--4.

\bibitem{Scarselli2009-GNN}
F.~Scarselli, M.~Gori, A.~C. Tsoi, M.~Hagenbuchner, and G.~Monfardini, ``The
  graph neural network model,'' \emph{{IEEE} Trans. Neural Netw.}, vol.~20,
  no.~1, pp. 61--80, 2009.

\bibitem{Vapnik91-ERM}
V.~N. Vapnik, ``Principles of risk minimization for learning theory,'' in
  \emph{Conf. Neural Inform. Process. Syst.}, 1991, pp. 831--838.

\bibitem{ying2018graph}
R.~Ying, R.~He, K.~Chen, P.~Eksombatchai, W.~L. Hamilton, and J.~Leskovec,
  ``Graph convolutional neural networks for web-scale recommender systems,'' in
  \emph{ACM Int. Conf. Knowl. Discov. Data Min.}, 2018, pp. 974--983.

\bibitem{stokes2020deep}
J.~M. Stokes, K.~Yang, K.~Swanson, W.~Jin, A.~Cubillos-Ruiz, N.~M. Donghia,
  C.~R. MacNair, S.~French, L.~A. Carfrae, Z.~Bloom-Ackermann \emph{et~al.},
  ``A deep learning approach to antibiotic discovery,'' \emph{Cell}, vol. 180,
  no.~4, pp. 688--702, 2020.

\bibitem{senior2020improved}
A.~W. Senior, R.~Evans, J.~Jumper, J.~Kirkpatrick, L.~Sifre, T.~Green, C.~Qin,
  A.~{\v{Z}}{\'\i}dek, A.~W. Nelson, A.~Bridgland \emph{et~al.}, ``Improved
  protein structure prediction using potentials from deep learning,''
  \emph{Nature}, vol. 577, no. 7792, pp. 706--710, 2020.

\bibitem{gainza2020deciphering}
P.~Gainza, F.~Sverrisson, F.~Monti, E.~Rodola, D.~Boscaini, M.~Bronstein, and
  B.~Correia, ``Deciphering interaction fingerprints from protein molecular
  surfaces using geometric deep learning,'' \emph{Nature Methods}, vol.~17,
  no.~2, pp. 184--192, 2020.

\bibitem{derrow2021eta}
A.~Derrow-Pinion, J.~She, D.~Wong, O.~Lange, T.~Hester, L.~Perez, M.~Nunkesser,
  S.~Lee, X.~Guo, B.~Wiltshire \emph{et~al.}, ``{ETA} prediction with graph
  neural networks in {Google} maps,'' in \emph{{ACM} Int. Conf. Inf. Knowl.
  Manag.}, 2021, pp. 3767--3776.

\bibitem{moura2020pooling}
M.~Cheung, J.~Shi, O.~Wright, L.~Y. Jiang, X.~Liu, and J.~M.~F. Moura, ``Graph
  signal processing and deep learning: Convolution, pooling, and topology,''
  \emph{{IEEE} Signal Process. Mag.}, vol.~37, no.~6, pp. 139--149, 2020.

\bibitem{defferrard2016chebnets}
M.~Defferrard, X.~Bresson, and P.~Vandergheynst, ``Convolutional neural
  networks on graphs with fast localized spectral filtering,'' in \emph{Conf.
  Neural Inform. Process. Syst.}, 2016, pp. 3844--3858.

\bibitem{Gama2019-Archit}
F.~Gama, A.~G.~Marques, G.~Leus, and A.~Ribeiro, ``{Convolutional Neural
  Network Architectures for Signals Supported on Graphs},'' \emph{{IEEE} Trans.
  Signal Process.}, vol.~67, no.~4, pp. 1034--1049, 2018.

\bibitem{Bruna2014-SpectralGNN}
J.~Bruna, W.~Zaremba, A.~Szlam, and Y.~LeCun, ``Spectral networks and deep
  locally connected networks on graphs,'' in \emph{Int. Conf. Learning
  Representations}, 2014, pp. 1--14.

\bibitem{he2022convolutional}
M.~He, Z.~Wei, and J.-R. Wen, ``Convolutional neural networks on graphs with
  {Chebyshev} approximation, revisited,'' \emph{Adv. Neural Inform. Process.
  Syst.}, vol.~35, pp. 7264--7276, 2022.

\bibitem{he2021bernnet}
M.~He, Z.~Wei, H.~Xu \emph{et~al.}, ``Bernnet: Learning arbitrary graph
  spectral filters via bernstein approximation,'' \emph{Advances in Neural
  Information Processing Systems}, vol.~34, pp. 14\,239--14\,251, 2021.

\bibitem{wang2022powerful}
X.~Wang and M.~Zhang, ``How powerful are spectral graph neural networks,'' in
  \emph{International Conference on Machine Learning}.\hskip 1em plus 0.5em
  minus 0.4em\relax PMLR, 2022, pp. 23\,341--23\,362.

\bibitem{Kipf2017-GCN}
T.~N. Kipf and M.~Welling, ``Semi-supervised classification with graph
  convolutional networks,'' in \emph{Int. Conf. Learning Representations},
  2017, pp. 1--14.

\bibitem{nt2019revisiting}
H.~Nt and T.~Maehara, ``Revisiting graph neural networks: All we have is
  low-pass filters,'' \emph{arXiv preprint arXiv:1905.09550}, 2019.

\bibitem{Chen2020-Oversmoothing}
D.~Chen, Y.~Lin, W.~Li, P.~Li, J.~Zhou, and X.~Sun, ``Measuring and relieving
  the over-smoothing problem for graph neural networks from the topological
  view,'' in \emph{{AAAI} Conf. Artif. Intell.}, vol.~34, no.~4, 2020, pp.
  3438--3445.

\bibitem{Weinberger2019-SGC}
F.~Wu, T.~Zhang, A.~H. de~Souza~Jr., C.~Fifty, T.~Yu, and K.~Q. Weinberger,
  ``Simplifying graph convolutional networks,'' in \emph{Int. Conf. Mach.
  Learning}, vol.~97, 2019, pp. 6861--6871.

\bibitem{Jegelka2019-GIN}
K.~Xu, W.~Hu, J.~Leskovec, and S.~Jegelka, ``How powerful are graph neural
  networks?'' in \emph{Int. Conf. Learning Representations}, 2019, pp. 1--17.

\bibitem{Hamilton2017-GraphSAGE}
W.~Hamilton, R.~Ying, and J.~Leskovec, ``Inductive representation learning on
  large graphs,'' in \emph{Conf. Neural Inform. Process. Syst.}, 2017, pp.
  1--11.

\bibitem{xu2018representation}
K.~Xu, C.~Li, Y.~Tian, T.~Sonobe, K.-i. Kawarabayashi, and S.~Jegelka,
  ``Representation learning on graphs with jumping knowledge networks,'' in
  \emph{Int. Conf. Mach. Learning}, 2018, pp. 5453--5462.

\bibitem{Velickovic2018-GAT}
P.~Veli{\v{c}}kovi{\'{c}}, G.~Cucurull, A.~Casanova, A.~Romero, P.~Li{\`{o}},
  and Y.~Bengio, ``Graph attention networks,'' in \emph{Int. Conf. Learning
  Representations}, 2018, pp. 1--12.

\bibitem{pim2021natural}
D.~Pim, T.~S. Cohen, and M.~Welling, ``Natural graph convolutions,'' 2021, {US
  Patent App. 17/239,580}.

\bibitem{Gama2019-Scattering}
F.~Gama, J.~Bruna, and A.~Ribeiro, ``Stability of graph scattering
  transforms,'' in \emph{Conf. Neural Inform. Process. Syst.}, 2019, pp.
  8038--8048.

\bibitem{Lerman2020-Scattering}
D.~Zou and G.~Lerman, ``Graph convolutional neural networks via scattering,''
  \emph{Appl. Comput. Harmonic Anal.}, vol.~49, no.~3, pp. 1046--1074, 2020.

\bibitem{Wolf2019-GeometricScattering}
F.~Gao, G.~Wolf, and M.~Hirn, ``Geometric scattering for graph data analysis,''
  in \emph{Int. Conf. Mach. Learning}, vol.~97, 2019, pp. 2122--2131.

\bibitem{ioannidis20scattering}
V.~N. Ioannidis, S.~Chen, and G.~B. Giannakis, ``Efficient and stable graph
  scattering transforms via pruning,'' \emph{{IEEE} Trans. Pattern Anal. Mach.
  Intell.}, vol.~44, no.~3, pp. 1232--1246, 2020.

\bibitem{wang2021speech}
T.~Wang, H.~Guo, X.~Yan, and Z.~Yang, ``Speech signal processing on graphs: The
  graph frequency analysis and an improved graph {Wiener} filtering method,''
  \emph{Speech Commun.}, vol. 127, pp. 82--91, 2021.

\bibitem{zheng2010graph}
M.~Zheng, J.~Bu, C.~Chen, C.~Wang, L.~Zhang, G.~Qiu, and D.~Cai, ``Graph
  regularized sparse coding for image representation,'' \emph{{IEEE} Trans.
  Image Process.}, vol.~20, no.~5, pp. 1327--1336, 2010.

\bibitem{ramamurthy2012learning}
K.~N. Ramamurthy, J.~J. Thiagarajan, P.~Sattigeri, and A.~Spanias, ``Learning
  dictionaries with graph embedding constraints,'' in \emph{Asilomar Conf.
  Signals, Systems and Comput.}, 2012, pp. 1974--1978.

\bibitem{yankelevsky2016dual}
Y.~Yankelevsky and M.~Elad, ``Dual graph regularized dictionary learning,''
  \emph{{IEEE} Trans. Signal Inform. Process. Networks}, vol.~2, no.~4, pp.
  611--624, 2016.

\bibitem{yankelevsky2019finding}
Y.~Yankelevsky and M.~Elad, ``Finding gems: Multi-scale dictionaries for
  high-dimensional graph signals,'' \emph{{IEEE} Trans. Signal Process.},
  vol.~67, no.~7, pp. 1889--1901, 2019.

\bibitem{thanou2018learning}
D.~Thanou and P.~Frossard, ``Learning of robust spectral graph dictionaries for
  distributed processing,'' \emph{EURASIP J. Adv. Signal Process}, vol. 2018,
  no.~1, pp. 1--17, 2018.

\bibitem{hu2016matched}
C.~Hu, J.~Sepulcre, K.~A. Johnson, G.~E. Fakhri, Y.~M. Lu, and Q.~Li, ``Matched
  signal detection on graphs: Theory and application to brain imaging data
  classification,'' \emph{NeuroImage}, vol. 125, pp. 587--600, 2016.

\bibitem{xiao2020anomalous}
Z.~Xiao, H.~Fang, and X.~Wang, ``Anomalous {IoT} sensor data detection: An
  efficient approach enabled by nonlinear frequency-domain graph analysis,''
  \emph{IEEE Internet Things J.}, vol.~8, no.~5, pp. 3812--3821, 2020.

\bibitem{xiao2020nonlinear}
Z.~Xiao, H.~Fang, and X.~Wang, ``Nonlinear polynomial graph filter for
  anomalous {IoT} sensor detection and localization,'' \emph{IEEE Internet
  Things J.}, vol.~7, no.~6, pp. 4839--4848, 2020.

\bibitem{ferrer2022volterra}
P.~Ferrer-Cid, J.~M. Barcelo-Ordinas, and J.~Garcia-Vidal, ``Volterra
  graph-based outlier detection for air pollution sensor networks,'' \emph{IEEE
  Trans. Netw. Sci. Eng.}, vol.~9, no.~4, pp. 2759--2771, 2022.

\bibitem{francisquini2022community}
R.~Francisquini, A.~C. Lorena, and M.~C. Nascimento, ``Community-based anomaly
  detection using spectral graph filtering,'' \emph{Appl. Soft Comput.}, vol.
  118, p. 108489, 2022.

\bibitem{egilmez2014spectral}
H.~E. Egilmez and A.~Ortega, ``Spectral anomaly detection using graph-based
  filtering for wireless sensor networks,'' in \emph{{IEEE} Int. Conf. Acoust.,
  Speech and Signal Process.}, 2014, pp. 1085--1089.

\bibitem{isufi2018blind}
E.~Isufi, A.~S. Mahabir, and G.~Leus, ``Blind graph topology change
  detection,'' \emph{{IEEE} Signal Process. Lett.}, vol.~25, no.~5, pp.
  655--659, 2018.

\bibitem{dempster_cov_selec}
A.~P. Dempster, ``Covariance selection,'' \emph{Biometrics}, vol.~28, no.~1,
  pp. 157--175, 1972.

\bibitem{meinshausen06}
N.~Meinshausen and P.~Buhlmann, ``High-dimensional graphs and variable
  selection with the {Lasso},'' \emph{Ann. Stat.}, vol.~34, pp. 1436--1462,
  2006.

\bibitem{DongLaplacianLearning}
X.~Dong, D.~Thanou, P.~Frossard, and P.~Vandergheynst, ``Learning {L}aplacian
  matrix in smooth graph signal representations,'' \emph{{IEEE} Trans. Signal
  Process.}, vol.~64, no.~23, pp. 6160--6173, 2016.

\bibitem{Kalofolias2016inference_smoothAISTATS16}
V.~Kalofolias, ``How to learn a graph from smooth signals,'' in \emph{Int.
  Conf. Artificial Intell., Statist.}\hskip 1em plus 0.5em minus 0.4em\relax J
  Mach. Learn. Res., 2016, pp. 920--929.

\bibitem{segarra2017topoidTSIPN}
S.~Segarra, A.~Marques, G.~Mateos, and A.~Ribeiro, ``Network topology inference
  from spectral templates,'' \emph{{IEEE} Trans. Signal Inform. Process.
  Networks}, vol.~3, no.~3, pp. 467--483, 2017.

\bibitem{shafipourdirectedTopoID2018}
R.~Shafipour, S.~Segarra, A.~G. Marques, and G.~Mateos, ``Directed network
  topology inference via graph filter identification,'' in \emph{{IEEE} Data
  Sci. Workshop}, 2018, pp. 210--214.

\bibitem{thanou17}
D.~Thanou, X.~Dong, D.~Kressner, and P.~Frossard, ``Learning heat diffusion
  graphs,'' \emph{{IEEE} Trans. Signal Inform. Process. Networks}, vol.~3,
  no.~3, pp. 484--499, 2017.

\bibitem{coutino2020state}
M.~Coutino, E.~Isufi, T.~Maehara, and G.~Leus, ``State-space network topology
  identification from partial observations,'' \emph{{IEEE} Trans. Signal
  Inform. Process. Networks}, vol.~6, pp. 211--225, 2020.

\bibitem{zhuconsensusinference2020}
Y.~Zhu, M.~T. Schaub, A.~Jadbabaie, and S.~Segarra, ``Network inference from
  consensus dynamics with unknown parameters,'' \emph{{IEEE} Trans. Signal
  Inform. Process. Networks}, vol.~6, pp. 300--315, 2020.

\bibitem{egilmezTopoID2019}
H.~E. Egilmez, E.~Pavez, and A.~Ortega, ``Graph learning from filtered signals:
  Graph system and diffusion kernel identification,'' \emph{{IEEE} Trans.
  Signal Inform. Process. Networks}, vol.~5, no.~2, pp. 360--374, 2019.

\bibitem{lezoray2012image}
O.~L{\'e}zoray and L.~Grady, \emph{Image Processing and Analysis with
  Graphs}.\hskip 1em plus 0.5em minus 0.4em\relax CRC Press Boca Raton, 2012.

\bibitem{cheung2018graph}
G.~Cheung, E.~Magli, Y.~Tanaka, and M.~K. Ng, ``Graph spectral image
  processing,'' \emph{Proc. IEEE}, vol. 106, no.~5, pp. 907--930, 2018.

\bibitem{liu2014progressive}
X.~Liu, D.~Zhai, D.~Zhao, G.~Zhai, and W.~Gao, ``Progressive image denoising
  through hybrid graph {Laplacian} regularization: A unified framework,''
  \emph{{IEEE} Trans. Image Process.}, vol.~23, no.~4, pp. 1491--1503, 2014.

\bibitem{pang2017graph}
J.~Pang and G.~Cheung, ``Graph {Laplacian} regularization for image denoising:
  Analysis in the continuous domain,'' \emph{{IEEE} Trans. Image Process.},
  vol.~26, no.~4, pp. 1770--1785, 2017.

\bibitem{elmoataz2008nonlocal}
A.~Elmoataz, O.~Lezoray, and S.~Bougleux, ``Nonlocal discrete regularization on
  weighted graphs: {A} framework for image and manifold processing,''
  \emph{{IEEE} Trans. Image Process.}, vol.~17, no.~7, pp. 1047--1060, 2008.

\bibitem{yaugan2016spectral}
A.~C. Ya{\u{g}}an and M.~T. {\"O}zgen, ``A spectral graph {Wiener} filter in
  graph {Fourier} domain for improved image denoising,'' in \emph{{IEEE} Global
  Conf. Signal and Inform. Process.}, 2016, pp. 450--454.

\bibitem{tian2014chebyshev}
D.~Tian, H.~Mansour, A.~Knyazev, and A.~Vetro, ``Chebyshev and conjugate
  gradient filters for graph image denoising,'' in \emph{IEEE Int. Conf.
  Multimed. Expo Workshops}, 2014, pp. 1--6.

\bibitem{gadde2013bilateral}
A.~Gadde, S.~K. Narang, and A.~Ortega, ``Bilateral filter: Graph spectral
  interpretation and extensions,'' in \emph{{IEEE} Int. Conf. Image Process.},
  2013, pp. 1222--1226.

\bibitem{onuki2016graph}
M.~Onuki, S.~Ono, M.~Yamagishi, and Y.~Tanaka, ``Graph signal denoising via
  trilateral filter on graph spectral domain,'' \emph{{IEEE} Trans. Signal
  Inform. Process. Networks}, vol.~2, no.~2, pp. 137--148, 2016.

\bibitem{knyazev2015accelerated}
A.~Knyazev and A.~Malyshev, ``Accelerated graph-based spectral polynomial
  filters,'' in \emph{IEEE Int. Workshop Mach. Learn. Signal Process.}, 2015,
  pp. 1--6.

\bibitem{huang2019fast}
Q.~Huang, R.~Li, Z.~Jiang, W.~Feng, S.~Lin, H.~Feng, and B.~Hu, ``Fast
  color-guided depth denoising for {RGB-D} images by graph filtering,'' in
  \emph{Asilomar Conf. Signals, Systems and Comput.}, 2019, pp. 1811--1815.

\bibitem{sadreazami2017data_a}
H.~Sadreazami, A.~Asif, and A.~Mohammadi, ``Data-driven image stylization using
  graph-based filtering,'' in \emph{Can. Conf. Electr. Comput. Eng}, 2017, pp.
  1--4.

\bibitem{sadreazami2017data}
H.~Sadreazami, A.~Asif, and A.~Mohammadi, ``Data-adaptive color image denoising
  and enhancement using graph-based filtering,'' in \emph{IEEE Int. Symp.
  Circuits Syst.}, 2017, pp. 1--4.

\bibitem{lu2022dct}
K.-S. Lu, A.~Ortega, D.~Mukherjee, and Y.~Chen, ``{DCT and DST} filtering with
  sparse graph operators,'' \emph{{IEEE} Trans. Signal Process.}, 2022.

\bibitem{salembier2018ship}
P.~Salembier, S.~Liesegang, and C.~L{\'o}pez-Mart{\'\i}nez, ``Ship detection in
  {SAR} images based on maxtree representation and graph signal processing,''
  \emph{IEEE Trans. Geosci. Remote Sens.}, vol.~57, no.~5, pp. 2709--2724,
  2018.

\bibitem{olfati2005consensus}
R.~Olfati-Saber and J.~S. Shamma, ``Consensus filters for sensor networks and
  distributed sensor fusion,'' in \emph{{IEEE} Conf. Decision Control}, 2005,
  pp. 6698--6703.

\bibitem{sandryhaila2014finite}
A.~Sandryhaila, S.~Kar, and J.~M. Moura, ``Finite-time distributed consensus
  through graph filters,'' in \emph{{IEEE} Int. Conf. Acoust., Speech and
  Signal Process.}, 2014, pp. 1080--1084.

\bibitem{safavi2014revisiting}
S.~Safavi and U.~A. Khan, ``Revisiting finite-time distributed algorithms via
  successive nulling of eigenvalues,'' \emph{{IEEE} Signal Process. Lett.},
  vol.~22, no.~1, pp. 54--57, 2014.

\bibitem{coutino2018limits}
M.~Coutino, E.~Isufi, T.~Maehara, and G.~Leus, ``On the limits of finite-time
  distributed consensus through successive local linear operations,'' in
  \emph{Asilomar Conf. Signals, Systems and Comput.}, 2018, pp. 993--997.

\bibitem{yi2019average}
J.-W. Yi, L.~Chai, and J.~Zhang, ``Average consensus by graph filtering: New
  approach, explicit convergence rate, and optimal design,'' \emph{IEEE Trans.
  Automat. Control.}, vol.~65, no.~1, pp. 191--206, 2019.

\bibitem{apers2016accelerating}
S.~Apers and A.~Sarlette, ``Accelerating consensus by spectral clustering and
  polynomial filters,'' \emph{IEEE Trans. Control Netw. Syst.}, vol.~4, no.~3,
  pp. 544--554, 2016.

\bibitem{ran2021fast}
Q.~Ran, J.-w. Yi, and L.~Chai, ``Fast consensus of multi-agent systems by
  second-order graph filters,'' in \emph{Chinese Conf. Control}, 2021, pp.
  5222--5227.

\bibitem{charalambous2018laplacian}
T.~Charalambous and C.~N. Hadjicostis, ``Laplacian-based matrix design for
  finite-time average consensus in digraphs,'' in \emph{{IEEE} Conf. Decision
  Control}, 2018, pp. 3654--3659.

\bibitem{li2021fast}
K.~Li, J.-W. Yi, and L.~Chai, ``Fast consensus of multi-agent systems on
  digraphs by graph filtering,'' in \emph{Chinese Conf. Control Decision},
  2021, pp. 5279--5284.

\bibitem{kruzick2018optimal}
S.~Kruzick and J.~F. Moura, ``Optimal filter design for consensus on random
  directed graphs,'' in \emph{IEEE Stat. Signal Process. Workshop}, 2018, pp.
  16--20.

\bibitem{ran2020group}
Q.~Ran, J.-W. Yi, L.~Chai, Y.-W. Wang, and X.~Chen, ``Group consensus of
  multi-agent systems by graph filtering,'' in \emph{{IEEE} Int. Conf. Control
  Autom.}, 2020, pp. 1290--1295.

\bibitem{chen2014signal}
S.~Chen, A.~Sandryhaila, J.~M. Moura, and J.~Kovacevic, ``Signal denoising on
  graphs via graph filtering,'' in \emph{{IEEE} Global Conf. Signal and Inform.
  Process.}, 2014, pp. 872--876.

\bibitem{emirov2020polynomial}
N.~Emirov, C.~Cheng, J.~Jiang, and Q.~Sun, ``Polynomial graph filter of
  multiple shifts and distributed implementation of inverse filtering,''
  \emph{arXiv preprint arXiv:2003.11152}, 2020.

\bibitem{romero2020fast}
D.~Romero, S.~Mollaebrahim, B.~Beferull-Lozano, and C.~Asensio-Marco, ``Fast
  graph filters for decentralized subspace projection,'' \emph{{IEEE} Trans.
  Signal Process.}, vol.~69, pp. 150--164, 2020.

\bibitem{isufi2017filtering}
E.~Isufi, A.~Loukas, A.~Simonetto, and G.~Leus, ``Filtering random graph
  processes over random time-varying graphs,'' \emph{{IEEE} Trans. Signal
  Process.}, vol.~65, no.~16, pp. 4406--4421, 2017.

\bibitem{gao2021stability}
Z.~Gao, E.~Isufi, and A.~Ribeiro, ``Stability of graph convolutional neural
  networks to stochastic perturbations,'' \emph{Signal Process.}, p. 108216,
  2021.

\bibitem{gao2021stochastic}
Z.~Gao, E.~Isufi, and A.~Ribeiro, ``Stochastic graph neural networks,''
  \emph{{IEEE} Trans. Signal Process.}, vol.~69, pp. 4428--4443, 2021.

\bibitem{saad2020accurate}
L.~B. Saad and B.~Beferull-Lozano, ``Accurate graph filtering in wireless
  sensor networks,'' \emph{IEEE Internet Things J.}, vol.~7, no.~12, pp.
  11\,431--11\,445, 2020.

\bibitem{gao2022learning}
Z.~Gao and E.~Isufi, ``Learning stochastic graph neural networks with
  constrained variance,'' \emph{{IEEE} Trans. Signal Process.}, vol.~71, pp.
  358--371, 2023.

\bibitem{baudet1978asynchronous}
G.~M. Baudet, ``Asynchronous iterative methods for multiprocessors,'' \emph{J.
  ACM}, vol.~25, no.~2, pp. 226--244, 1978.

\bibitem{teke2020iir}
O.~Teke and P.~P. Vaidyanathan, ``{IIR} filtering on graphs with random
  node-asynchronous updates,'' \emph{{IEEE} Trans. Signal Process.}, vol.~68,
  pp. 3945--3960, 2020.

\bibitem{teke2020node}
O.~Teke and P.~Vaidyanathan, ``Node-asynchronous implementation of filter banks
  on graphs,'' in \emph{Asilomar Conf. Signals, Systems and Comput.}, 2020, pp.
  460--464.

\bibitem{coutino2019asynchronous}
M.~Coutino and G.~Leus, ``Asynchronous distributed edge-variant graph
  filters,'' in \emph{{IEEE} Data Sci. Workshop}, 2019, pp. 115--119.

\bibitem{chamon2017finite}
L.~F. Chamon and A.~Ribeiro, ``Finite-precision effects on graph filters,'' in
  \emph{{IEEE} Global Conf. Signal and Inform. Process.}, 2017, pp. 603--607.

\bibitem{saad2021quantization}
L.~B. Saad, B.~Beferull-Lozano, and E.~Isufi, ``Quantization analysis and
  robust design for distributed graph filters,'' \emph{{IEEE} Trans. Signal
  Process.}, 2021.

\bibitem{nobre2019optimized}
I.~C.~M. Nobre and P.~Frossard, ``Optimized quantization in distributed graph
  signal processing,'' in \emph{{IEEE} Int. Conf. Acoust., Speech and Signal
  Process.}, 2019, pp. 5376--5380.

\bibitem{li2021task}
P.~Li, N.~Shlezinger, H.~Zhang, B.~Wang, and Y.~C. Eldar, ``Task-based graph
  signal compression,'' \emph{arXiv preprint arXiv:2110.12387}, 2021.

\bibitem{ding2021minimum}
K.~Ding, J.~Wu, and L.~Xie, ``Minimum-degree distributed graph filter design,''
  \emph{{IEEE} Trans. Signal Process.}, vol.~69, pp. 1083--1096, 2021.

\bibitem{elias2020adaptive}
V.~R. Elias, V.~C. Gogineni, W.~A. Martins, and S.~Werner, ``Adaptive graph
  filters in reproducing kernel hilbert spaces: Design and performance
  analysis,'' \emph{{IEEE} Trans. Signal Inform. Process. Networks}, vol.~7,
  pp. 62--74, 2020.

\bibitem{scardapane2020distributed}
S.~Scardapane, I.~Spinelli, and P.~Di~Lorenzo, ``Distributed training of graph
  convolutional networks,'' \emph{{IEEE} Trans. Signal Inform. Process.
  Networks}, vol.~7, pp. 87--100, 2020.

\bibitem{chen2014semi}
S.~Chen, F.~Cerda, P.~Rizzo, J.~Bielak, J.~H. Garrett, and
  J.~Kova{\v{c}}evi{\'c}, ``Semi-supervised multiresolution classification
  using adaptive graph filtering with application to indirect bridge structural
  health monitoring,'' \emph{{IEEE} Trans. Signal Process.}, vol.~62, no.~11,
  pp. 2879--2893, 2014.

\bibitem{fan2022graph}
J.~Fan, C.~Tepedelenlioglu, and A.~Spanias, ``Graph-based classification with
  multiple shift matrices,'' \emph{{IEEE} Trans. Signal Inform. Process.
  Networks}, 2022.

\bibitem{berberidis2018adaptive}
D.~Berberidis, A.~N. Nikolakopoulos, and G.~B. Giannakis, ``Adaptive diffusions
  for scalable learning over graphs,'' \emph{{IEEE} Trans. Signal Process.},
  vol.~67, no.~5, pp. 1307--1321, 2018.

\bibitem{song2022graph}
Z.~Song, X.~Yang, Z.~Xu, and I.~King, ``Graph-based semi-supervised learning: A
  comprehensive review,'' \emph{IEEE Transactions on Neural Networks and
  Learning Systems}, 2022.

\bibitem{tremblay2020approximating}
N.~Tremblay and A.~Loukas, ``Approximating spectral clustering via sampling:
  {A} review,'' \emph{Sampling Techniques for Supervised or Unsupervised
  Tasks}, pp. 129--183, 2020.

\bibitem{tremblay2016compressive}
N.~Tremblay, G.~Puy, R.~Gribonval, and P.~Vandergheynst, ``Compressive spectral
  clustering,'' in \emph{Int. Conf. Mach. Learning}, 2016, pp. 1002--1011.

\bibitem{boutsidis2015spectral}
C.~Boutsidis, P.~Kambadur, and A.~Gittens, ``Spectral clustering via the power
  method-provably,'' in \emph{Int. Conf. Mach. Learning}, 2015, pp. 40--48.

\bibitem{teke2020nodeSC}
O.~Teke and P.~P. Vaidyanathan, ``Node-asynchronous spectral clustering on
  directed graphs,'' in \emph{{IEEE} Int. Conf. Acoust., Speech and Signal
  Process.}, 2020, pp. 5325--5329.

\bibitem{wai_blind_2020}
H.-T. Wai, S.~Segarra, A.~E. Ozdaglar, A.~Scaglione, and A.~Jadbabaie, ``Blind
  community detection from low-rank excitations of a graph filter,''
  \emph{{IEEE} Trans. Signal Process.}, vol.~68, pp. 436--451, 2020.

\bibitem{roddenberry_blind_2020}
T.~M. Roddenberry, M.~T. Schaub, H.-T. Wai, and S.~Segarra, ``Exact blind
  community detection from signals on multiple graphs,'' \emph{{IEEE} Trans.
  Signal Process.}, vol.~68, pp. 5016--5030, 2020.

\bibitem{roddenberry_blind_centrality_2021}
T.~M. Roddenberry and S.~Segarra, ``Blind inference of eigenvector centrality
  rankings,'' \emph{{IEEE} Trans. Signal Process.}, vol.~69, pp. 3935--3946,
  2021.

\bibitem{he_blind_centrality_2022}
Y.~He and H.-T. Wai, ``Detecting central nodes from low-rank excited graph
  signals via structured factor analysis,'' \emph{{IEEE} Trans. Signal
  Process.}, vol.~70, pp. 2416--2430, 2022.

\bibitem{kaushik_change_point_2021}
C.~Kaushik, T.~M. Roddenberry, and S.~Segarra, ``Network topology change-point
  detection from graph signals with prior spectral signatures,'' in
  \emph{{IEEE} Int. Conf. Acoust., Speech and Signal Process.}, 2021, pp.
  5395--5399.

\bibitem{shaked_change_point_2021}
S.~Shaked and T.~Routtenberg, ``Identification of edge disconnections in
  networks based on graph filter outputs,'' \emph{{IEEE} Trans. Signal Inform.
  Process. Networks}, vol.~7, pp. 578--594, 2021.

\bibitem{marenco_change_point_2022}
B.~Marenco, P.~Bermolen, M.~Fiori, F.~Larroca, and G.~Mateos, ``Online change
  point detection for weighted and directed random dot product graphs,''
  \emph{{IEEE} Trans. Signal Inform. Process. Networks}, vol.~8, pp. 144--159,
  2022.

\bibitem{chen2022review}
Z.~Chen and S.~Wang, ``A review on matrix completion for recommender systems,''
  \emph{Knowl. Inf. Syst.}, pp. 1--34, 2022.

\bibitem{natarajan2014inductive}
N.~Natarajan and I.~S. Dhillon, ``Inductive matrix completion for predicting
  gene--disease associations,'' \emph{Bioinformatics}, vol.~30, no.~12, pp.
  i60--i68, 2014.

\bibitem{weng2012low}
Z.~Weng and X.~Wang, ``Low-rank matrix completion for array signal
  processing,'' in \emph{{IEEE} Int. Conf. Acoust., Speech and Signal
  Process.}, 2012, pp. 2697--2700.

\bibitem{lee2001survey}
S.-g. Lee and H.-g. Seol, ``A survey on the matrix completion problem,''
  \emph{Trends in Mathematics}, vol.~4, no.~1, pp. 38--43, 2001.

\bibitem{gao2021graph}
C.~Gao, Y.~Zheng, N.~Li, Y.~Li, Y.~Qin, J.~Piao, Y.~Quan, J.~Chang, D.~Jin,
  X.~He, and Y.~Li, ``A survey of graph neural networks for recommender
  systems: Challenges, methods, and directions,'' \emph{ACM Trans. Rec. Syst.},
  vol.~1, no.~1, pp. 1--51, 2023.

\bibitem{gu2010collaborative}
Q.~Gu, J.~Zhou, and C.~Ding, ``Collaborative filtering: Weighted nonnegative
  matrix factorization incorporating user and item graphs,'' in \emph{{SIAM}
  Int. Conf. Data Min.}, 2010, pp. 199--210.

\bibitem{du2011user}
L.~Du, X.~Li, and Y.-D. Shen, ``User graph regularized pairwise matrix
  factorization for item recommendation,'' in \emph{Int. Conf. Adv. Data Min.
  Appl.}, 2011, pp. 372--385.

\bibitem{kalofolias2014matrix}
V.~Kalofolias, X.~Bresson, M.~Bronstein, and P.~Vandergheynst, ``Matrix
  completion on graphs,'' \emph{arXiv preprint arXiv:1408.1717}, 2014.

\bibitem{rao2015collaborative}
N.~Rao, H.-F. Yu, P.~K. Ravikumar, and I.~S. Dhillon, ``Collaborative filtering
  with graph information: Consistency and scalable methods,'' \emph{Adv. Neural
  Inform. Process. Syst.}, vol.~28, 2015.

\bibitem{isufi2021accuracy}
E.~Isufi, M.~Pocchiari, and A.~Hanjalic, ``Accuracy-diversity trade-off in
  recommender systems via graph convolutions,'' \emph{Inf. Process. Manag.},
  vol.~58, no.~2, p. 102459, 2021.

\bibitem{nikolakopoulos2019personalized}
A.~N. Nikolakopoulos, D.~Berberidis, G.~Karypis, and G.~B. Giannakis,
  ``Personalized diffusions for top-n recommendation,'' in \emph{{ACM} Conf.
  Rec. Syst.}, 2019, pp. 260--268.

\bibitem{chen2020revisiting}
L.~Chen, L.~Wu, R.~Hong, K.~Zhang, and M.~Wang, ``Revisiting graph based
  collaborative filtering: A linear residual graph convolutional network
  approach,'' in \emph{{AAAI} Conf. Artif. Intell.}, vol.~34, no.~01, 2020, pp.
  27--34.

\bibitem{he2020lightgcn}
X.~He, K.~Deng, X.~Wang, Y.~Li, Y.~Zhang, and M.~Wang, ``{LightGCN}:
  Simplifying and powering graph convolution network for recommendation,'' in
  \emph{ACM/SIGIR Conf. Res. Dev. Inf. Retr.}, 2020, pp. 639--648.

\bibitem{shen2021powerful}
Y.~Shen, Y.~Wu, Y.~Zhang, C.~Shan, J.~Zhang, B.~K. Letaief, and D.~Li, ``How
  powerful is graph convolution for recommendation?'' in \emph{{ACM} Int. Conf.
  Inf. Knowl. Manag.}, 2021, pp. 1619--1629.

\bibitem{schulz2018tutorial}
E.~Schulz, M.~Speekenbrink, and A.~Krause, ``A tutorial on {Gaussian} process
  regression: {Modelling}, exploring, and exploiting functions,'' \emph{J.
  Math. Psychol.}, vol.~85, pp. 1--16, 2018.

\bibitem{wang2020intuitive}
J.~Wang, ``An intuitive tutorial to {Gaussian} processes regression,''
  \emph{arXiv preprint arXiv:2009.10862}, 2020.

\bibitem{venkitaraman2020gaussian}
A.~Venkitaraman, S.~Chatterjee, and P.~Handel, ``Gaussian processes over
  graphs,'' in \emph{{IEEE} Int. Conf. Acoust., Speech and Signal Process.},
  2020, pp. 5640--5644.

\bibitem{zhi2020gaussian}
Y.-C. Zhi, Y.~C. Ng, and X.~Dong, ``Gaussian processes on graphs via spectral
  kernel learning,'' \emph{IEEE Trans. Signal Inf. Process. Netw.}, vol.~9, pp.
  304--314, 2023.

\bibitem{borovitskiy2021matern}
V.~Borovitskiy, I.~Azangulov, A.~Terenin, P.~Mostowsky, M.~Deisenroth, and
  N.~Durrande, ``Mat{\'e}rn {Gaussian} processes on graphs,'' in \emph{Int.
  Conf. Artificial Intell., Statist.}, 2021, pp. 2593--2601.

\bibitem{lozes2015pde}
F.~Lozes, A.~Elmoataz, and O.~L{\'e}zoray, ``{PDE}-based graph signal
  processing for {3-D} color point clouds: {O}pportunities for cultural
  heritage,'' \emph{IEEE Signal Process. Mag.}, vol.~32, no.~4, pp. 103--111,
  2015.

\bibitem{chen20203d}
S.~Chen, B.~Liu, C.~Feng, C.~Vallespi-Gonzalez, and C.~Wellington, ``{3D} point
  cloud processing and learning for autonomous driving: {I}mpacting map
  creation, localization, and perception,'' \emph{IEEE Signal Process. Mag.},
  vol.~38, no.~1, pp. 68--86, 2020.

\bibitem{hu2021graph}
W.~Hu, J.~Pang, X.~Liu, D.~Tian, C.-W. Lin, and A.~Vetro, ``Graph signal
  processing for geometric data and beyond: {T}heory and applications,''
  \emph{{IEEE} Trans. Multimedia}, 2021.

\bibitem{jiao2022graph}
L.~Jiao, J.~Chen, F.~Liu, S.~Yang, C.~You, X.~Liu, L.~Li, and B.~Hou, ``Graph
  representation learning meets computer vision: {A} survey,'' \emph{IEEE
  Trans. Artificial Intell.}, 2022.

\bibitem{khasanova2017graph}
R.~Khasanova and P.~Frossard, ``Graph-based classification of omnidirectional
  images,'' in \emph{{IEEE} Int. Conf. Comput. Vision}, 2017, pp. 869--878.

\bibitem{cui2018spectral}
Z.~Cui, Y.~Cai, W.~Zheng, C.~Xu, and J.~Yang, ``Spectral filter tracking,''
  \emph{IEEE Trans. Image Process.}, vol.~28, no.~5, pp. 2479--2489, 2018.

\bibitem{gao2019graph}
J.~Gao, T.~Zhang, and C.~Xu, ``Graph convolutional tracking,'' in \emph{Conf.
  Comput. Vision and Pattern Recognition}, 2019, pp. 4649--4659.

\bibitem{weng2020gnn3dmot}
X.~Weng, Y.~Wang, Y.~Man, and K.~M. Kitani, ``{GNN3DMOT: Graph} neural network
  for {3D} multi-object tracking with {2D-3D} multi-feature learning,'' in
  \emph{Conf. Comput. Vision and Pattern Recognition}, 2020, pp. 6499--6508.

\bibitem{thanou2016graph}
D.~Thanou, P.~A. Chou, and P.~Frossard, ``Graph-based compression of dynamic
  {3D} point cloud sequences,'' \emph{IEEE Trans. Image Process.}, vol.~25,
  no.~4, pp. 1765--1778, 2016.

\bibitem{liang2020learning}
M.~Liang, B.~Yang, R.~Hu, Y.~Chen, R.~Liao, S.~Feng, and R.~Urtasun, ``Learning
  lane graph representations for motion forecasting,'' in \emph{Eur. Conf.
  Comput. Vision}, 2020, pp. 541--556.

\bibitem{wang2019linkage}
Z.~Wang, L.~Zheng, Y.~Li, and S.~Wang, ``Linkage based face clustering via
  graph convolution network,'' in \emph{Conf. Comput. Vision and Pattern
  Recognition}, 2019, pp. 1117--1125.

\bibitem{lassance2021improved}
C.~Lassance, Y.~Latif, R.~Garg, V.~Gripon, and I.~Reid, ``Improved visual
  localization via graph filtering,'' \emph{J. Imaging}, vol.~7, no.~2, p.~20,
  2021.

\bibitem{wang2019dynamic}
Y.~Wang, Y.~Sun, Z.~Liu, S.~E. Sarma, M.~M. Bronstein, and J.~M. Solomon,
  ``Dynamic graph {CNN} for learning on point clouds,'' \emph{ACM Trans.
  Graphics}, vol.~38, no.~5, pp. 1--12, 2019.

\bibitem{giraldo2021emerging}
J.~H. Giraldo, S.~Javed, M.~Sultana, S.~K. Jung, and T.~Bouwmans, ``The
  emerging field of graph signal processing for moving object segmentation,''
  in \emph{Int. Workshop Frontiers Comput. Vision}, 2021, pp. 31--45.

\bibitem{cai2019exploiting}
Y.~Cai, L.~Ge, J.~Liu, J.~Cai, T.-J. Cham, J.~Yuan, and N.~M. Thalmann,
  ``Exploiting spatial-temporal relationships for {3D} pose estimation via
  graph convolutional networks,'' in \emph{IEEE Conf. Comput. Vision}, 2019,
  pp. 2272--2281.

\bibitem{choi2020pose2mesh}
H.~Choi, G.~Moon, and K.~M. Lee, ``{Pose2Mesh: Graph} convolutional network for
  {3D} human pose and mesh recovery from a {2D} human pose,'' in \emph{Eur.
  Conf. Comput. Vision}, 2020, pp. 769--787.

\bibitem{liu20213d}
J.~Liu, Y.~Zhao, S.~Chen, and Y.~Zhang, ``A {3D} mesh-based
  lifting-and-projection network for human pose transfer,'' \emph{{IEEE} Trans.
  Multimedia}, 2021.

\bibitem{chen2017fast}
S.~Chen, D.~Tian, C.~Feng, A.~Vetro, and J.~Kova{\v{c}}evi{\'c}, ``Fast
  resampling of three-dimensional point clouds via graphs,'' \emph{IEEE Trans.
  Signal Process.}, vol.~66, no.~3, pp. 666--681, 2017.

\bibitem{dinesh2020point}
C.~Dinesh, G.~Cheung, and I.~V. Baji{\'c}, ``Point cloud denoising via feature
  graph {L}aplacian regularization,'' \emph{{IEEE} Trans. Image Process.},
  vol.~29, pp. 4143--4158, 2020.

\bibitem{bayram2017spectral}
E.~Bayram, ``Spectral graph based approach for analysis of {3D LIDAR} point
  clouds,'' Master's thesis, Middle East Technical University, 2017.

\bibitem{akyazi2018graph}
P.~Akyazi and P.~Frossard, ``Graph-based inpainting of disocclusion holes for
  zooming in {3D} scenes,'' in \emph{Eur. Signal Process. Conf.}, 2018, pp.
  867--871.

\bibitem{yamamoto2016deblurring}
K.~Yamamoto, M.~Onuki, and Y.~Tanaka, ``Deblurring of point cloud attributes in
  graph spectral domain,'' in \emph{Int. Conf. Image Process.}, 2016, pp.
  1559--1563.

\bibitem{dinesh20193d}
C.~Dinesh, G.~Cheung, and I.~V. Baji{\'c}, ``{3D} point cloud color denoising
  using convex graph-signal smoothness priors,'' in \emph{IEEE Int. Workshop
  Multimedia Signal Process.}, 2019, pp. 1--6.

\bibitem{perraudin2014gspbox}
N.~Perraudin, J.~Paratte, D.~Shuman, L.~Martin, V.~Kalofolias,
  P.~Vandergheynst, and D.~K. Hammond, ``{GSPBOX}: A toolbox for signal
  processing on graphs,'' \emph{arXiv preprint arXiv:1408.5781}, 2014.

\bibitem{girault2017grasp}
B.~Girault, S.~S. Narayanan, A.~Ortega, P.~Gon{\c{c}}alves, and E.~Fleury,
  ``Grasp: A matlab toolbox for graph signal processing,'' in \emph{2017 IEEE
  International Conference on Acoustics, Speech and Signal Processing
  (ICASSP)}.\hskip 1em plus 0.5em minus 0.4em\relax IEEE, 2017, pp. 6574--6575.

\bibitem{fey2019fast}
M.~Fey and J.~E. Lenssen, ``Fast graph representation learning with {PyTorch
  Geometric},'' \emph{arXiv preprint arXiv:1903.02428}, 2019.

\bibitem{isufi20162}
E.~Isufi, G.~Leus, and P.~Banelli, ``2-dimensional finite impulse response
  graph-temporal filters,'' in \emph{{IEEE} Global Conf. Signal and Inform.
  Process.}, 2016, pp. 405--409.

\bibitem{grassi2017time}
F.~Grassi, A.~Loukas, N.~Perraudin, and B.~Ricaud, ``A time-vertex signal
  processing framework: Scalable processing and meaningful representations for
  time-series on graphs,'' \emph{{IEEE} Trans. Signal Process.}, vol.~66,
  no.~3, pp. 817--829, 2017.

\bibitem{das2022graph}
B.~Das and E.~Isufi, ``Graph filtering over expanding graphs,'' in \emph{{IEEE}
  Data Sci. Learning Workshop}, 2022, pp. 1--8.

\bibitem{li2020federated}
T.~Li, A.~K. Sahu, A.~Talwalkar, and V.~Smith, ``Federated learning:
  Challenges, methods, and future directions,'' \emph{IEEE Signal Process.
  Mag.}, vol.~37, no.~3, pp. 50--60, 2020.

\bibitem{zhou2022bridging}
X.~Zhou, S.~Liu, W.~Xu, K.~Xin, Y.~Wu, and F.~Meng, ``Bridging hydraulics and
  graph signal processing: A new perspective to estimate water distribution
  network pressures,'' \emph{Water Res.}, vol. 217, p. 118416, 2022.

\bibitem{cardoso2020algorithms}
J.~V. d.~M. Cardoso, J.~Ying, and D.~P. Palomar, ``Algorithms for learning
  graphs in financial markets,'' \emph{arXiv preprint arXiv:2012.15410}, 2020.

\bibitem{bick2021higher}
C.~Bick, E.~Gross, H.~A. Harrington, and M.~T. Schaub, ``What are higher-order
  networks?'' \emph{arXiv preprint arXiv:2104.11329}, 2021.

\bibitem{stanley2020multiway}
J.~S. Stanley, E.~C. Chi, and G.~Mishne, ``Multiway graph signal processing on
  tensors: Integrative analysis of irregular geometries,'' \emph{{IEEE} Signal
  Process. Mag.}, vol.~37, no.~6, pp. 160--173, 2020.

\bibitem{barbarossa2020topological}
S.~Barbarossa and S.~Sardellitti, ``Topological signal processing: Making sense
  of data building on multiway relations,'' \emph{{IEEE} Signal Process. Mag.},
  vol.~37, no.~6, pp. 174--183, 2020.

\bibitem{schaub2021signal}
M.~T. Schaub, Y.~Zhu, J.-B. Seby, T.~M. Roddenberry, and S.~Segarra, ``Signal
  processing on higher-order networks: Livin'on the edge... and beyond,''
  \emph{Signal Process.}, vol. 187, p. 108149, 2021.

\bibitem{yang2022simplicial}
M.~Yang, E.~Isufi, M.~T. Schaub, and G.~Leus, ``Simplicial convolutional
  filters,'' \emph{arXiv preprint arXiv:2201.11720}, 2022.

\bibitem{roddenberry_icml_21}
T.~M. Roddenberry, N.~Glaze, and S.~Segarra, ``Principled simplicial neural
  networks for trajectory prediction,'' in \emph{Int. Conf. Mach. Learning},
  M.~Meila and T.~Zhang, Eds., vol. 139, 2021, pp. 9020--9029.

\bibitem{barbarossa2016introduction}
S.~Barbarossa and M.~Tsitsvero, ``An introduction to hypergraph signal
  processing,'' in \emph{{IEEE} Int. Conf. Acoust., Speech and Signal
  Process.}, 2016, pp. 6425--6429.

\bibitem{zhang2019introducing}
S.~Zhang, Z.~Ding, and S.~Cui, ``Introducing hypergraph signal processing:
  Theoretical foundation and practical applications,'' \emph{IEEE Internet
  Things J.}, vol.~7, no.~1, pp. 639--660, 2019.

\bibitem{leus2021topological}
G.~Leus, M.~Yang, M.~Coutino, and E.~Isufi, ``Topological {Volterra} filters,''
  in \emph{{IEEE} Int. Conf. Acoust., Speech and Signal Process.}, 2021, pp.
  5385--5399.

\end{thebibliography}

\end{document}